\documentclass[12pt]{article}

\usepackage{amsfonts,amsthm,amsmath,amssymb,upgreek,bm}
\usepackage{dsfont}
\usepackage[paper=letterpaper,margin=.85in]{geometry}
\usepackage{graphicx}
\usepackage{units}
\usepackage{float}
\usepackage[export]{adjustbox}
\usepackage{bbold}
\usepackage[shortlabels]{enumitem}

\usepackage{color}
\usepackage[dvipsnames]{xcolor}
\definecolor{darkblue}{rgb}{0.1,0.1,.7}
\definecolor{purple}{rgb}{0.6,0,0.6}
\definecolor{orange}{rgb}{0.9,0.6,0}
\usepackage[colorlinks, linkcolor=darkblue, citecolor=darkblue, urlcolor=darkblue, linktocpage]{hyperref} 
\usepackage{physics}
\usepackage{tensor}
\usepackage{psfrag}
\usepackage{footmisc}
\usepackage{url}
\usepackage{mathtools}
\usepackage{tikz}
\usepackage[]{graphicx}
\usepackage[]{latexsym}
\usepackage{geometry}
\usepackage{amscd}
\usepackage{subcaption}
\usepackage{cleveref}

\renewcommand{\ket}[1]{|{#1}\rangle}
\renewcommand{\bra}[1]{\langle{#1}|}

\usepackage{tikz}
\usepackage[compat=1.1.0]{tikz-feynman}
\usepackage{luatex85}
\tikzfeynmanset{warn luatex=false}
\usetikzlibrary{
    calc,
    positioning,
    decorations.pathmorphing,
    decorations.pathreplacing,
    arrows.meta
}

\tikzset{
    graviton photon/.style={
        draw,
        decorate,
        decoration={
            complete sines,
            amplitude=1mm,
            segment length=2mm
        }
    }
}

\tikzfeynmanset{
    graviton/.style={
        /tikz/draw=none,
        /tikz/decoration={name=none},
        /tikz/postaction={
            /tikz/decorate=true,
            /tikz/decoration={
                show path construction,
                moveto code={},
                lineto code={%
                    \draw[graviton photon]
                        ($(\tikzinputsegmentfirst)!1.2pt!90:(\tikzinputsegmentlast)$)
                        --
                        ($(\tikzinputsegmentlast)!1.2pt!-90:(\tikzinputsegmentfirst)$);%
                    \draw[graviton photon]
                        ($(\tikzinputsegmentfirst)!1.2pt!-90:(\tikzinputsegmentlast)$)
                        --
                        ($(\tikzinputsegmentlast)!1.2pt!90:(\tikzinputsegmentfirst)$);%
                }
            }
        }
    }
}

\numberwithin{equation}{section}

\begin{document}
\thispagestyle{empty}
\vspace*{2.5cm}
\begin{center}

{\LARGE\bfseries 
    Eikonal Scattering of Two D0-branes with Graviton Emission
}
\begin{center}
\vspace{1cm}
 {\bf Paolo Di Vecchia${}^{1,2}$, Dawid Maskalaniec${}^3$}\\
\bigskip 
\rm
\bigskip 
${}^1$The Niels Bohr Institute, Blegdamsvej 17, DK-2100 Copenhagen, Denmark
\rm
\\[0.2cm]
${}^2$NORDITA, KTH Royal Institute of Technology and Stockholm University, \\
Hannes Alfv{\'{e}}ns v{\"{a}}g 12, SE-11419 Stockholm, Sweden  
\rm
\\[0.2cm]
${}^3$ Leinweber Institute for Theoretical Physics and Department of Physics, 
\\
University of California, Berkeley, CA 94720, USA
\rm
\end{center}
\vspace{2.5cm}
{\bf Abstract}
\end{center}

\begin{quotation}
\noindent
We study the emission of a massless closed-string state in the scattering of two D0-branes using the boundary--state formalism. We compute the corresponding five--point amplitudes for graviton and dilaton emission in bosonic string theory and in type II superstring theory, and analyze their field--theory limit. For nonzero impact parameter, the resulting amplitudes are reproduced by an independent calculation in the worldline formulation of the corresponding low--energy theory.
\end{quotation}

\setcounter{page}{0}
\setcounter{tocdepth}{2}
\setcounter{footnote}{0}
\newpage

\newpage
\tableofcontents
\newpage

\section{Introduction}
\label{Introduction}
The waveform of gravitational waves emitted in the scattering of two spinless black holes can be computed from the five--point amplitude involving four scalar particles and one graviton. This quantity has been studied using several approaches~\cite{Goldberger:2016iau,Luna:2017dtq,Mogull:2020sak,DiVecchia:2020ymx}. In particular, in Ref.~\cite{DiVecchia:2020ymx}, the five--point amplitude with massive scalars was obtained by taking the field--theory limit of an amplitude involving Kaluza--Klein gravitons whose polarizations lie in the extra dimensions. In that setup, the scalar particles were perturbative string states.

In this paper, we instead consider non--perturbative string states, namely D0-branes, and study the emission of a graviton, or more generally a massless closed--string state, in the scattering of two D0-branes. The D0-branes are described using boundary states, while the emitted closed--string state is represented by the insertion of the corresponding vertex operator. This process was previously studied in Ref.~\cite{Hussain:1997hz}, extending earlier analyses of elastic D0-brane scattering in the open--string channel~\cite{Bachas:1995kx}\footnote{See also the recent review by Bachas~\cite{Bachas:2023ixh}.} and in the closed--string channel~\cite{Billo:1997eg}\footnote{See also Ref.~\cite{DiVecchia:1997vef} and the reviews~\cite{DiVecchia:1999mal,DiVecchia:1999fje}.}. We revisit this calculation and analyze its field--theory limit, comparing the result with an independent computation in the worldline quantum field theory formalism~\cite{Mogull:2020sak}.

We compute the five-point amplitudes involving four D0-branes and the emission of a graviton or a dilaton in both bosonic string theory and type-II superstring theory, and then analyze their field--theory limits. The emission amplitude for a Kalb-Ramond 2-form  vanishes identically. The field-theory limit of the bosonic-string amplitude agrees with that of the NS-NS sector of the superstring. For nonzero impact parameter, $b\neq 0$, the resulting amplitudes are reproduced by an independent calculation in NS-NS gravity using the worldline formalism. By keeping in the field theory limit of the string amplitude \eqref{CBC5} only the terms in \eqref{AppD_eq1}, we did not find agreement with the worldline calculation for the terms localized at $b=0$. Since additional terms in \eqref{CBC5} may contribute to these terms, we do not write them in this paper.

The paper is organized as follows. In Sect.~\ref{sec2} we present the string amplitudes and their final expressions, discussing the bosonic string in Subsect.~\ref{bosonic} and the superstring in Subsect.~\ref{NSNS}. In Sect.~\ref{Bosonic3} we derive the field-theory limit of the bosonic--string amplitude, which coincides with the field-theory limit of the NS--NS sector of the superstring. In Sect.~\ref{superstring} we derive the field--theory limit of the R--R sector of the superstring. In Sect.~\ref{sec:FieldTheory} we perform the corresponding field--theory calculation using the worldline formalism in both the NS--NS and R--R sectors, for the emission of either a graviton or a dilaton. Finally, Sect.~\ref{conclu} contains our conclusions and outlook.

The details of the calculations are collected in several appendices. In App.~\ref{AppA} we perform the complete bosonic--string calculation, while Apps.~\ref{AppB} and~\ref{AppC} contain the corresponding calculations in the NS-NS and R-R sectors of the superstring, respectively. In App.~\ref{AppD} we derive the field--theory limits of these string amplitudes: Subapp.~\ref{AppD1} treats the bosonic string, Subapp.~\ref{AppD2} the NS-NS sector of the superstring, and Subapp.~\ref{AppD3} the R-R sector. Finally, in App.~\ref{AppInte} we evaluate the field-theory limits of the various integrals needed in our calculations.

\section{String amplitudes}
\setcounter{equation}{0}
\label{sec2}
In this section, we present the string amplitudes describing the scattering of two D0-branes with the emission of a massless closed--string state, both in bosonic string theory and in type--II superstring theory. The detailed derivations are deferred to the appendices, while here we collect the final expressions. In Subsect.~\ref{bosonic} we discuss the bosonic string, and in Subsect.~\ref{NSNS} we extend the analysis to the superstring.

\subsection{Bosonic string amplitude}
\label{bosonic}

The scattering amplitude for eikonal scattering of two D0-branes with the emission of a graviton is given in the bosonic string theory by the following expression
\begin{equation}
    A_5=\frac{C_{S^2}}{\pi}  \left( \frac{\alpha' \kappa_D}{8\pi}\right)^2 \frac{\kappa_D}{2\pi} \frac{1}{2\pi}\int \frac{d^2w}{|w|^2} \int d^2z\,\langle  V_1, y_1| W^{\mu \nu}  (z, {\bar{z}}) w^{L_0-1} {\bar{w}}^{{\bar{L}}_0-1}| V_2, y_2\rangle. \label{ANN1}
\end{equation}
Here, $C_{S^2}=\frac{8\pi}{\alpha'} (\frac{\kappa_D}{2\pi})^{-2}$ denotes the normalization of closed--string amplitudes on the sphere (See Eq. (D.13) of \cite{DiVecchia:2023frv}). The factors $\alpha'\kappa_D/(8\pi)$ are associated with the two closed--string propagators, while $\kappa_D/(2\pi)$ is the normalization of the emitted closed--string vertex operator. The term $1/(2\pi)$ is the symmetry factor associated with the residual $U(1)$ symmetry of the annulus. There is no additional factor of $1/2$ associated with exchanging the two boundaries, since the two D0-branes have different velocities and their exchange is not a symmetry of the configuration.\footnote{Compared with the normalization used in Appendix~D of Ref.~\cite{DiVecchia:2023frv}, we include an additional overall factor of $1/\pi$. We find that this factor is required to reproduce the correctly normalized field--theory result from the string amplitude.}
The two integrals are computed in the intervals $0\leq |w|\leq 1, |w|\leq |z|\leq1$. The operators $L_0$ and $\bar L_0$ are the normally--ordered Virasoro zero modes of the free closed bosonic string. With creation and annihilation operators normalized as
\begin{equation}
[a_m^\mu,a_{-n}^\nu]=\eta^{\mu\nu}\delta_{m,n},\qquad
[\tilde a_m^\mu,\tilde a_{-n}^\nu]=\eta^{\mu\nu}\delta_{m,n},
\qquad m,n>0,
\end{equation}
they are given by
\begin{equation}\label{bosonic_virasoro_zero_modes}
    L_0=\frac{\alpha'}{4}\hat p^2+\sum_{n=1}^{\infty}n\,a_{-n}\cdot a_n, \qquad 
    \bar L_0=\frac{\alpha'}{4}\hat p^2+\sum_{n=1}^{\infty}n\,\tilde a_{-n}\cdot \tilde a_n .
\end{equation}
In the case of the bosonic string, the two boundary states describing the two stacks of D0-branes are given by
\begin{align}
    | V_2, y_2\rangle_{\text{B}} =&\; \frac{N_2 T_0}{2} \sqrt{1-V_2^2}\,\delta ({\hat{q}}^1 - V_2 {\hat{q}}^0 )\bigg[\prod_{i=2}^{D-1}\delta ( {\hat{q}}^i - y_2^i)\bigg] \nonumber 
    \\ 
    & \times \bigg[\prod_{k=2}^{D-1}  e^{\sum_{n=1}^{\infty} a_{n, k}^\dagger\, {\tilde{a}}_{n, k}^\dagger}\bigg]e^{-\sum_{n=1}^{\infty} a_n^\dagger \cdot M (V_2) \cdot {\tilde{a}}_n^\dagger}|0\rangle _{a} |0\rangle_{{\tilde{a}}} |p=0\rangle
\label{ANN2}
\end{align}
and 
\begin{align}
    {}_{\text{B}}\langle  V_1, y_1| = &\;\frac{N_1 T_0}{2}  \langle p=0| {}_a\langle 0| {}_{\tilde{a}}\langle 0|  \sqrt{1-V_1^2}\,\delta(q^1-V_1 q^0 )\bigg[\prod_{i=2}^{D-1}\delta ( {\hat{q}}^i-y_1^i)\bigg]\nonumber 
    \\
    & \times  \bigg[\prod_{k=2}^{D-1} e^{\sum_{n=1}^{\infty} a_{n, k}\, \tilde{a}_{n, k}} \bigg] e^{-\sum_{n=1}^{\infty} a_n \cdot M(V_1)\cdot  {\tilde{a}}_n}
\end{align}
where $N_1$ and $N_2$ are the numbers of coincident D0-branes participating in the scattering and $T_0$ is the tension of a single D0-brane. We assume that the branes move along the spatial direction $y^1$, while they are located at positions $y_1^i$ and $y_2^i$ along the remaining space directions $i=2,\dots,D-1$. The moving boundary state is obtained from the static one by performing a Lorentz boost, as described in \cite{Billo:1997eg}. It is convenient to parametrize velocities $V_1$, $V_2$ of the two D0-branes by $v_1$ and $v_2$, according to
\begin{align}
    V_{1,2}=\tanh(v_{1,2}).
    \label{A1b}
\end{align}
The matrices $M(V)$ appearing in the boundary states are
\begin{equation}
    M(V) = \left( \begin{array}{cc} \frac{1+V^2}{1-V^2} & \frac{2V}{1-V^2} \\ \frac{2V}{1-V^2} &\frac{1+V^2}{1-V^2} \end{array}\right)= \left( \begin{array}{cc}  \cosh (2v) & \sinh (2v) \\ \sinh(2v) & \cosh (2v)\end{array} \right),
\end{equation}
with $V=V_{1,2}$ and $v=v_{1,2}$.

The vertex operator corresponding to a massless closed-string state is given by
\begin{equation}
\epsilon_\mu  {\tilde{\epsilon}}_\nu W^{\mu \nu} (z, {\bar{z}}) = 
\frac{2i}{\sqrt{\alpha'}} \epsilon_\mu\partial_z X^\mu \frac{2i}{\sqrt{\alpha'}} {\tilde{\epsilon}}_\nu \partial_{{\bar{z}}} X^\nu e^{ik\cdot X},\qquad k^2=\epsilon\cdot k={\tilde{\epsilon}}\cdot k=0
\label{ANN3}
\end{equation}
Here, $\epsilon_{\mu}$, $\tilde{\epsilon}_{\nu}$ are polarization vectors corresponding to the polarization tensor $\epsilon_{\mu\nu} = \epsilon_{\mu}\tilde{\epsilon}_{\nu}$. The embedding coordinate of the closed string is given by
\begin{align}
 X^\mu (z, {\bar{z}}) =&\; \hat{q}^\mu - i\frac{\alpha'}{2} {\hat{p}}^\mu \log |z|^2+ i\sqrt{\frac{\alpha'}{2}}\sum_{n=1}^\infty \frac{1}{\sqrt{n}} \bigg( a_n^\mu z^{-n}- a_n^{\dagger \mu} z^n \bigg) 
\nonumber \\
&+ i \sqrt{\frac{\alpha'}{2}}\sum_{n=1}^\infty \frac{1}{\sqrt{n}} \bigg( {\tilde{a}}_n^\mu {\bar{z}}^{-n}- {\tilde{a}}_n^{\dagger \mu} {\bar{z}}^n \bigg).
\label{ANN4}
\end{align}
In order to simplify the calculation, it is convenient to exponentiate also the polarization-dependent terms and use the following expression for the vertex,
\begin{align}
    \tfrac{1}{2}W^{\mu \nu} (z, {\bar{z}}) =&\; e^{i k\cdot \hat{q} +\frac{\alpha'}{2}k \cdot \hat{p} \log |z|^2  + \frac{\sqrt{2\alpha'}}{2}\hat{p}\cdot (\frac{\epsilon}{z}+ \frac{\tilde{\epsilon}}{\bar{z}} )}\nonumber 
    \\
    & \times e^{-\frac{\sqrt{2\alpha'}}{2}k\cdot \sum_{n=1}^\infty \frac{1}{\sqrt{n}}(a_n z^{-n}- a_n^\dagger z^n)} e^{\epsilon\cdot \sum_{n=1}^\infty \sqrt{n}(a_n z^{-n-1}+ a_n^\dagger z^{n-1})}\nonumber \\
    & \times e^{-\frac{\sqrt{2\alpha'}}{2}k\cdot \sum_{n=1}^\infty \frac{1}{\sqrt{n}}({\tilde{a}}_n {\bar{z}}^{-n}- {\tilde{a}}_n^\dagger {\bar{z}}^n)} e^{{\tilde{\epsilon}}\cdot \sum_{n=1}^\infty \sqrt{n}({\tilde{a}}_n {\bar{z}}^{-n-1}+ {\tilde{a}}_n^\dagger {\bar{z}}^{n-1})}\label{ANN7}
\end{align}
The calculation is performed in App.~\ref{AppA}.
In particular, the contribution of the zero modes is computed in App.~\ref{AppA1}, while that of the nonzero modes is computed in App.~\ref{AppA2}.

The complete result for the bosonic string, concerning the emission of a massless closed--string state, is given by the product of the contribution of the zero modes in \eqref{BZMX3}
and the contribution of the nonzero modes  in \eqref{RSVE4}. It can be written as:
\begin{align}
    {A}_5 =&\;  \frac{\kappa_D }{2\pi} \frac{\alpha' N_1 N_2T_0^2}{8\pi|\sinh (v_{12})|}  \int \frac{d^{D-2}q}{(2\pi)^{D-2}}  e^{iq\cdot b} \int_0^1 \frac{d^2 w}{|w|^{4}}
    \int_{|w|}^{1}  d^2 z\, |z|^{\frac{\alpha'}{2} (r^2-q^2)} |w|^{\frac{\alpha'}{2} q^2} e^{\frac{\sqrt{2\alpha'}}{2}{q}(\frac{\epsilon}{z} +\frac{{\tilde{\epsilon}}}{{\bar{z}}})}\nonumber 
    \\ 
    &\times \prod_{n=1}^\infty (1 - 2|w|^{2n}\cosh(2v_{21}) + |w|^{4n})^{-1}(1 - |w|^{2n})^{4 - D + 2\alpha'\sinh^2(nv_{12})(k_0^2 - k_1^2)}\nonumber 
    \\
    &\times \prod_{n=0}^\infty (1-|z|^2|w|^{2n})^{-2\tilde{A}_{v_1 + nv_{12}}A_{v_1 + nv_{12}}-{\tilde{A}}\cdot {A}}(1-|\frac{w}{z}|^2|w|^{2n})^{-2\tilde{A}_{v_2 + nv_{21}}A_{v_2 + nv_{21}}-{\tilde{A}}\cdot{A}}\nonumber 
    \\
    &\times \prod_{n=1}^\infty e^{\sqrt{\frac{\alpha'}{2}} \frac{4|w|^{2n}}{1 - |w|^{2n}} \cosh(n v_{12})\left[\frac{1}{z}(\epsilon_{nv_{1}}k_{nv_{2}} - \epsilon_{nv_{2}}k_{nv_{1}}) + \frac{1}{\bar{z}}(\tilde{\epsilon}_{nv_{1}}k_{nv_{2}} - \tilde{\epsilon}_{nv_{2}}k_{nv_{1}})\right]}.
    \label{CBC5}
\end{align}
Here $r=q+k$ and $q\cdot b = \sum_{i=2}^{D-1}q^i b^i$, where $b^i = y_2^i - y_1^i$ is the impact parameter. We use the following shorthand: for any Lorentz vector $X^\mu$,
\begin{equation}
    X_v \equiv X^0\cosh v - X^1\sinh v .
\end{equation}
This convention applies, in particular, to $X=k,\epsilon,\tilde{\epsilon},A,\tilde{A}$. The differential operators $A$, $\tilde{A}$ appearing in \eqref{CBC5} are defined by
\begin{align}
    A = \sqrt{\frac{\alpha'}{2}}k + \epsilon\, \partial_z,\qquad \tilde{A} = \sqrt{\frac{\alpha'}{2}}k + \tilde{\epsilon}\,\partial_{\bar{z}} .
\end{align}
The five-point amplitude is obtained from  \eqref{CBC5}
by expanding the exponentials containing the polarizations $\epsilon$ and ${\tilde{\epsilon}}$ and keeping only the terms linear in the two polarizations. We will carry out this expansion only in the field-theory limit.

As a quick check of the normalization, we extract from the above result the field-theory limit of the elastic amplitude, with no graviton emission. The elastic amplitude follows immediately from the previous expression by removing the graviton vertex and hence all dependence on $z$. In this way, after performing the integration over $w$ and taking the $\alpha'\to0$ limit, we obtain
\begin{equation}
    A_{\rm elastic}= \frac{8\pi N_1 N_2 G_D m_0^2}{\sqrt{\gamma^2-1}}\frac{1}{q^2}\left(\frac{D-4}{4}+\gamma^2-\frac{1}{2}\right),
\end{equation}
where we have used Eq. \eqref{TGV4}. Here $\gamma = \cosh(v_{12})$. By Fourier--transforming this amplitude to the impact parameter space, we get
\begin{equation}
    {\tilde{A}}_{\rm elastic}= \frac{2\pi N_1N_2G_D m_0^2\Gamma (\frac{D-4}{2})(\pi b^2)^{\frac{4-D}{2}}}{\sqrt{\gamma^2-1}} \bigg( \frac{D-4}{4}+\gamma^2-\frac{1}{2}\bigg).
\end{equation}
In the ultrarelativistic limit, $\gamma\to \infty$, graviton exchange is expected to dominate over the dilaton contribution. We may therefore compare our result with the leading eikonal for two massive scalars in Einstein gravity, given by Eq.~(3.10) of Ref.~\cite{DiVecchia:2023frv}. The two expressions agree upon identifying $m_1=N_1m_0$ and $m_2=N_2m_0$.

\subsection{Superstring amplitude}
\label{NSNS}

In the superstring case, the boundary state also contains a fermionic part. In the NS-NS sector, this fermionic contribution is given by \cite{Billo:1997eg,DiVecchia:1999fje}
\begin{equation}
    |V, \eta\rangle_{\text{NS}} =\prod_{r=1/2}^\infty \exp\bigg[{i\eta \begin{pmatrix} b_r^0{}^\dagger & b_r^1{}^\dagger\end{pmatrix}M(v)SM(v)\begin{pmatrix} \tilde{b}_r^0{}^\dagger \\ \tilde{b}_r^1{}^\dagger \end{pmatrix}}\bigg]e^{-i\eta b_r^{\perp}{}^\dagger\cdot \tilde{b}_r^\perp{}^\dagger} \ket{0},
    \label{NS_bdry_state}
\end{equation}
where $S = \begin{pmatrix}
    -1 & 0 \\ 0 & -1
\end{pmatrix}$ in Lorentzian signature and $S = \begin{pmatrix}
    +1 & 0\\ 0 & -1
\end{pmatrix}$ in Euclidean signature. The superscript $\perp$ denotes the components transverse to the D0-brane worldvolume; for instance, $b_r^\perp$ is the vector with components $b_r^i$, where $i=2,\ldots,D-1$. The parameter $\eta = \pm 1$ specifies the spin structure. In the R-R sector, the boundary state for the nonzero fermionic modes is \cite{Billo:1997eg,DiVecchia:1999fje}
\begin{equation}
    |V, \eta\rangle_{\text{R}}= \prod_{n=1}^\infty \exp\bigg[{i\eta \begin{pmatrix}
    d_n^0{}^\dagger & d_n^1{}^\dagger\end{pmatrix}M(v)SM(v)\begin{pmatrix}
        \tilde{d}_n^0{}^\dagger \\ \tilde{d}_n^1{}^\dagger
    \end{pmatrix}}\bigg]e^{-i\eta d_n^{\perp}{}^\dagger\cdot \tilde{d}_n^\perp{}^\dagger} \ket{0},
    \label{RR_bdry_state}
\end{equation}
while the boundary state for fermionic zero modes is:
\begin{equation}\label{RR_bdry_state_zero_modes}
    \left(C(\cosh(v)\Gamma^0 + \sinh(v)\Gamma^1)\frac{1 + i\eta \Gamma_{11}}{1 + i\eta}\right)_{AB}\ket{A}\ket{\tilde{B}},
\end{equation}
where $\Gamma^{\mu}$ are the 10-dimensional gamma matrices, $\Gamma_{11} = \Gamma^0\Gamma^1\cdots \Gamma^{9}$, and $C$ is the charge-conjugation matrix. Indices $A,B$ run from $1$ to $32$.

The amplitude for the scattering of two D0-branes  with the emission of a graviton is
\begin{align}
    A(\eta_1, \eta_2) = &\;\frac{\kappa_D }{2\pi} \frac{\alpha' N_1 N_2T_0^2}{16\pi} \int \frac{d^2 w}{|w|^2}\int d^2 z  \nonumber 
    \\
    &\times {}_{\text{B}}\langle V_1, y_1|{}_{\text{F}}\langle V_1, \eta_1| W_{\epsilon}(z, {\bar{z}}) w^{L_0^B-1}{\bar{w}}^{{\bar{L}}_0^B-1}w^{L_0^F-a}{\bar{w}}^{{\bar{L}}_0^F-a}|V_2,y_2\rangle_{\text{B}} |V_2, \eta_2\rangle_{\text{F}}
\label{EHK6}
\end{align}
where $a=\frac{1}{2}$ in the NS sector and $a=0$ in the R sector and $\ket{V_{1,2},\eta}_{\text{F}}$ is the fermionic boundary state, given by \eqref{NS_bdry_state} in the NS sector and by the tensor product of \eqref{RR_bdry_state} and \eqref{RR_bdry_state_zero_modes} in the R sector. $L_0^F$, $\bar{L}_0^F$ are the Virasoro zero modes in the appropriate free fermion CFT. Bosonic operators $L_0^B$, $\bar{L}_0^B$ are defined in Eq.~\eqref{bosonic_virasoro_zero_modes}. The vertex operator is defined as
\begin{align}
    &W_{\epsilon}(z, {\bar{z}})=\epsilon_{\mu}\tilde{\epsilon}_{\nu}\frac{2i}{\sqrt{\alpha'}} \left(\partial_z X^\mu  + \frac{\alpha'}{2}(k\cdot \psi) \psi^\mu\right)\frac{2i}{\sqrt{\alpha'}} \left(  \partial_{{\bar{z}}}X^\nu+\frac{\alpha'}{2} (k\cdot {\bar{\psi}}){\bar{\psi}}^\nu\right)  e^{ik\cdot X}, 
    \label{vertexope}
\end{align}
where $k^2= \epsilon\cdot k={\tilde{\epsilon}}\cdot k=0.$ 

As in the case of bosonic string, it is convenient to move the polarization--dependent terms into exponents. In the superstring, however, we implement this step in a slightly different way. We introduce the Grassmann--valued vertex operators
\begin{align}
    &V(z; \theta, \phi) = \exp\left[ i\left(k + \sqrt{\tfrac{2}{\alpha'}}\phi\theta\,\epsilon \,\partial_z\right)\cdot X(z)\right]\exp\left[-\left(\sqrt{\tfrac{\alpha'}{2}}k\theta + \epsilon\phi\right)\cdot \psi\right],
    \\
    &\tilde{V} ({\bar{z}};{\tilde{\theta}}, {\tilde{\phi}})= \exp\left[i\left(k + \sqrt{\tfrac{2}{\alpha'}}{\tilde{\phi}}{\bar{\theta}}\,\tilde \epsilon\,\partial_{\bar{z}}\right)\cdot \bar{X}({\bar{z}})\right]\exp\left[-\left(\sqrt{\tfrac{\alpha'}{2}}k{\tilde{\theta}} + {\tilde{\epsilon}}{\tilde{\phi}}\right)\cdot {\tilde{\psi}}\right].
\end{align}
The original vertex operator $W_{\epsilon}(z,\bar z)$ is then replaced by the product
\begin{equation}
    W_{\epsilon}(z,\bar z)\longrightarrow 2V(z;\theta,\phi)\,\tilde{V}(\bar z;\tilde\theta,\tilde\phi).
\end{equation}
The physical vertex is recovered by extracting the term linear in both polarizations. We do this by integrating over the Grassmann variables in the following order: $\int d\theta \int d \phi \int d{\tilde{\theta}}\int d {\tilde{\phi}}$. In this way, we automatically select all terms linear in the two polarizations.

The fields $X(z)$ and $\bar{X}(\bar{z})$ are defined in Eq.~\eqref{ANN4}. The fermionic fields are taken to be dimensionless and are expanded as
\begin{equation}
    i\psi^\mu (z)= \sum_{r\in \mathbb{Z} + \nu}b_r^\mu z^{-r-1/2},\quad  i{\bar{\psi}}^\mu({\bar{z}})=\sum_{r\in\mathbb{Z} + \nu} \tilde{b}_r^\mu\bar{z}^{-r-1/2}.
\end{equation}
Here $b_{-r}^\mu = (b_{r}^\mu)^\dagger$, with $\nu = \frac{1}{2}$ in the NS-NS sector and $\nu = 0$ in the R-R sector. For $r,s\neq 0$, the modes $b_r$ and $\tilde b_r$ satisfy the anticommutation relations $\{b_{r}^\mu, (b_{s}^\nu)^\dagger\} = \eta^{\mu\nu}\delta_{rs}.$ The zero modes are Hermitian, $b_0^\mu = (b_0^\mu)^\dagger$, and generate the Clifford algebra, $\{b_0^\mu, b_0^\nu\} = \eta^{\mu\nu}$.

The complete calculation of \eqref{EHK6} is performed in Appendices \ref{AppB} in the NS-NS sector and \ref{AppC} in the R-R sector. The NS-NS sector amplitude is given by:
\begin{align} A_{\text{NS}}(\eta_1, \eta_2)=\;
    \nonumber &\frac{\kappa_D }{2\pi} \frac{\alpha' N_1 N_2T_0^2}{8\pi|\sinh v_{12}|} \int \frac{d^{D-2}q}{(2\pi)^{D-2}}e^{iq\cdot b} \int_0^1 \frac{d^2 w}{|w|^{3}}
    \int_{|w|}^{1}  d^2z \int d\theta \int d
    \phi \int d\tilde{\theta}
    \int  d\tilde{\phi}
    \\
    \nonumber &\times   |z|^{\frac{\alpha'}{2} (r^2-q^2)}|w|^{\frac{\alpha'}{2} q^2} e^{\frac{\sqrt{2\alpha'}}{2}{q}(\frac{\epsilon}{z}\phi\theta +\frac{{\tilde{\epsilon}}}{{\bar{z}}}\tilde{\phi}\tilde{\theta})} 
    \\
    \nonumber &\times \prod_{n=1}^\infty (1 - 2|w|^{2n}\cosh(2v_{21}) + |w|^{4n})^{-1}(1 - |w|^{2n})^{4 - D + 2\alpha'\sinh^2(nv_{12})(k_0^2 - k_1^2)}
    \\
    \nonumber &\times \prod_{n=0}^\infty (1-|z|^2|w|^{2n})^{-2\tilde{A}_{v_1 + nv_{12}}A_{v_1 + nv_{12}}-{\tilde{A}}\cdot {A}}
    \\
    \nonumber &\times \prod_{n=0}^\infty(1-|\frac{w}{z}|^2|w|^{2n})^{-2\tilde{A}_{v_2 + nv_{21}}A_{v_2 + nv_{21}}-{\tilde{A}}\cdot{A}}
    \\
    \nonumber &\times \prod_{r=1/2}^\infty (1+2\eta_1\eta_2|w|^{2r}\cosh(2v_{21}) + |w|^{4r})(1+\eta_1\eta_2|w|^{2r})^{D - 4}
    \\
    \nonumber &\times \prod_{n=0}^\infty e^{-i\eta_1(\tilde{R}\cdot R + 2\tilde{R}_{v_1 + nv_{12}}R_{v_1 + nv_{12}})\frac{(-\eta_1 \eta_2 |w|)^n}{1 - |z|^2|w|^{2n}}} 
    \\
    \nonumber &\times \prod_{n=0}^\infty e^{-\frac{i\eta_2|w|}{|z|^2}(\tilde{R}\cdot R + 2\tilde{R}_{v_2 + nv_{21}}R_{v_2 + nv_{21}})\frac{(-\eta_1 \eta_2 |w|)^n}{1 - |\frac{w}{z}|^2|w|^{2n}}}
    \\
    &\times \prod_{n=1}^\infty
    e^{-4\cosh(nv_{12})\sqrt{\frac{\alpha'}{2}}
    \frac{(-\eta_1 \eta_2|w|)^n}{1 - |w|^{2n}}
    \left[\frac{1}{z}(k_{nv_1}\epsilon_{nv_2}-k_{nv_2}\epsilon_{nv_1})\,
    \theta\phi + \frac{1}{\bar z}(k_{nv_1}\tilde{\epsilon}_{nv_2}-k_{nv_2}\tilde{\epsilon}_{nv_1})\,
    \tilde{\theta}\tilde{\phi} \right]},
    \label{NSNSFINAL}
\end{align}
and the R-R sector amplitude is
\begin{align}
    \nonumber  A_{\text{R}}(\eta_1, \eta_2)=&\;\frac{\kappa_D }{2\pi} \frac{\alpha' N_1 N_2T_0^2}{8\pi|\sinh (v_{12})|} \int \frac{d^{D-2}q}{(2\pi)^{D-2}}  e^{iq\cdot b} \int_0^1 \frac{d^2w}{|w|^{2}} \int_{|w|}^{1}  d^2 z \int d\theta  \int d \phi \int d\tilde{\theta}\int  d\tilde{\phi}
    \\
    \nonumber &\times |z|^{\frac{\alpha'}{2} (r^2-q^2)}|w|^{\frac{\alpha'}{2} q^2} e^{\sqrt{\frac{\alpha'}{2}}q\cdot (\frac{\epsilon}{z}\phi\theta +\frac{{\tilde{\epsilon}}}{{\bar{z}}}\tilde{\phi}\tilde{\theta})} 
    \\
    \nonumber &\times \prod_{n=1}^\infty (1 - 2|w|^{2n}\cosh(2v_{21}) + |w|^{4n})^{-1}(1 - |w|^{2n})^{2 - D+2\alpha'\sinh^2(nv_{12})(k_0^2 - k_1^2)}
    \\
    \nonumber &\times \prod_{n=0}^\infty (1-|z|^2|w|^{2n})^{-2\tilde{A}_{v_1 + nv_{12}}A_{v_1 + nv_{12}}-{\tilde{A}}\cdot {A}}
    \\
    \nonumber &\times \prod_{n=0}^\infty(1-|\frac{w}{z}|^2|w|^{2n})^{-2\tilde{A}_{v_2 + nv_{21}}A_{v_2 + nv_{21}}-{\tilde{A}}\cdot{A}}
    \\
    \nonumber & \times \prod_{n=1}^\infty (1+2\eta_1\eta_2|w|^{2n}\cosh(2v_{21}) + |w|^{4n})(1+\eta_1\eta_2|w|^{2n})^{D - 4}
    \\
    \nonumber &\times \prod_{n=0}^\infty e^{-i\eta_1|z|(-2\tilde{R}_{v_1 + nv_{12}}R_{v_1 + nv_{12}} + \tilde{R}\cdot R)\frac{(-\eta_1 \eta_2 |w|^2)^n}{1 - |z|^2|w|^{2n}}} 
    \\
    \nonumber &\times \prod_{n=0}^\infty e^{- \frac{i\eta_2|w|^2}{|z|^3}(-2\tilde{R}_{v_2 + nv_{21}}R_{v_2 + nv_{21}} + \tilde{R}\cdot R)\frac{(-\eta_1 \eta_2 |w|^2)^n}{1 - |\frac{w}{z}|^2|w|^{2n}}}
    \\
    \nonumber &\times \prod_{n=1}^\infty
    e^{-4\cosh(nv_{12})\sqrt{\frac{\alpha'}{2}}
    \frac{(-\eta_1 \eta_2|w|^2)^n}{1 - |w|^{2n}}
    \left[\frac{1}{z}(k_{nv_1}\epsilon_{nv_2}-k_{nv_2}\epsilon_{nv_1})\,
    \theta\phi + \frac{1}{\bar z}(k_{nv_1}\tilde{\epsilon}_{nv_2}-k_{nv_2}\tilde{\epsilon}_{nv_1})\,
    \tilde{\theta}\tilde{\phi} \right]}
    \\
    \nonumber&\times 8(\eta_1 \eta_2 - 1)\Big[\cosh(v_{12}) + \frac{i\eta_1}{2}(\cosh(v_{12})(\epsilon\cdot \tilde{\epsilon})\phi\tilde{\phi} + R_{v_1}\tilde{R}_{v_2} + R_{v_2}\tilde{R}_{v_1})
    \\
    \nonumber& \qquad \qquad \qquad +\frac{1}{2}\sqrt{\frac{\alpha'}{2}}\left(\frac{1}{z}(k_{v_1} \epsilon_{v_2} - k_{v_2} \epsilon_{v_1})\theta\phi + \frac{1}{\bar{z}}(k_{v_1} \tilde{\epsilon}_{v_2} - k_{v_2} \tilde{\epsilon}_{v_1})\tilde{\theta}\tilde{\phi}\right)
    \\
    & \qquad \qquad \qquad + \frac{\alpha'}{4|z|^2} k_{v_1}k_{v_2}(\tilde{\epsilon}\cdot \epsilon)\theta\phi\tilde{\theta}\tilde{\phi}\Big],
    \label{RRFINAL}
\end{align}
where
\begin{align}
R= \sqrt{\frac{\alpha'}{2}}k \theta+\epsilon \phi,\quad {\tilde{R}}= \sqrt{\frac{\alpha'}{2}}k {\tilde{\theta}}+{\tilde{\epsilon}} {\tilde{\phi}}.
\end{align}
The final amplitude is obtained by first performing the GSO projection separately in the NS-NS and R-R sectors, and then summing the contributions from the two sectors. In each sector, the GSO projection amounts to summing over spin structures according to
\begin{equation}
    A = \frac{1}{4}\left[A(++)+ A(--) + sA(+-) + sA(-+)\right],
\end{equation}
with $s= -1$ in the NS-NS sector and $s=+1$ in the R-R sector of type IIA theory. We carry out this procedure only in the field-theory limit.

\section{Field--theory limit of the bosonic string amplitude}
\label{Bosonic3}
In this section, we discuss the field-theory limit, $\alpha'\to 0$, of the five-point amplitude in the bosonic string. The complete string calculation was carried out in Appendix~\ref{AppA}, leading to Eq.~\eqref{CBC5}. The field-theory limit is then analyzed in Appendix~\ref{AppD1}. As can be seen from Eq.~\eqref{CBC5}, the string amplitude contains the open-string partition function for strings stretched between the two branes, $|w|^{-2}\prod_{n=1}^\infty(1 - 2|w|^{2n}\cosh(2v_{12}) + |w|^{4n})^{-1} (1-|w|^{2n})^{4-D}$. The field-theory limit corresponds to the degeneration in which the string worldsheet becomes a long tube between the branes, so that it approaches the worldline of a particle exchanged between them. This corresponds to the $|w|\to 0$ limit of the partition function:
\begin{align}
    &\frac{1}{|w|^2}\prod_{n=1}^\infty\frac{1}{(1-2 |w|^{2n} \cosh(2v_{12})+ |w|^{4n})(1-|w|^{2n})^{D-4}}\nonumber 
    \\
    & \sim \frac{1}{|w|^2}+ (D-4+2\cosh (2v_{12}))+{\cal{O}}(|w|^2).
    \label{expa2}
\end{align}
The first term originates from the tachyon in the closed-string spectrum, while the second term corresponds to the massless closed-string states. Higher-order terms correspond to massive string excitations and do not contribute in the field-theory limit. Following the procedure used in \cite{DiVecchia:1996uq} (See sections~4 and 5), in order to compute the field--theory limit, we should consider both the contribution of the tachyon and that of the massless states.  

The vertex insertion is integrated over the region $|z|\in (|w|,1)$, so the expansion must be performed with both $x=|z|^2$ and $y=|w/z|^2$ kept finite. The region near $x=0$, $y=1$ corresponds to emission directly from the first D-particle, while the region near $x=1$, $y=0$ corresponds to emission from the second D-particle. The region where $x\to0$ and $y\to0$ corresponds to emission from the intermediate particle exchanged between the branes, far from both worldvolumes. 

There is, however, an important caveat. From the calculation in Appendix~\ref{AppD1} one can see that the $\alpha'\to0$ limit also receives contributions from the region $x\to1$, $y\to1$. In this region, $|w|^2=xy\to1$, so the usual suppression of massive closed-string states is lost, suggesting that an infinite tower of massive states may contribute at the same order to \eqref{expa2}. In momentum space, these contributions are independent of the transferred momentum $q$ and therefore correspond to terms proportional to $\delta(b)$ in impact-parameter space. By retaining in the field-theory limit only the terms in Eq.~\eqref{AppD_eq1}, we do not find agreement with the $q$-independent contributions obtained from the worldline calculation. We have not analyzed the remaining contributions from the region $x\to1$, $y\to1$ systematically enough to determine their precise field-theory limit. A more careful treatment of this integration region is required to determine whether these additional terms reproduce the corresponding $q$-independent terms of the worldline calculation. Note, however, that agreement with the worldline calculation is not necessarily expected, since this region may be sensitive to short-distance physics in which additional open-string degrees of freedom become relevant~\cite{Shenker:1995xq,Danielsson:1996uw,Kabat:1996cu,Douglas:1996yp}.

For this reason, in this paper, we exclude these terms from our field-theory limit and leave a more detailed investigation of this issue for future work. In what follows, we restrict to $b\neq0$ and, correspondingly, drop all terms in the momentum-space amplitudes that are independent of $q.$\footnote{The final result takes the form of a Fourier transform over the transferred momentum $q$, which naturally suggests interpreting the integrand as a momentum-space D0-brane scattering amplitude. From this perspective, the $q$-independent terms would give finite constant contributions and should not be discarded. The identification of the integrand with the full momentum-space amplitude is, however, not immediate, since the D0-branes are non-perturbative, infinitely heavy objects following prescribed trajectories at fixed impact parameter, rather than ordinary asymptotic momentum eigenstates in perturbative string theory.} Following this procedure, we have obtained the two contributions in Eqs.~\eqref{FinalMassless8} and \eqref{Ftachyon2}, which we rewrite here as
\begin{align}
    A_{5}^{\mathrm{massless}}
    =&\;\frac{\kappa_D}{2\pi}\frac{\pi N_1N_2T_0^2}{8\lvert\sinh(v_{12})\rvert}(D-4+2\cosh(2v_{12}))\int \frac{\mathrm{d}^{D-2}q}{(2\pi)^{D-2}} e^{i  q\cdot b} \nonumber
    \\
    &\times\Bigg\{\frac{8(q\cdot\epsilon)(q\cdot\tilde{\epsilon})}{q^2r^2} - 4(\epsilon\cdot \tilde{\epsilon})\left(\frac1{q^2}+\frac1{r^2}\right) \nonumber
    \\
    &\qquad\;-\frac4{k_{v_1}^2q^2}\Bigl(k_{v_1}\epsilon_{v_1}(q\cdot\tilde{\epsilon})+k_{v_1}\tilde{\epsilon}_{v_1}(q\cdot\epsilon)-(\tilde{\epsilon}_{v_1}\epsilon_{v_1})(k\cdot q)\Bigr)\nonumber 
    \\
    &\qquad\;+\frac4{k_{v_2}^2r^2}\Bigl(k_{v_2}\epsilon_{v_2}(q\cdot\tilde{\epsilon})+k_{v_2}\tilde{\epsilon}_{v_2}(q\cdot\epsilon)-(\tilde{\epsilon}_{v_2}\epsilon_{v_2})(k\cdot q)\Bigr) \nonumber
    \\
    &\qquad\; +(\epsilon\cdot {\tilde{\epsilon}})\bigg(\frac{r^2}{k_{v_1}^2 q^2}+\frac{q^2}{k_{v_2}^2 r^2}\bigg) \Bigg\},
    \label{Fmassless1}
\end{align}
and 
\begin{align}
    A_{5}^{\mathrm{tachyonic}}
    =&\frac{\kappa_D}{2\pi}\frac{\pi N_1N_2T_0^2}{8\lvert\sinh(v_{12})\rvert}\int \frac{\mathrm{d}^{D-2}q}{(2\pi)^{D-2}} e^{i  q\cdot b} \nonumber 
    \\
    &\times \bigg\{ \frac{32k_{v_1}^2 k_{v_2}^2}{q^2 r^2} \bigg( \frac{\epsilon_{v_1}}{k_{v_1}}-\frac{\epsilon_{v_2}}{k_{v_2}}\bigg)\bigg( \frac{{\tilde{\epsilon}}_{v_1}}{k_{v_1}}-\frac{{\tilde{\epsilon}}_{v_2}}{k_{v_2}}\bigg)\nonumber 
    \\
    &\qquad + \frac{32 \cosh(v_{12}) k_{v_1} k_{v_2}}{q^2 r^2}\bigg[ \bigg( \frac{\epsilon_{v_1}}{k_{v_1}}-\frac{\epsilon_{v_2}}{k_{v_2}}\bigg)(q\cdot {\tilde{\epsilon}})+  \bigg( \frac{{\tilde{\epsilon}}_{v_1}}{k_{v_1}}-\frac{{\tilde{\epsilon}}_{v_2}}{k_{v_2}}\bigg)(q\cdot {\epsilon})\bigg]\nonumber 
    \\
    &\qquad -\frac{16 \cosh(v_{12}) k_{v_1}k_{v_2}}{q^2}\bigg[ \bigg( \frac{\epsilon_{v_1}}{k_{v_1}}-\frac{\epsilon_{v_2}}{k_{v_2}}\bigg)\frac{ {\tilde{\epsilon}}_{v_1}}{k_{v_1}}+ \bigg( \frac{{\tilde{\epsilon}}_{v_1}}{k_{v_1}}-\frac{{\tilde{\epsilon}}_{v_2}}{k_{v_2}}\bigg)\frac{ {{\epsilon}}_{v_1}}{k_{v_1}}\bigg] \nonumber 
    \\
    &\qquad +\frac{16 \cosh(v_{12}) k_{v_1}k_{v_2}}{r^2}\bigg[ \bigg( \frac{\epsilon_{v_1}}{k_{v_1}}-\frac{\epsilon_{v_2}}{k_{v_2}}\bigg)\frac{ {\tilde{\epsilon}}_{v_2}}{k_{v_2}}+ \bigg( \frac{{\tilde{\epsilon}}_{v_1}}{k_{v_1}}-\frac{{\tilde{\epsilon}}_{v_2}}{k_{v_2}}\bigg)\frac{ {{\epsilon}}_{v_2}}{k_{v_2}}\bigg]\nonumber 
    \\
    &\qquad + 16(\epsilon\cdot {\tilde{\epsilon}})\bigg(\frac{k_{v_1}^2+k_{v_2}^2}{q^2 r^2}+ \frac{\cosh(v_{12})(\cosh(v_{12}) k_{v_1} - k_{v_2})}{k_{v_1}q^2}\nonumber 
    \\
    &\qquad \qquad \qquad\qquad \qquad \quad \;\, +\frac{\cosh(v_{12}) (\cosh(v_{12}) k_{v_2} - k_{v_2})}{k_{v_2}r^2}\bigg)\bigg\}
    \label{Ftachyon1}
\end{align}
We now introduce the velocities of the two incoming D-particles participating in the scattering,
\begin{equation}
    u_1^\mu = (\cosh v_{1}, \sinh v_1, 0_\perp),\quad u_2^\mu  = (\cosh v_2, \sinh v_2, 0_\perp)
\end{equation}
and define the relative boost factor
\begin{equation}
    \gamma = - u_1 \cdot u_2=\cosh (v_{12}).
\end{equation}
With these conventions, we have
\begin{align}
    k_{v_1}= - u_1\cdot k,\quad k_{v_2}= - u_2\cdot k,\quad \epsilon_{v_1}= -u_1\cdot\epsilon,\quad \epsilon_{v_2}=- \epsilon\cdot u_2.
\end{align}
It is also convenient to define the following two vectors
\begin{equation}\label{P12Q12definitions}
    P_{12}^\mu= (k\cdot u_1)u_2^\mu - (k\cdot u_2)u_1^\mu,\quad  Q_{12}^\mu=2q^\mu -\frac{r^2 }{u_1\cdot k}u_1^\mu + \frac{q^2}{u_2\cdot k}u_2^\mu.
\end{equation}
They are gauge invariant in the sense that
\begin{equation}
    k_\mu P_{12}^\mu=k_\mu Q_{12}^\mu=0. \label{gaugeinv}
\end{equation}
In terms of these quantities, Eqs.~\eqref{Fmassless1} and \eqref{Ftachyon1} can be written in a more compact form:
\begin{align}
    A_{5}^{\mathrm{massless}}
    =&\;\frac{\kappa_D^3N_1N_2m_0^2}{4\sqrt{\gamma^2 - 1}}\left(\frac{D-6}{4}+\gamma^2\right)\int \frac{\mathrm{d}^{D-2}q}{(2\pi)^{D-2}} e^{i  q\cdot b} \nonumber
    \\
    &\times \bigg\{\frac{2 (\epsilon\cdot Q_{12})({\tilde{\epsilon}}\cdot Q_{12})}{q^2 r^2}+ (\epsilon \cdot{\tilde{\epsilon}})\bigg(\frac{r^2}{k_{v_1}^2 q^2}+\frac{q^2}{k_{v_2}^2 r^2}-\frac{4}{q^2}-\frac{4}{r^2}\bigg)\bigg\},
    \label{A5massless1}
\end{align}
where we have used the relation $\cosh (2 v_{12})=2\gamma^2-1$.

The tachyonic contribution becomes 
\begin{align}
    A_{5}^{\mathrm{tachyonic}} =&\;\frac{\kappa_D^3N_1N_2m_0^2}{16\sqrt{\gamma^2 - 1}}\int \frac{\mathrm{d}^{D-2}q}{(2\pi)^{D-2}} e^{i  q\cdot b}\nonumber 
    \\
    &\times \bigg\{ \frac{16 [2(\epsilon\cdot P_{12})(\tilde{\epsilon}\cdot P_{12}) - \gamma (\epsilon\cdot P_{12})(\tilde{\epsilon}\cdot Q_{12})-\gamma (\tilde\epsilon\cdot P_{12})({\epsilon}\cdot Q_{12})]}{q^2 r^2} \nonumber 
    \\
    &\qquad  + 16(\epsilon\cdot {\tilde{\epsilon}})\bigg(\frac{k_{v_1}^2+k_{v_2}^2}{q^2 r^2}+ \frac{\gamma(\gamma k_{v_1} - k_{v_2})}{k_{v_1}q^2} +\frac{\gamma (\gamma k_{v_2} - k_{v_1})}{k_{v_2}r^2}\bigg)\bigg\}.
    \label{A5tach1}
\end{align}
The final result for the bosonic string is obtained by adding the two contributions:
\begin{align}
    A_{5}
    =&\;\frac{\kappa_D^3 N_1N_2 m_0^2 }{4\sqrt{\gamma^2-1}}\int \frac{\mathrm{d}^{D-2}q}{(2\pi)^{D-2}} e^{i  q\cdot b} \nonumber 
    \\
    &\times \bigg\{ \frac{4 [2(\epsilon\cdot P_{12})(\tilde{\epsilon}\cdot P_{12}) - \gamma (\epsilon\cdot P_{12})(\tilde{\epsilon}\cdot Q_{12})-\gamma (\tilde\epsilon\cdot P_{12})({\epsilon}\cdot Q_{12})]}{q^2 r^2}\nonumber 
    \\
    &\qquad+\left(\frac{D-6}{4} + \gamma^2\right)\left[\frac{2 (\epsilon \cdot Q_{12})(\tilde{\epsilon}\cdot Q_{12})}{q^2r^2} + (\epsilon \cdot {\tilde{\epsilon}})\bigg(\frac{r^2}{k_{v_1}^2 q^2}+\frac{q^2}{k_{v_2}^2 r^2}-\frac{4}{q^2}-\frac{4}{r^2}\bigg)\right] \nonumber 
    \\
    &\qquad + 4(\epsilon\cdot {\tilde{\epsilon}})\bigg(\frac{k_{v_1}^2+k_{v_2}^2}{q^2 r^2}+ \frac{\gamma(\gamma k_{v_1} - k_{v_2})}{k_{v_1}q^2} +\frac{\gamma (\gamma k_{v_2} - k_{v_1})}{k_{v_2}r^2}\bigg)\bigg\}
    \label{FinalFinal}
\end{align}
where we have used the relation
\begin{equation}
T_{0}^2 = 8\pi G_N m_{0}^2 = \kappa_{D}^2 m_{0}^2
    \label{TGV4}
\end{equation}
which follows from the D0-brane mass and the expression for Newton's constant in $D$ dimensions,
\begin{align}
    & m_0 = \frac{1}{\sqrt{\alpha'}g_s},\qquad T_0= \sqrt{\pi}\,2^{\frac{10-D}{4}}(2\pi \sqrt{\alpha'})^{\frac{D-4}{2}},\nonumber 
    \\[0.1cm]
    & G_D= 2^{-3-\frac{D-10}{2}}(2\pi \sqrt{\alpha'})^{D-4}\alpha' g_s^2.
    \label{0brane_tension}
\end{align}

Let us now extract from \eqref{FinalFinal} the amplitudes for graviton and dilaton emission. There is no emission of the Kalb-Ramond field $B_{\mu\nu}$, since the result is symmetric under the exchange of $\epsilon$ and $\tilde{\epsilon}$. For the graviton we get
\begin{align}
    A_{5}^{\text{graviton}}
    =&\;\epsilon_{\mu}\tilde{\epsilon}_{\nu}\, \frac{\kappa_D^3 N_1N_2 m_0^2}{2\sqrt{\gamma^2-1}} \int \frac{\mathrm{d}^{D-2}q}{(2\pi)^{D-2}} e^{i  q\cdot b} \nonumber 
    \\
    &\times \bigg\{ \frac{4P_{12}^{(\mu}(P_{12}^{\nu)} - \gamma Q_{12}^{\nu)})}{q^2 r^2} + \left(\frac{D-6}{4} + \gamma^2\right)\frac{Q_{12}^\mu Q_{12}^\nu}{q^2r^2} \bigg\}
     \label{grav_amp_field_theory_lim}
\end{align}

The correctly normalized amplitude for the dilaton is obtained by the following substitution
\begin{equation}
    \epsilon^{\mu} {\tilde{\epsilon}}^\nu \longrightarrow \frac{1}{\sqrt{D-2}}\left(\eta^{\mu \nu}- \frac{k^\mu {\bar{k}}^\nu + k^{\nu}{\bar{k}}^\mu}{k\cdot {\bar{k}}}\right)
 \end{equation}
where $\bar{k}$ is an auxiliary null vector ${\bar{k}}^2=0$ different from $k$ and such that $k \cdot {\bar{k}}\neq 0$. The dependence on ${\bar{k}}$ drops out because of Eqs. \eqref{gaugeinv} and one gets
\begin{align}
    A_5^{\text{dilaton}} =&\;\frac{\kappa_D^3 N_1N_2 m_0^2 }{4\sqrt{\gamma^2-1}} \frac{D-4}{\sqrt{D-2}}\int \frac{\mathrm{d}^{D-2}q}{(2\pi)^{D-2}} e^{i  q\cdot b}\nonumber 
    \\
    &\times \bigg\{4\bigg(\frac{k_{v_1}^2 +k_{v_2}^2}{q^2 r^2} + \frac{\gamma(\gamma k_{v_1} -  k_{v_2})}{k_{v_1}q^2}+ \frac{\gamma(\gamma k_{v_2} - k_{v_1})}{k_{v_2}r^2}\bigg)\nonumber 
    \\
    &\qquad  
    +\left(\frac{D-6}{4} +\gamma^2\right)\bigg(\frac{r^2}{q^2 k_{v_1}^2}+\frac{q^2}{r^2 k_{v_2}^2}-\frac{4}{q^2} -\frac{4}{r^2}\bigg)\bigg\}\label{dil_amp_field_theory_lim}.
\end{align}
In the derivation of the above formula we used the following identities:
\begin{subequations}
\begin{align}
    & P_{12} \cdot P_{12}= -k_{v_1}^2-k_{v_2}^2 +2\gamma k_{v_1} k_{v_2} \label{P12P12}
    \\
    &Q_{12} \cdot Q_{12}=4q^2 +4r^2- \frac{(r^2)^2}{k_{v_1}^2} -\frac{(q^2)^2}{k_{v_2}^2}+\frac{2\gamma q^2 r^2}{k_{v_1} k_{v_2}} \label{Q12Q12}
    \\
    & P_{12}\cdot Q_{12}=2k_{v_1}k_{v_2}+\gamma q^2 + \gamma r^2 - \frac{r^2 k_{v_2}}{k_{v_1}}-\frac{q^2 k_{v_1}}{k_{v_2}}
    \label{P12Q12}
\end{align}
\end{subequations}

\section{Field--theory limit of the superstring amplitude}
\label{superstring}
In this section, we discuss the field-theory limit of the five-point amplitude in type II superstring theory. The starting point is the superstring amplitude, obtained by adding the NS-NS contribution, given in Eq.~\eqref{NSNSFINAL}, and the R-R contribution, given in Eq.~\eqref{RRFINAL}. The field-theory limit is obtained by taking $\alpha'$ to zero while focusing on the degeneration in which the closed string worldsheet becomes a long tube exchanged between the two D-particles. In terms of the annulus modulus, this corresponds to the small-$|w|$ limit. Since the vertex insertion is integrated over the region $|w|\leq |z|\leq 1$, the expansion must be performed with both $|z|$ and $|w/z|$ kept finite, as in the bosonic string calculation. Strictly speaking, however, the $\alpha'\to0$ limit also receives contributions from the region in which $|w|\to1$. As in the bosonic-string calculation, these terms are localized at $b=0$ in impact-parameter space and will not be included in the present field-theory limit.

The details of the limiting procedure, together with the GSO projection, are worked out in Appendix~\ref{AppD2} for the NS-NS sector and in Appendix~\ref{AppD3} for the R-R sector.

The field--theory limit  of the NS-NS calculation agrees with the field--theory limit of the bosonic string amplitude. This agreement is expected, since the massless sector of the bosonic string contains the same NS-NS gravitational fields: the graviton, the dilaton, and the Kalb-Ramond field $B_{\mu\nu}$. In the present amplitude, the $B$-field does not contribute, as expected, since it does not couple to the D0-branes.

The R-R sector contribution to the GSO projection of the field--theory limit of the superstring amplitude is:
\begin{align}
    \nonumber  A_{\text{R-R}}\simeq&\;\frac{\kappa_{D}}{2\pi}\frac{\pi N_1N_2 T_0^2}{16|\sinh(v_{12})|}\cdot 8\int\frac{ d^{D-2}{q}}{(2\pi)^{D-2}}\,e^{iq\cdot b}\times 
    \\
    &\nonumber \times \Bigg\{-16\cosh(v_{12})\Big[(q\cdot \epsilon)(q\cdot\tilde{\epsilon})\frac{1}{q^2 r^2}+\frac{1}{2q^2}(\tilde{\epsilon}\cdot\epsilon) + \frac{1}{2 r^2}(\tilde{\epsilon}\cdot\epsilon)
    \\
    \nonumber &\qquad -\frac{1}{2k_{v_1}^2q^2}(k_{v_1}\epsilon_{v_1}(q\cdot\tilde{\epsilon}) + k_{v_1}\tilde{\epsilon}_{v_1}(q\cdot\epsilon) - (q\cdot k)\tilde{\epsilon}_{v_1}\epsilon_{v_1} + \frac{1}{2}(q\cdot k)(\tilde{\epsilon}\cdot\epsilon))
    \\
    \nonumber &\qquad + \frac{1}{2k_{v_2}^2 r^2}(k_{v_2}\epsilon_{v_2}(q\cdot\tilde{\epsilon}) + k_{v_2}\tilde{\epsilon}_{v_2}(q\cdot\epsilon) - (q\cdot k)\tilde{\epsilon}_{v_2}\epsilon_{v_2} + \frac{1}{2}(q\cdot k)(\tilde{\epsilon}\cdot\epsilon))\Big]
    \\
    &\nonumber +\frac{4(\tilde{\epsilon}\cdot \epsilon)}{q^2}\Big[\cosh(v_{12}) + \frac{k_{v_2}}{k_{v_1}}\Big] + \frac{4(\tilde{\epsilon}\cdot \epsilon)}{r^2}\Big[\cosh(v_{12}) + \frac{k_{v_1}}{k_{v_2}}\Big]
    \\
    \nonumber & - \frac{8k_{v_1}k_{v_2}(\tilde{\epsilon}\cdot \epsilon)}{q^2 r^2} - \frac{8\cosh(v_{12}) (\tilde{\epsilon}\cdot \epsilon)}{q^2} - \frac{8 \cosh(v_{12})(\tilde{\epsilon}\cdot \epsilon)}{r^2}
    \\
    \nonumber & + (k_{v_1} \tilde{\epsilon}_{v_2} - k_{v_2} \tilde{\epsilon}_{v_1})\left(\frac{8(q\cdot \epsilon)}{q^2 r^2}-\frac{4\epsilon_{v_1}}{k_{v_1}q^2} + \frac{4\epsilon_{v_2}}{k_{v_2} r^2}\right)
    \\
    &  + (k_{v_1} \epsilon_{v_2} - k_{v_2} \epsilon_{v_1})\left(\frac{8(q\cdot \tilde{\epsilon})}{q^2 r^2} - \frac{4\tilde{\epsilon}_{v_1}}{k_{v_1}q^2} + \frac{4\tilde{\epsilon}_{v_2}}{k_{v_2}r^2}\right)
    \Bigg\}
    \label{FINALR-R-sec4}
\end{align}
Using the transverse vectors $P_{12}$ and $Q_{12}$ defined in Eq.~\eqref{P12Q12definitions},  we can recast this expression as
\begin{align}
    \nonumber  A_{\text{R-R}} =&\; \epsilon_{\mu}\tilde{\epsilon}_{\nu}\,\frac{\kappa_D^3 N_1N_2 m_0^2}{4\sqrt{\gamma^2 - 1}}\int\frac{ d^{D-2}{q}}{(2\pi)^{D-2}}\,e^{iq\cdot b}\times 
    \\
    &\nonumber \times \Bigg\{-2\gamma \Big[\frac{2Q_{12}^\mu Q_{12}^\nu}{q^2r^2} +\frac{r^2}{k_{v_1}^2q^2}\eta^{\mu\nu} + \frac{q^2}{k_{v_2}^2r^2}\eta^{\mu\nu} - \frac{4\eta^{\mu\nu}}{q^2} - \frac{4\eta^{\mu\nu}}{r^2}\Big]
    \\
    &\qquad -\frac{4\eta^{\mu\nu}}{k_{v_1}q^2}(\gamma k_{v_1} - k_{v_2}) - \frac{4\eta^{\mu\nu}}{k_{v_2}r^2}(\gamma k_{v_2} - k_{v_1})- \frac{8k_{v_1}k_{v_2}\eta^{\mu\nu}}{q^2r^2} + \frac{8 P_{12}^{(\mu}Q_{12}^{\nu)}}{q^2 r^2}\Bigg\}.
    \label{TRULYFINALR-R}
\end{align}
We can now extract the amplitudes for the emission of a graviton and a dilaton. The amplitude for the emission of the $B$-field vanishes, since the result above is symmetric under the exchange of $\epsilon$ and $\tilde{\epsilon}$. To obtain the graviton amplitude, we impose $\epsilon\cdot\tilde{\epsilon}=0$, which amounts to dropping the terms in the integrand proportional to $\eta^{\mu\nu}$. Thus,
\begin{align}
    A_{\text{R-R}}^{\text{graviton}} = &\;\epsilon_{\mu}\tilde{\epsilon}_{\nu}\,\frac{ \kappa_D^3 N_1N_2 m_0^2}{2\sqrt{\gamma^2 - 1}}\int\frac{ d^{D-2}{q}}{(2\pi)^{D-2}}\,e^{iq\cdot b} \Bigg(-\frac{2\gamma Q_{12}^\mu Q_{12}^\nu}{q^2r^2}  + \frac{4 P_{12}^{(\mu}Q_{12}^{\nu)}}{q^2 r^2} \Bigg).
    \label{TRULYFINALR-R-graviton}
\end{align}
To obtain the dilaton amplitude, we replace $\epsilon^\mu\tilde{\epsilon}^\nu$ by 
$\frac{1}{\sqrt{D-2}}\left(\eta^{\mu\nu} - (k^\mu \bar{k}^\nu + k^\nu \bar{k}^\mu)/(k\cdot \bar{k})\right)$, where $\bar{k}$ is an auxiliary null vector different from $k$. Using \cref{P12P12,Q12Q12,P12Q12}, we find
\begin{align}
    A_{\text{R-R}}^{\text{dilaton}} = \;& \frac{\kappa_D^3 N_1N_2 m_0^2}{\sqrt{\gamma^2 - 1}}\frac{D-4}{\sqrt{D-2}}\int\frac{ d^{D-2}{q}}{(2\pi)^{D-2}}\,e^{iq\cdot b}\times \nonumber 
    \\
    &\times \left\{-\frac{2k_{v_1}k_{v_2}}{q^2 r^2} + \frac{1}{q^2}\left(\gamma + \frac{k_{v_2}}{k_{v_1}} - \frac{\gamma r^2}{2k_{v_1}^2}\right) +  \frac{1}{r^2}\left(\gamma + \frac{k_{v_1}}{k_{v_2}} - \frac{\gamma q^2}{2k_{v_2}^2}\right)\right\}
    \label{RR_dilaton_emission}
\end{align}
Adding to the above expression the NS--NS contribution, given by \eqref{dil_amp_field_theory_lim}, we obtain the amplitude for dilaton emission in the field--theory limit of the superstring theory,
\begin{align}
    A^{\text{dilaton}}\equiv &\,A^{\text{dilaton}}_{\text{NS-NS}}+ A^{\text{dilaton}}_{\text{R-R}}\nonumber 
    \\[0.2cm]
    =&\,\frac{\kappa_D^3 N_1N_2 m_0^2}{4\sqrt{\gamma^2 - 1}}\frac{D-4}{\sqrt{D-2}}\int\frac{ d^{D-2}{q}}{(2\pi)^{D-2}}\,e^{iq\cdot b} \nonumber 
    \\
    &\,\times\bigg\{ \frac{4(k_{v_1}-k_{v_2})^2}{q^2 r^2} +\frac{4(k_{v_2}-k_{v_1})}{q^2 k_{v_1}}+\frac{4 (k_{v_1}-k_{v_2})}{r^2 k_{v_2}}+ \frac{\gamma(\gamma k_{v_1}-k_{v_2})}{q^2 k_{v_1}}+ \frac{\gamma(\gamma k_{v_2}-k_{v_1})}{r^2 k_{v_2}}\nonumber 
    \\
    &\qquad +\bigg( \frac{D-10}{4}+(\gamma-1)^2\bigg)\bigg(\frac{r^2}{q^2k_{v_1}^2}+\frac{q^2}{r^2k_{v_2}^2}-\frac{4}{q^2}-\frac{4}{r^2}\bigg)\bigg\}.
\end{align}
The amplitude for graviton emission in the field--theory limit of the superstring theory is obtained by summing together \eqref{grav_amp_field_theory_lim} and \eqref{TRULYFINALR-R-graviton}. More explicitly,
\begin{align}
    A^{\rm graviton}=&\, \epsilon_{\mu}\tilde{\epsilon}_{\nu}\,\frac{ \kappa_D^3 N_1N_2 m_0^2}{2\sqrt{\gamma^2 - 1}}\cdot  \int\frac{ d^{D-2}{q}}{(2\pi)^{D-2}}\,e^{iq\cdot b}\nonumber 
    \\
    &\,\times \bigg\{ \frac{4P_{12}^{(\mu}(P_{12}^{\nu)} - (\gamma-1) Q_{12}^{\nu)})}{q^2 r^2} +\left(\frac{D-10}{4} + (\gamma-1)^2 \right)\frac{Q_{12}^\mu Q_{12}^\nu}{q^2r^2} \bigg\}.
\end{align}
A configuration of two static D-branes in $D=10$ dimensions is BPS and, in particular, stable. Therefore, the emission amplitudes are expected to vanish in the static limit. Indeed, taking $v_1\to v_2$, one finds that the full graviton amplitude  and the full dilaton amplitude  vanish for $D=10$ and $\gamma=1$.

\section{Field--theory calculation}\label{sec:FieldTheory}
In this section, we show that the field-theory limit of the string five-point amplitude agrees with the five-point amplitude computed from the Lagrangian where the D-particles are described by the worldline. For simplicity, we focus only on the tensor structure of the momentum-space amplitudes and do not attempt to match the overall normalization constant of the amplitudes. We use the worldline formalism developed in \cite{Mogull:2020sak}, treating the D0-branes as heavy worldlines following the trajectories:
\begin{equation}
    X^\mu(\tau) = b_i^\mu + u_i^\mu \tau + z_i^\mu(\tau),\qquad i=1,2,
\end{equation}
where $z_i^\mu(\tau)$ describes a perturbation of the straight trajectory $b_i^\mu + u_i^\mu\tau$ of the $i$th D-particle. We set $b_1^\mu = 0$ and identify $b_2^\mu\equiv -b^\mu$ with the impact parameter. As in the string theory calculation, we set
\begin{equation}
    u_i^\mu = (\cosh(v_i),\sinh(v_i),\vec{0}),\qquad i=1,2.
\end{equation}
Following \cite{Mogull:2020sak}, we treat the perturbations $z^\mu(\tau)$ as dynamical quantum fields with an appropriate propagator and interaction vertices.

As the action describing the scattering process, we take the $D$-dimensional Einstein-Maxwell-dilaton theory action $S_{\text{grav}}$ coupled to two D0-brane worldline actions $S_{\text{D0}}^{(i)}$:
\begin{equation}
    S = S_{\text{grav}} + S_{\text{D0}}^{(1)}  + S_{\text{D0}}^{(2)}.
\end{equation}
In string frame, the gravitational action is:
\begin{equation}
    S_{\rm grav} = \frac{1}{2\kappa_{D}^2}\int d^Dx\sqrt{-G}\left[e^{-2\Phi}\left(R(G)  + 4G^{\mu\nu}(\partial_\mu \Phi)(\partial_\nu \Phi)\right) - \frac{1}{4}G^{\mu\rho}G^{\nu\sigma}F_{\mu\nu}F_{\rho\sigma}\right],
\end{equation}
where $F_{\mu\nu} = 2\partial_{[\mu}C_{\nu]}$, and the D-particle action is:
\begin{equation}
    S_{\text{D0}}^{(i)} = -m_i\int d\tau\,e^{-\Phi(X)}\sqrt{-G_{\mu\nu}(X)\dot{X}^\mu \dot{X}^\nu} + m_i\int d\tau\,\dot{X}^\mu C_{\mu}(X).
\end{equation}
We expand $\Phi$ around zero, absorbing the usual finite part of the dilaton VEV, namely the string coupling, by rescaling the $D$-dimensional Newton constant $\kappa_D^2 = 8\pi G_N^{(D)}$, the brane masses $m_i$, and the RR 1-form field $C_\mu.$ We expect to recover the results \eqref{grav_amp_field_theory_lim} and \eqref{dil_amp_field_theory_lim} from an EFT without the NS-NS 2-form and RR 3-form, since these fields do not couple to abelian D-particle worldvolume actions.

To simplify the Feynman rules, it is useful to pass to the Einstein frame by rescaling the metric as $G_{\mu\nu}\to G_{\mu\nu}e^{\frac{4}{D-2}\Phi}$. In Einstein frame, the field-theory action is:
\begin{align}\label{EFT_Einstein_action}
    S_{\text{grav}} = \frac{1}{2\kappa_{D}^2}\int d^Dx\sqrt{-G}\left[R(G) - \frac{4}{D-2}G^{\mu\nu}(\partial_\mu \Phi)(\partial_\nu \Phi) - \frac{1}{4}e^{-2a\Phi}G^{\mu\rho}G^{\nu\sigma}F_{\mu\nu}F_{\rho\sigma}\right],
\end{align}
where $a = -\frac{D-4}{D-2}$, and the D-particle worldvolume action is:
\begin{equation}\label{Einst_frame_Dpart_action}
    S_{\text{D0}}^{(i)} = -m_i\int d\tau\,e^{a\Phi(X)}\sqrt{-G_{\mu\nu}(X)\dot{X}^\mu\dot{X}^\nu} + m_i\int d\tau\,\dot{X}^\mu C_\mu(X).
\end{equation}

In \cite{Mogull:2020sak}, it was argued that, after gauge fixing the worldline reparametrization symmetry, the point-particle action $-m\int d\tau \sqrt{-G_{\mu\nu}\dot{X}^\mu\dot{X}^\nu}$ reduces to the Polyakov form, $\frac{m}{2}\int d\tau [G_{\mu\nu}\dot{X}^\mu \dot{X}^\nu - 1].$ Analogously, rewriting \eqref{Einst_frame_Dpart_action} gives:
\begin{equation}\label{Polyakov_worldline_action}
    S_{\text{D0}}^{(i)} = \frac{m_i}{2}\int d\tau\left[G_{\mu\nu}(X)\dot{X}^\mu \dot{X}^\nu - e^{2a\Phi(X)}\right] + m_i\int d\tau\,\dot{X}^\mu C_\mu(X).
\end{equation}

Expanding $S_{\text{D0}}^{(i)}$ in the trajectory fluctuations $z_i^\mu(\tau) = X^\mu(\tau) - b_i^\mu - u_i^\mu \tau$, and in the fields $G_{\mu\nu},$ $\Phi,$ $C_\mu$ around the flat background $G_{\mu\nu} = \eta_{\mu\nu},$ $\Phi = 0,$ $C_\mu = 0$, we can read off the Feynman rules for $z_i^\mu$, $\Phi$, $C_\mu$, and the metric fluctuations $h_{\mu\nu}$, which we define via $G_{\mu\nu} = \eta_{\mu\nu} + 2\kappa_D h_{\mu\nu}$. A detailed derivation of the worldline Feynman rules was presented in \cite{Mogull:2020sak}.

Before listing the Feynman rules, let us fix the diagrammatic conventions used below. Double wiggly lines denote gravitons $h_{\mu\nu}$, single wiggly lines denote RR photons $C_\mu$, and dashed lines denote dilatons $\Phi$. We will represent worldline fluctuations $z_i^\mu(\omega)$ by solid oriented lines carrying the Fourier frequency $\omega$. Whenever this propagator appears as an internal line, one should integrate over $\omega$ with measure $\frac{d\omega}{2\pi}$. The dotted horizontal segments indicate the undeformed D-particle trajectory only; they are not propagating fields and are included solely as a visual aid. Momentum attached to bulk field is taken to be outgoing from the corresponding vertex and will be written in the argument of that field.

The coupling of the graviton to the fixed worldline (without the fluctuation $z_i^\mu$) is given by:
\begin{equation}
{\begin{tikzpicture}[baseline=(b1.mid)]
    \begin{feynman}
        \vertex (a1);
        \vertex[right=0.7cm of a1] (a2);
        \vertex[left=0.7cm of a1] (a3);
        \vertex[below=0.8cm of a1] (a4);
        \vertex[below=0.6cm of a1] (b1);
        \diagram* {
            {(a1) -- [graviton] (a4) ,},
            {(a2) -- [dotted] (a3) ,}
        };
        \node[below=0.01cm of a4] {\footnotesize \(h_{\alpha\beta}(\ell)\)};
    \end{feynman}
    \end{tikzpicture}}\;=i\kappa_D m_i u_i^\alpha u_i^\beta e^{-i\ell \cdot b_i}\cdot 2\pi \delta (u_i\cdot \ell),
\end{equation}
with outgoing momentum $\ell$. The two-point worldline-graviton, $z^\mu(\omega)h_{\alpha\beta}(\ell)$-vertex is:
\begin{equation}
    {\begin{tikzpicture}[baseline=(b1.mid)]
    \begin{feynman}
        \vertex (a1);
        \vertex[right=0.7cm of a1] (a2);
        \vertex[left=0.7cm of a1] (a3);
        \vertex[below=0.8cm of a1] (a4);
        \vertex[below=0.6cm of a1] (b1);
        \diagram* {
            {(a1) -- [graviton] (a4) ,},
            {(a1) -- [dotted] (a3) ,},
            {(a1) -- [fermion] (a2) ,}
        };
        \node[below=0.01cm of a4] {\footnotesize \(h_{\alpha\beta}(\ell)\)};
        \node[right=0.01cm of a2] {\footnotesize \(z_i^\mu(\omega)\)};
    \end{feynman}
    \end{tikzpicture}}\;=m_i \kappa_D e^{-i\ell \cdot b_i}\cdot 2\pi\delta(u_i\cdot \ell - \omega)(2\omega u_i^{(\alpha}\eta^{\beta)\mu} - u_i^\alpha u_i^\beta \ell^\mu).
\end{equation}
Similarly, the coupling of the worldline to the dilaton field and to the RR 1-form is given by (we denote the dilaton by a straight dashed line, and the RR photon by a single wiggly line):
\begin{subequations}
\begin{align}
{\begin{tikzpicture}[baseline=(b1.mid)]
    \begin{feynman}
        \vertex (a1);
        \vertex[right=0.7cm of a1] (a2);
        \vertex[left=0.7cm of a1] (a3);
        \vertex[below=0.8cm of a1] (a4);
        \vertex[below=0.6cm of a1] (b1);
        \diagram* {
            {(a1) -- [dashed] (a4) ,},
            {(a2) -- [dotted] (a3) ,}
        };
        \node[below=0.01cm of a4]{\footnotesize \(\Phi(\ell)\)};
        \node[right=0.01cm of a2] {\(\qquad\)};
    \end{feynman}
    \end{tikzpicture}}\;&=-im_i a\, e^{-i\ell\cdot b_i} \cdot 2\pi \delta(u_i\cdot \ell),
    \\[0.2cm]
    {\begin{tikzpicture}[baseline=(b1.mid)]
    \begin{feynman}
        \vertex (a1);
        \vertex[right=0.7cm of a1] (a2);
        \vertex[left=0.7cm of a1] (a3);
        \vertex[below=0.8cm of a1] (a4);
        \vertex[below=0.6cm of a1] (b1);
        \diagram* {
            {(a1) -- [photon] (a4) ,},
            {(a2) -- [dotted] (a3) ,}
        };
        \node[below=0.01cm of a4]{\footnotesize \( C_\alpha(\ell)\)};
        \node[right=0.01cm of a2] {\(\qquad\)};
    \end{feynman}
    \end{tikzpicture}}\;&= im_iu_i^\alpha \,e^{-i\ell \cdot b_i} \cdot 2\pi \delta(u_i\cdot \ell),
    \\[0.2cm]
    {\begin{tikzpicture}[baseline=(b1.mid)]
    \begin{feynman}
        \vertex (a1);
        \vertex[right=0.7cm of a1] (a2);
        \vertex[left=0.7cm of a1] (a3);
        \vertex[below=0.8cm of a1] (a4);
        \vertex[below=0.6cm of a1] (b1);
        \diagram* {
            {(a1) -- [dashed] (a4) ,},
            {(a1) -- [dotted] (a3) ,},
            {(a1) -- [fermion] (a2) ,}
        };;
        \node[below=0.01cm of a4]{\footnotesize \(\Phi(\ell)\)};
        \node[right=0.01cm of a2] {\footnotesize \(z_i^\mu(\omega)\)};
    \end{feynman}
    \end{tikzpicture}}\;&= m_ia\ell^\mu\,e^{-i\ell\cdot b_i}\cdot 2\pi \delta(u_i\cdot \ell  - \omega),
    \\[0.2cm]
    {\begin{tikzpicture}[baseline=(b1.mid)]
    \begin{feynman}
        \vertex (a1);
        \vertex[right=0.7cm of a1] (a2);
        \vertex[left=0.7cm of a1] (a3);
        \vertex[below=0.8cm of a1] (a4);
        \vertex[below=0.6cm of a1] (b1);
        \diagram* {
            {(a1) -- [photon] (a4) ,},
            {(a1) -- [dotted] (a3) ,},
            {(a1) -- [fermion] (a2) ,}
        };;
        \node[below=0.01cm of a4]{\footnotesize \( C_\alpha(\ell)\)};
        \node[right=0.01cm of a2] {\footnotesize \(z_i^\mu(\omega)\)};
    \end{feynman}
    \end{tikzpicture}}\;&=  m_i e^{-i\ell\cdot b_i}\cdot 2\pi \delta(u_i\cdot \ell - \omega) (\omega \eta^{\mu\alpha} - u_i^\alpha \ell^\mu).
\end{align}
\end{subequations}
In what follows, we will also need a seagull $\Phi\Phi$-worldline vertex:
\begin{equation}
    {\begin{tikzpicture}[baseline=(b1.mid)]
    \begin{feynman}
        \vertex (a1);
        \vertex[right=0.7cm of a1] (a2);
        \vertex[left=0.7cm of a1] (a3);
        \vertex[below=0.7cm of a2] (c2);
        \vertex[below=0.7cm of a3] (c3);
        \vertex[below=0.6cm of a1] (b1);
        \diagram* {
            {(a1) -- [dashed] (c2) ,},
            {(a1) -- [dashed] (c3) ,},
            {(a2) -- [dotted] (a3) ,}
        };
        \node[below=0.01cm of c2]{\footnotesize \( \Phi(\ell_1)\)};
        \node[below=0.01cm of c3]{\footnotesize \( \Phi(\ell_2)\)};
    \end{feynman}
    \end{tikzpicture}}\;= -2im_ia^2\,e^{-i(\ell_1 + \ell_2)\cdot b_i} \cdot 2\pi\delta(u_i\cdot (\ell_1 + \ell_2)).
\end{equation}
Note that, because \eqref{Polyakov_worldline_action} is linear in $h_{\mu\nu}$ and $C_\mu$, there are no seagull vertices involving graviton or RR photon lines.

The propagator for the worldline fluctuations $z_i^\mu(\tau)$ is:
\begin{equation}
    {\begin{tikzpicture}[baseline=(b1.mid)]
    \begin{feynman}
        \vertex (a1);
        \vertex[below=0.1cm of a1] (b1);
        \vertex[right=1.5cm of a1] (a2);
        \diagram* {
            {(a1) -- [fermion] (a2) ,}
        };;
        \node[left=0.01cm of a1] {\footnotesize \(z^\nu_i(-\omega)\)};
        \node[right=0.01cm of a2] {\footnotesize \(z^\mu_i(\omega)\)};
    \end{feynman}
    \end{tikzpicture}}=\frac{i\eta^{\mu\nu}}{m_i\omega^2}.
\end{equation}
For our purposes, we do not need to specify the $i\epsilon$-prescription for the $\omega^{-2}$ factor. In harmonic gauge for $h_{\mu\nu}$ and Feynman gauge for the RR photon, the propagators for the bulk fields are:
\begin{subequations}
\begin{align}
    {\begin{tikzpicture}[baseline=(b1.mid)]
    \begin{feynman}
        \vertex (a1);
        \vertex[below=0.1cm of a1] (b1);
        \vertex[right=1.5cm of a1] (a2);
        \diagram* {
            {(a1) -- [graviton] (a2) ,}
        };;
        \node[left=0.01cm of a1] {\footnotesize \(h_{\rho\sigma}(-\ell)\)};
        \node[right=0.01cm of a2] {\footnotesize \(h_{\mu\nu}(\ell)\)};
    \end{feynman}
    \end{tikzpicture}} &= \frac{-i}{\ell^2}\cdot\frac{1}{2}\left(\eta^{\mu\rho}\eta^{\nu\sigma} + \eta^{\mu\sigma}\eta^{\nu\rho} - \frac{2}{D-2}\eta^{\mu\nu}\eta^{\rho\sigma}\right),
    \\
    {\begin{tikzpicture}[baseline=(b1.mid)]
    \begin{feynman}
        \vertex (a1);
        \vertex[below=0.1cm of a1] (b1);
        \vertex[right=1.5cm of a1] (a2);
        \diagram* {
            {(a1) -- [dashed] (a2) ,}
        };;
        \node[left=0.01cm of a1] {\footnotesize \(\Phi(-\ell)\)};
        \node[right=0.01cm of a2] {\footnotesize \(\Phi(\ell)\)};
    \end{feynman}
    \end{tikzpicture}} &= \frac{-i\kappa_D^2(D-2)}{4\ell^2},
    \\
    {\begin{tikzpicture}[baseline=(b1.mid)]
    \begin{feynman}
        \vertex (a1);
        \vertex[below=0.1cm of a1] (b1);
        \vertex[right=1.5cm of a1] (a2);
        \diagram* {
            {(a1) -- [photon] (a2) ,}
        };;
        \node[left=0.01cm of a1] {\footnotesize \(C_{\nu}(-\ell)\)};
        \node[right=0.01cm of a2] {\footnotesize \(C_{\mu}(\ell)\)};
    \end{feynman}
    \end{tikzpicture}} &= \frac{-2i\kappa_D^2}{\ell^2}\eta_{\mu\nu}.
\end{align}
\end{subequations}
Finally, we list the Feynman rules for the three-point vertices of the bulk fields:
\begin{subequations}
\begin{align}
    {\begin{tikzpicture}[baseline=(d1.mid)]
    \begin{feynman}
        \vertex (a1);
        \vertex[right=0.8cm of a1] (a2);
        \vertex[below=1.2cm of a1] (b1);
        \vertex[below=0.6cm of a1] (c1);
        \vertex[right=0.8 of c1] (c2);
        \vertex[below=0.1cm of c1] (d1);
        \diagram* {
            {(c1) -- [dashed] (a1),},
            {(b1) -- [dashed] (c1),},
            {(c1) -- [graviton] (c2),}
        };
        \node[above=0.01cm of a1] {\footnotesize \(\Phi(-p_1)\)};
        \node[right=0.01cm of c2] {\footnotesize \(h_{\mu\nu}(p_1+p_2)\)};
        \node[below=0.01cm of b1] {\footnotesize \(\Phi(-p_2)\)};
    \end{feynman}
    \end{tikzpicture}} &= \frac{-8i}{\kappa_D(D-2)}\left[p_1^{(\mu}p_{1}^{\nu)} - \frac{1}{2}\eta^{\mu\nu}(p_1\cdot p_2)\right],
    \\[0.2cm]
    {\begin{tikzpicture}[baseline=(d1.mid)]
    \begin{feynman}
        \vertex (a1);
        \vertex[right=0.8cm of a1] (a2);
        \vertex[below=1.2cm of a1] (b1);
        \vertex[below=0.6cm of a1] (c1);
        \vertex[right=0.8 of c1] (c2);
        \vertex[below=0.1cm of c1] (d1);
        \diagram* {
            {(a1) -- [photon] (b1),},
            {(c1) -- [graviton] (c2),}
        };
        \node[above=0.01cm of a1] {\footnotesize \(C_\alpha(-p_1)\)};
        \node[right=0.01cm of c2] {\footnotesize \(h_{\mu\nu}(p_1+p_2)\)};
        \node[below=0.01cm of b1] {\footnotesize \(C_\beta(-p_2)\)};
    \end{feynman}
    \end{tikzpicture}} &= -\frac{i}{2\kappa_D}\big[(p_1\cdot p_2)(\eta^{\alpha\mu}\eta^{\beta\nu} + \eta^{\alpha\nu}\eta^{\beta\mu}-\eta^{\alpha\beta}\eta^{\mu\nu})\nonumber 
    \\
    & \qquad\qquad\quad  + \eta^{\alpha\beta}(p_1^\mu p_2^\nu + p_1^\nu p_2^\mu) + \eta^{\mu\nu}p_1^\alpha p_2^\beta \nonumber 
    \\
    & \qquad\qquad\quad  -(p_1^\mu\eta^{\nu\beta} + p_1^\nu\eta^{\mu\beta})p_2^\alpha  - (p_2^\mu\eta^{\nu\alpha} + p_2^\nu\eta^{\mu\alpha})p_1^\beta \big].
    \\[0.2cm]
    {\begin{tikzpicture}[baseline=(d1.mid)]
    \begin{feynman}
        \vertex (a1);
        \vertex[right=0.8cm of a1] (a2);
        \vertex[below=1.2cm of a1] (b1);
        \vertex[below=0.6cm of a1] (c1);
        \vertex[right=0.8 of c1] (c2);
        \vertex[below=0.1cm of c1] (d1);
        \diagram* {
            {(b1) -- [photon] (a1),},
            {(c1) -- [dashed] (c2),}
        };
        \node[above=0.01cm of a1] {\footnotesize \(C_\alpha(-p_1)\)};
        \node[right=0.01cm of c2] {\footnotesize \(\Phi(p_1+p_2)\;\;\)};
        \node[below=0.01cm of b1] {\footnotesize \(C_\beta(-p_2)\)};
    \end{feynman}
    \end{tikzpicture}} &= -\frac{2ia}{2\kappa_D^2} \left[(p_1\cdot p_2)\eta^{\alpha\beta} - p_1^\beta p_2^\alpha \right].
\end{align}
\end{subequations}
Note that in Einstein frame there is no $\Phi^3$-vertex. In principle, the three-graviton vertex is also required. However, the corresponding scattering amplitude in pure gravity in $D$ dimensions has already been calculated in \cite{Luna:2017dtq} (see Eq.~4.21 therein), and thus we focus on calculating diagrams involving dilaton or RR photon lines.

\subsection{Graviton emission amplitude}
In this subsection we will calculate the graviton emission amplitude.
\subsubsection{NS--NS sector}
First, we will focus on the dilaton gravity sector of \eqref{EFT_Einstein_action}, and argue that it recovers the field--theory limit of bosonic string result and NS-NS sector of the superstring result. The amplitude is given by the following sum of Feynman diagrams:
\begin{align}
    A_{\text{NS-NS}}^{\mu\nu} =& {\begin{tikzpicture}[baseline=(d1.mid)]
    \begin{feynman}
        \vertex (a1);
        \vertex[right=1.2cm of a1] (a2);
        \vertex[left=0.6cm of a1] (a3);
        \vertex[right=0.6cm of a1] (a4);
        \vertex[below=1.2cm of a1] (b1);
        \vertex[right=1.2cm of b1] (b2);
        \vertex[left=0.6cm of b1] (b3);
        \vertex[right=0.6 of b1] (b4);
        \vertex[below=0.6cm of a1] (c1);
        \vertex[right=1.2cm of c1] (c2);
        \vertex[below=0.1cm of c1] (d1);
        \diagram* {
            {(a1) -- [graviton] (b1),},
            {(a1) -- [fermion] (a4),},
            {(a2) -- [dotted] (a3),},
            {(b2) -- [dotted] (b3),},
            {(a4) -- [graviton] (c2),}
        };
    \end{feynman}
    \end{tikzpicture}}
    \,+
    {\begin{tikzpicture}[baseline=(d1.mid)]
    \begin{feynman}
        \vertex (a1);
        \vertex[right=1.2cm of a1] (a2);
        \vertex[left=0.6cm of a1] (a3);
        \vertex[right=0.6cm of a1] (a4);
        \vertex[below=1.2cm of a1] (b1);
        \vertex[right=1.2cm of b1] (b2);
        \vertex[left=0.6cm of b1] (b3);
        \vertex[right=0.6 of b1] (b4);
        \vertex[below=0.6cm of a1] (c1);
        \vertex[right=1.2cm of c1] (c2);
        \vertex[below=0.1cm of c1] (d1);
        \diagram* {
            {(a1) -- [graviton] (b1),},
            {(b1) -- [fermion] (b4),},
            {(a2) -- [dotted] (a3),},
            {(b2) -- [dotted] (b3),},
            {(b4) -- [graviton] (c2),}
        };
    \end{feynman}
    \end{tikzpicture}}
    \,+
    {\begin{tikzpicture}[baseline=(d1.mid)]
    \begin{feynman}
        \vertex (a1);
        \vertex[right=1.2cm of a1] (a2);
        \vertex[left=0.6cm of a1] (a3);
        \vertex[right=0.6cm of a1] (a4);
        \vertex[below=1.2cm of a1] (b1);
        \vertex[right=1.2cm of b1] (b2);
        \vertex[left=0.6cm of b1] (b3);
        \vertex[right=0.6 of b1] (b4);
        \vertex[below=0.6cm of a1] (c1);
        \vertex[right=1.2cm of c1] (c2);
        \vertex[below=0.1cm of c1] (d1);
        \diagram* {
            {(a1) -- [dashed] (b1),},
            {(a1) -- [fermion] (a4),},
            {(a2) -- [dotted] (a3),},
            {(b2) -- [dotted] (b3),},
            {(a4) -- [graviton] (c2),}
        };
    \end{feynman}
    \end{tikzpicture}} \nonumber 
    \\[0.2cm]
    &+
    {\begin{tikzpicture}[baseline=(d1.mid)]
    \begin{feynman}
        \vertex (a1);
        \vertex[right=1.2cm of a1] (a2);
        \vertex[left=0.6cm of a1] (a3);
        \vertex[right=0.6cm of a1] (a4);
        \vertex[below=1.2cm of a1] (b1);
        \vertex[right=1.2cm of b1] (b2);
        \vertex[left=0.6cm of b1] (b3);
        \vertex[right=0.6 of b1] (b4);
        \vertex[below=0.6cm of a1] (c1);
        \vertex[right=1.2cm of c1] (c2);
        \vertex[below=0.1cm of c1] (d1);
        \diagram* {
            {(a1) -- [dashed] (b1),},
            {(b1) -- [fermion] (b4),},
            {(a2) -- [dotted] (a3),},
            {(b2) -- [dotted] (b3),},
            {(b4) -- [graviton] (c2),}
        };
    \end{feynman}
    \end{tikzpicture}} + {\begin{tikzpicture}[baseline=(d1.mid)]
    \begin{feynman}
        \vertex (a1);
        \vertex[right=1.2cm of a1] (a2);
        \vertex[left=0.6cm of a1] (a3);
        \vertex[right=0.6cm of a1] (a4);
        \vertex[below=1.2cm of a1] (b1);
        \vertex[right=1.2cm of b1] (b2);
        \vertex[left=0.6cm of b1] (b3);
        \vertex[right=0.6 of b1] (b4);
        \vertex[below=0.6cm of a1] (c1);
        \vertex[right=1.2cm of c1] (c2);
        \vertex[below=0.1cm of c1] (d1);
        \diagram* {
            {(a1) -- [graviton] (b1),},
            {(a2) -- [dotted] (a3),},
            {(b2) -- [dotted] (b3),},
            {(c1) -- [graviton] (c2),}
        };
    \end{feynman}
    \end{tikzpicture}}
    \,+
    {\begin{tikzpicture}[baseline=(d1.mid)]
    \begin{feynman}
        \vertex (a1);
        \vertex[right=1.2cm of a1] (a2);
        \vertex[left=0.6cm of a1] (a3);
        \vertex[right=0.6cm of a1] (a4);
        \vertex[below=1.2cm of a1] (b1);
        \vertex[right=1.2cm of b1] (b2);
        \vertex[left=0.6cm of b1] (b3);
        \vertex[right=0.6 of b1] (b4);
        \vertex[below=0.6cm of a1] (c1);
        \vertex[right=1.2cm of c1] (c2);
        \vertex[below=0.1cm of c1] (d1);
        \diagram* {
            {(a1) -- [dashed] (b1),},
            {(a2) -- [dotted] (a3),},
            {(b2) -- [dotted] (b3),},
            {(c1) -- [graviton] (c2),}
        };
    \end{feynman}
    \end{tikzpicture}}
\end{align}
Pure gravity contribution, given by
\begin{equation}
{\begin{tikzpicture}[baseline=(d1.mid)]
    \begin{feynman}
        \vertex (a1);
        \vertex[right=1.2cm of a1] (a2);
        \vertex[left=0.6cm of a1] (a3);
        \vertex[right=0.6cm of a1] (a4);
        \vertex[below=1.2cm of a1] (b1);
        \vertex[right=1.2cm of b1] (b2);
        \vertex[left=0.6cm of b1] (b3);
        \vertex[right=0.6 of b1] (b4);
        \vertex[below=0.6cm of a1] (c1);
        \vertex[right=1.2cm of c1] (c2);
        \vertex[below=0.1cm of c1] (d1);
        \diagram* {
            {(a1) -- [graviton] (b1),},
            {(a1) -- [fermion] (a4),},
            {(a2) -- [dotted] (a3),},
            {(b2) -- [dotted] (b3),},
            {(a4) -- [graviton] (c2),}
        };
    \end{feynman}
    \end{tikzpicture}}
    \,+
    {\begin{tikzpicture}[baseline=(d1.mid)]
    \begin{feynman}
        \vertex (a1);
        \vertex[right=1.2cm of a1] (a2);
        \vertex[left=0.6cm of a1] (a3);
        \vertex[right=0.6cm of a1] (a4);
        \vertex[below=1.2cm of a1] (b1);
        \vertex[right=1.2cm of b1] (b2);
        \vertex[left=0.6cm of b1] (b3);
        \vertex[right=0.6 of b1] (b4);
        \vertex[below=0.6cm of a1] (c1);
        \vertex[right=1.2cm of c1] (c2);
        \vertex[below=0.1cm of c1] (d1);
        \diagram* {
            {(a1) -- [graviton] (b1),},
            {(b1) -- [fermion] (b4),},
            {(a2) -- [dotted] (a3),},
            {(b2) -- [dotted] (b3),},
            {(b4) -- [graviton] (c2),}
        };
    \end{feynman}
    \end{tikzpicture}}\,+
    {\begin{tikzpicture}[baseline=(d1.mid)]
    \begin{feynman}
        \vertex (a1);
        \vertex[right=1.2cm of a1] (a2);
        \vertex[left=0.6cm of a1] (a3);
        \vertex[right=0.6cm of a1] (a4);
        \vertex[below=1.2cm of a1] (b1);
        \vertex[right=1.2cm of b1] (b2);
        \vertex[left=0.6cm of b1] (b3);
        \vertex[right=0.6 of b1] (b4);
        \vertex[below=0.6cm of a1] (c1);
        \vertex[right=1.2cm of c1] (c2);
        \vertex[below=0.1cm of c1] (d1);
        \diagram* {
            {(a1) -- [graviton] (b1),},
            {(a2) -- [dotted] (a3),},
            {(b2) -- [dotted] (b3),},
            {(c1) -- [graviton] (c2),}
        };
    \end{feynman}
    \end{tikzpicture}}
\end{equation}
has been calculated in \cite{Luna:2017dtq}, and it is given by:
\begin{align}
    & \frac{im_1 m_2 \kappa_D^3}{2}\int \frac{d^D q}{(2\pi)^{D-2}}e^{ib\cdot q}\delta (u_1\cdot q) \delta(u_2\cdot r)\nonumber 
    \\
    & \times\bigg[\frac{4P_{12}^{(\mu}(P_{12}^{\nu)} - \gamma Q_{12}^{\nu)})}{q^2 r^2} + \left(\gamma^2 - \frac{1}{D-2}\right)\left(\frac{Q_{12}^\mu Q_{12}^\nu}{q^2 r^2} - \frac{P_{12}^\mu P_{12}^\nu}{k_{v_1}^2 k_{v_2}^2}\right)\bigg] \label{pure_gravity_amp}
\end{align}
where $A^{(\mu} B^{\nu)}=\frac{1}{2}( A^\mu B^\nu+A^\nu B^\mu)$.

We need to calculate only diagrams involving dilaton lines. Applying the above Feynman rules, we calculate:
\begin{align}
    {\begin{tikzpicture}[baseline=(d1.mid)]
    \begin{feynman}
        \vertex (a1);
        \vertex[right=1.2cm of a1] (a2);
        \vertex[left=0.6cm of a1] (a3);
        \vertex[right=0.6cm of a1] (a4);
        \vertex[below=1.2cm of a1] (b1);
        \vertex[right=1.2cm of b1] (b2);
        \vertex[left=0.6cm of b1] (b3);
        \vertex[right=0.6 of b1] (b4);
        \vertex[below=0.6cm of a1] (c1);
        \vertex[right=1.2cm of c1] (c2);
        \vertex[below=0.1cm of c1] (d1);
        \diagram* {
            {(a1) -- [dashed, reversed momentum'=\(q\)] (b1),},
            {(a1) -- [fermion] (a4),},
            {(a2) -- [dotted] (a3),},
            {(b2) -- [dotted] (b3),},
            {(a4) -- [graviton] (c2),}
        };
        \node[below=0.01cm of c2]{\footnotesize \(h_{\mu\nu}(-k)\)};
    \end{feynman}
    \end{tikzpicture}} = \int\frac{d\omega}{2\pi}&[-m_1 \kappa_D (2\pi)\delta(u_1\cdot k+ \omega)(2\omega u_1^{(\mu}\eta^{\nu)\rho}-u_1^\mu u_1^\nu (-k)^{\rho})]\nonumber
    \\
    \times &\frac{i\eta_{\rho\sigma}}{m_1\omega^2}[m_1 a (-q)^\rho (2\pi)\delta(u_1\cdot q - \omega)]\nonumber 
    \\[0.2cm]
    \times &\frac{-i\kappa_D^2(D-2)}{4q^2}[-im_2\kappa_De^{iq\cdot b}(2\pi)\delta(u_2\cdot q)].
\end{align}
Integrating the delta functions over $\omega$ sets $\omega = -u_1\cdot k = k_{v_1}$, and thus, the above expression reduces to:
\begin{equation}\label{dil_gravem1-v2}
    2\pi \delta(u_1\cdot r)\,2\pi\delta(u_2\cdot q)\,\frac{im_1 m_2 \kappa_D^3 a^2(D-2)}{4k_{v_1}^2q^2}\left[2k_{v_1}u_1^{(\mu}q^{\nu)} + u_1^\mu u_1^\nu (k\cdot q)\right].
\end{equation}
Similarly,
\begin{equation}
    {\begin{tikzpicture}[baseline=(d1.mid)]
    \begin{feynman}
        \vertex (a1);
        \vertex[right=1.2cm of a1] (a2);
        \vertex[left=0.6cm of a1] (a3);
        \vertex[right=0.6cm of a1] (a4);
        \vertex[below=1.2cm of a1] (b1);
        \vertex[right=1.2cm of b1] (b2);
        \vertex[left=0.6cm of b1] (b3);
        \vertex[right=0.6 of b1] (b4);
        \vertex[below=0.8cm of a1] (c1);
        \vertex[right=1.2cm of c1] (c2);
        \vertex[below=0.1cm of c1] (d1);
        \diagram* {
            {(a1) -- [dashed, reversed momentum' = \(r\)] (b1),},
            {(b1) -- [fermion] (b4),},
            {(a2) -- [dotted] (a3),},
            {(b2) -- [dotted] (b3),},
            {(b4) -- [graviton] (c2),}
        };
        \node[above=0cm of c2]{\footnotesize \(h_{\mu\nu}(-k)\)};
    \end{feynman}
    \end{tikzpicture}} = 2\pi \delta(u_1\cdot r)\,2\pi\delta(u_2\cdot q)e^{ib\cdot q}\,\frac{im_1 m_2 \kappa_D^3 a^2(D-2)}{4k_{v_2}^2 r^2}\left[2k_{v_2}u_2^{(\mu}q^{\nu)} + u_2^\mu u_2^\nu (k\cdot q)\right].\label{dil_gravem2-v2}
\end{equation}
The third diagram, representing emission of the graviton from the intermediate line, is given by:
\begin{align}
    {\begin{tikzpicture}[baseline=(d1.mid)]
    \begin{feynman}
        \vertex (a1);
        \vertex[right=0.6cm of a1] (a2);
        \vertex[left=0.6cm of a1] (a3);
        \vertex[below=1.4cm of a1] (b1);
        \vertex[right=0.6cm of b1] (b2);
        \vertex[left=0.6cm of b1] (b3);
        \vertex[right=0.6 of b1] (b4);
        \vertex[below=0.7cm of a1] (c1);
        \vertex[right=0.6cm of c1] (c2);
        \vertex[below=0.1cm of c1] (d1);
        \diagram* {
            {(c1) -- [dashed, momentum=\(r\)] (a1),},
            {(b1) -- [dashed, momentum=\(q\)] (c1),},
            {(a2) -- [dotted] (a3),},
            {(b2) -- [dotted] (b3),},
            {(c1) -- [graviton] (c2),}
        };
        \node[right=0.01 of c2]{\footnotesize \(h_{\mu\nu}(-k)\)};
    \end{feynman}
    \end{tikzpicture}} = &\;[-im_1\kappa_D (2\pi)\delta(u_1\cdot r))]\frac{-i\kappa_D^2(D-2)}{4r^2}\frac{-8i}{\kappa_D(D-2)}\times \nonumber 
    \\
    \times &\left[(-r)^{(\mu}q^{\nu)} + \frac{1}{2}\eta^{\mu\nu}(r\cdot q)\right]\frac{-i\kappa_D^2(D-2)}{4q^2}
    [-im_2\kappa_D e^{iq\cdot b} (2\pi)\delta(u_2\cdot q)].
\end{align}
Since graviton polarization is transverse and traceless, we can remove the terms proportional to $k^\mu$, $k^\nu$ and $\eta^{\mu\nu}$. Then, the above expression simplifies to:
\begin{equation}
    2\pi \delta(u_1\cdot r)\,2\pi\delta(u_2\cdot q)\,e^{ib\cdot q}\,\frac{im_1 m_2\kappa_D^3 a^2(D-2)}{4q^2r^2}\,2q^{(\mu}q^{\nu)}.\label{dil_gravem3-v2}
\end{equation}
Summing \eqref{dil_gravem1-v2}, \eqref{dil_gravem2-v2} and \eqref{dil_gravem3-v2}, and suppressing the common factor $2\pi \delta(u_1\cdot (k+q))\,2\pi\delta(u_2\cdot q)e^{ib\cdot q}$, we obtain:
\begin{align}
    \frac{i(D-2)a^2m_1m_2\kappa_D^3 }{2q^2 r^2}\left[q^{(\mu}q^{\nu)} + \frac{u_1^{(\mu}q^{\nu)}r^2}{k_{v_1}} + \frac{u_1^\mu u_1^\nu r^2(r^2 - q^2)}{4k_{v_1}^2} + \frac{u_2^{(\mu}q^{\nu)}q^2}{k_{v_2}} + \frac{u_2^\mu u_2^\nu q^2(r^2 - q^2)}{4k_{v_2}^2}\right].
\end{align}
One can easily check that this expression is equal to:
\begin{equation}\label{dil_contribution_to_grav_amp}
    \frac{i(D-2)a^2m_1m_2\kappa_D^3}{8} \left[\frac{Q_{12}^{\mu}Q_{12}^{\nu}}{q^2 r^2} - \frac{P_{12}^{\mu}P_{12}^{\nu}}{k_{v_1}^2 k_{v_2}^2}\right].
\end{equation}

Summing \eqref{pure_gravity_amp} and \eqref{dil_contribution_to_grav_amp} and remembering that the integral over the two $\delta$-functions gives just a factor $\frac{1}{\sqrt{\gamma^2-1}}$ we obtain 
\begin{align}
    -i A_{\text{NS-NS}}^{\mu \nu}=&\frac{m_1m_2 \kappa_D^3}{2\sqrt{\gamma^2-1}}\int \frac{d^{D-2} q}{(2\pi)^{D-2}}e^{ib\cdot q} \nonumber 
    \\
    &\times\left[\frac{4P_{12}^{(\mu}(P_{12}^{\nu)} - \gamma Q_{12}^{\nu)})}{q^2 r^2} + \left(\frac{D-6}{4}+\gamma^2\right)\left(\frac{Q_{12}^\mu Q_{12}^\nu}{q^2 r^2} - \frac{P_{12}^\mu P_{12}^\nu}{k_{v_1}^2 k_{v_2}^2}\right)\right]\label{GravitonNSNS}.
\end{align}
This formula agrees with \eqref{grav_amp_field_theory_lim} up to  terms proportional to $\delta(b)$ (i.e., the terms independent of $q$ as the last term in the previous equation) as already explained above.  

It is worth noting that, in general dimension $D\neq10$, \eqref{GravitonNSNS} differs from the corresponding graviton-emission amplitude for perturbative massive Kaluza--Klein scalar states~\cite{Luna:2017dtq,Mogull:2020sak,DiVecchia:2020ymx}. The pure-gravity contribution is universal, while the dilaton contribution depends on the coupling of the massive state to the dilaton. D0-branes and perturbative Kaluza--Klein scalar states have different dilaton couplings in general dimension, so their five-point amplitudes differ accordingly. In $D=10$, the D0-brane coupling agrees with that implied by its interpretation as a Kaluza--Klein state of eleven-dimensional gravity.

\subsubsection{RR sector}
From now on, we will always suppress the common factor of $2\pi\delta(r\cdot u_1)\,2\pi \delta(q\cdot u_2)\,e^{-q\cdot b}$ in all amplitudes. The contribution to the graviton emission amplitude from the RR sector consists of three diagrams:
\begin{equation}
A_{\text{R-R}}^{\mu\nu} = {\begin{tikzpicture}[baseline=(d1.mid)]
    \begin{feynman}
        \vertex (a1);
        \vertex[right=1.2cm of a1] (a2);
        \vertex[left=0.6cm of a1] (a3);
        \vertex[right=0.6cm of a1] (a4);
        \vertex[below=1.2cm of a1] (b1);
        \vertex[right=1.2cm of b1] (b2);
        \vertex[left=0.6cm of b1] (b3);
        \vertex[right=0.6 of b1] (b4);
        \vertex[below=0.6cm of a1] (c1);
        \vertex[right=1.2cm of c1] (c2);
        \vertex[below=0.1cm of c1] (d1);
        \diagram* {
            {(a1) -- [photon] (b1),},
            {(a1) -- [fermion] (a4),},
            {(a2) -- [dotted] (a3),},
            {(b2) -- [dotted] (b3),},
            {(a4) -- [graviton] (c2),}
        };
    \end{feynman}
    \end{tikzpicture}}
    \,+
    {\begin{tikzpicture}[baseline=(d1.mid)]
    \begin{feynman}
        \vertex (a1);
        \vertex[right=1.2cm of a1] (a2);
        \vertex[left=0.6cm of a1] (a3);
        \vertex[right=0.6cm of a1] (a4);
        \vertex[below=1.2cm of a1] (b1);
        \vertex[right=1.2cm of b1] (b2);
        \vertex[left=0.6cm of b1] (b3);
        \vertex[right=0.6 of b1] (b4);
        \vertex[below=0.6cm of a1] (c1);
        \vertex[right=1.2cm of c1] (c2);
        \vertex[below=0.1cm of c1] (d1);
        \diagram* {
            {(a1) -- [photon] (b1),},
            {(b1) -- [fermion] (b4),},
            {(a2) -- [dotted] (a3),},
            {(b2) -- [dotted] (b3),},
            {(b4) -- [graviton] (c2),}
        };
    \end{feynman}
    \end{tikzpicture}}\,+
    {\begin{tikzpicture}[baseline=(d1.mid)]
    \begin{feynman}
        \vertex (a1);
        \vertex[right=1.2cm of a1] (a2);
        \vertex[left=0.6cm of a1] (a3);
        \vertex[right=0.6cm of a1] (a4);
        \vertex[below=1.2cm of a1] (b1);
        \vertex[right=1.2cm of b1] (b2);
        \vertex[left=0.6cm of b1] (b3);
        \vertex[right=0.6 of b1] (b4);
        \vertex[below=0.6cm of a1] (c1);
        \vertex[right=1.2cm of c1] (c2);
        \vertex[below=0.1cm of c1] (d1);
        \diagram* {
            {(a1) -- [photon] (b1),},
            {(a2) -- [dotted] (a3),},
            {(b2) -- [dotted] (b3),},
            {(c1) -- [graviton] (c2),}
        };
    \end{feynman}
    \end{tikzpicture}}
\end{equation}
We have:
\begin{align}
    {\begin{tikzpicture}[baseline=(d1.mid)]
    \begin{feynman}
        \vertex (a1);
        \vertex[right=1.2cm of a1] (a2);
        \vertex[left=0.6cm of a1] (a3);
        \vertex[right=0.6cm of a1] (a4);
        \vertex[below=1.2cm of a1] (b1);
        \vertex[right=1.2cm of b1] (b2);
        \vertex[left=0.6cm of b1] (b3);
        \vertex[right=0.6 of b1] (b4);
        \vertex[below=0.6cm of a1] (c1);
        \vertex[right=1.2cm of c1] (c2);
        \vertex[below=0.1cm of c1] (d1);
        \diagram* {
            {(a1) -- [photon, reversed momentum'=\(q\)] (b1),},
            {(a1) -- [fermion] (a4),},
            {(a2) -- [dotted] (a3),},
            {(b2) -- [dotted] (b3),},
            {(a4) -- [graviton] (c2),}
        };
        \node[below=0.00cm of c2]{{\footnotesize\(h_{\mu\nu}(-k)\)}};
    \end{feynman}
    \end{tikzpicture}} = \frac{2i\kappa_D^3  m_1m_2}{q^2k_{v_1}^2}
    \Big[-2k_{v_1}^2 u_{1}^{(\mu}u_{2}^{\nu)} - 2\gamma k_{v_1}u_{1}^{(\mu}q^{\nu)} + (k_{v_1}k_{v_2} - \gamma(q\cdot k))u_1^\mu u_1^\nu \Big].
    \label{eq:EFT-RR-wl1}
\end{align}
and similarly,
\begin{equation}
    {\begin{tikzpicture}[baseline=(d1.mid)]
    \begin{feynman}
        \vertex (a1);
        \vertex[right=1.2cm of a1] (a2);
        \vertex[left=0.6cm of a1] (a3);
        \vertex[right=0.6cm of a1] (a4);
        \vertex[below=1.2cm of a1] (b1);
        \vertex[right=1.2cm of b1] (b2);
        \vertex[left=0.6cm of b1] (b3);
        \vertex[right=0.6 of b1] (b4);
        \vertex[below=0.8cm of a1] (c1);
        \vertex[right=1.2cm of c1] (c2);
        \vertex[below=0.1cm of c1] (d1);
        \diagram* {
            {(a1) -- [photon, reversed momentum'=\(r\)] (b1),},
            {(b1) -- [fermion] (b4),},
            {(a2) -- [dotted] (a3),},
            {(b2) -- [dotted] (b3),},
            {(b4) -- [graviton] (c2),}
        };
        \node[above=0.00cm of c2]{{\footnotesize\(h_{\mu\nu}(-k)\)}};
    \end{feynman}
    \end{tikzpicture}} = \frac{2i\kappa_D^3  m_1m_2}{r^2k_{v_2}^2} \Big[-2k_{v_2}^2 u_1^{(\mu}u_{2}^{\nu)} + 2\gamma k_{v_2}u_{2}^{(\mu}q^{\nu)} + (k_{v_1}k_{v_2} + \gamma (q\cdot k))u_2^\mu u_2^\nu \Big].
    \label{eq:EFT-RR-wl2}
\end{equation}
Emission of the graviton from the intermediate line gives:
\begin{align}
    {\begin{tikzpicture}[baseline=(d1.mid)]
    \begin{feynman}
        \vertex (a1);
        \vertex[right=0.6cm of a1] (a2);
        \vertex[left=0.6cm of a1] (a3);
        \vertex[right=0.6cm of a1] (a4);
        \vertex[below=1.2cm of a1] (b1);
        \vertex[right=0.6cm of b1] (b2);
        \vertex[left=0.6cm of b1] (b3);
        \vertex[right=0.6 of b1] (b4);
        \vertex[below=0.6cm of a1] (c1);
        \vertex[right=0.6cm of c1] (c2);
        \vertex[below=0.1cm of c1] (d1);
        \diagram* {
            {(c1) -- [photon, momentum=\(r\)] (a1),},
            {(b1) -- [photon, momentum=\(q\)] (c1),},
            {(a2) -- [dotted] (a3),},
            {(b2) -- [dotted] (b3),},
            {(c1) -- [graviton] (c2),}
        };
        \node[right=0.01cm of c2]{{\footnotesize\(h_{\mu\nu}(-k)\)}};
    \end{feynman}
    \end{tikzpicture}} = \frac{2i\kappa_D^3  m_1m_2}{q^2r^2}\left[(r^2+q^2)u_1^{(\mu}u_2^{\nu)}-2\gamma q^\mu q^\nu + 2k_{v_2}q^{(\mu}u_1^{\nu)} - 2k_{v_1}q^{(\mu}u_2^{\nu)}\right]
    \label{eq:EFT-RR-bulk}
\end{align}
Summing up the diagrams \eqref{eq:EFT-RR-wl1}, \eqref{eq:EFT-RR-wl2} and \eqref{eq:EFT-RR-bulk} and reintroducing the integral over $q$, we obtain:
\begin{equation}
    -iA^{\mu\nu}_{\text{R-R}} = -\frac{\kappa_D^3 m_1 m_2}{\sqrt{\gamma^2-1}}\int \frac{d^{D-2} q}{(2\pi)^{D-2}}e^{iqb}\left[\frac{\gamma Q_{12}^{\mu}Q_{12}^{\nu}-2P_{12}^{(\mu}Q_{12}^{\nu)}}{q^2 r^2} - \frac{\gamma P_{12}^{\mu}P_{12}^{\nu}}{k_{v_1}^2 k_{v_2}^2}\right].
    \label{GravitonRR}
\end{equation}
This result agrees with \eqref{TRULYFINALR-R-graviton} modulo $q$-independent terms as already explained.

The sum of the contributions of NS-NS and R-R is given by the sum of \eqref{GravitonNSNS} and \eqref{GravitonRR}
\begin{align}
    -i A^{\mu \nu}_{\rm total}=&\,\frac{\kappa_D^3 m_1 m_2}{2\sqrt{\gamma^2-1}}\int \frac{d^{D-2} q}{(2\pi)^{D-2}}e^{iqb}\nonumber 
    \\
    &\,\times \bigg\{ \frac{4P_{12}^\mu P_{12}^\nu}{q^2 r^2}- 2(\gamma-1) \frac{(P_{12}^\mu Q_{12}^\nu+P_{12}^\nu Q_{12}^\mu)}{q^2 r^2}\nonumber 
    \\
    &\qquad +\bigg(\frac{D-10}{4}+(\gamma-1)^2 \bigg) \bigg( \frac{Q_{12}^\mu  Q_{12}^\nu}{q^2 r^2}-\frac{P_{12}^\mu P_{12}^\nu}{k_{v_1}^2 k_{v_2}^2}\bigg) \bigg\}.
\end{align}

\subsection{Dilaton emission amplitude}
\subsubsection{NS-NS sector}
Let us list the diagrams contributing to the dilaton emission amplitude $A$ without the RR-photon lines (that is, the diagrams from the NS-NS sector):
\begin{align}
    A_{\text{NS-NS}}^{\text{dilaton}} =& {\begin{tikzpicture}[baseline=(d1.mid)]
    \begin{feynman}
        \vertex (a1);
        \vertex[right=1.2cm of a1] (a2);
        \vertex[left=0.6cm of a1] (a3);
        \vertex[right=0.6cm of a1] (a4);
        \vertex[below=1.2cm of a1] (b1);
        \vertex[right=1.2cm of b1] (b2);
        \vertex[left=0.6cm of b1] (b3);
        \vertex[right=0.6 of b1] (b4);
        \vertex[below=0.6cm of a1] (c1);
        \vertex[right=1.2cm of c1] (c2);
        \vertex[below=0.1cm of c1] (d1);
        \diagram* {
            {(a1) -- [graviton] (b1),},
            {(a1) -- [fermion] (a4),},
            {(a2) -- [dotted] (a3),},
            {(b2) -- [dotted] (b3),},
            {(a4) -- [dashed] (c2),}
        };
    \end{feynman}
    \end{tikzpicture}}
    \,+
    {\begin{tikzpicture}[baseline=(d1.mid)]
    \begin{feynman}
        \vertex (a1);
        \vertex[right=1.2cm of a1] (a2);
        \vertex[left=0.6cm of a1] (a3);
        \vertex[right=0.6cm of a1] (a4);
        \vertex[below=1.2cm of a1] (b1);
        \vertex[right=1.2cm of b1] (b2);
        \vertex[left=0.6cm of b1] (b3);
        \vertex[right=0.6 of b1] (b4);
        \vertex[below=0.6cm of a1] (c1);
        \vertex[right=1.2cm of c1] (c2);
        \vertex[below=0.1cm of c1] (d1);
        \diagram* {
            {(a1) -- [graviton] (b1),},
            {(b1) -- [fermion] (b4),},
            {(a2) -- [dotted] (a3),},
            {(b2) -- [dotted] (b3),},
            {(b4) -- [dashed] (c2),}
        };
    \end{feynman}
    \end{tikzpicture}}\,+
    {\begin{tikzpicture}[baseline=(d1.mid)]
    \begin{feynman}
        \vertex (a1);
        \vertex[right=1.2cm of a1] (a2);
        \vertex[left=0.6cm of a1] (a3);
        \vertex[right=0.6cm of a1] (a4);
        \vertex[below=1.2cm of a1] (b1);
        \vertex[right=1.2cm of b1] (b2);
        \vertex[left=0.6cm of b1] (b3);
        \vertex[right=0.6 of b1] (b4);
        \vertex[below=0.6cm of a1] (c1);
        \vertex[right=1.2cm of c1] (c2);
        \vertex[below=0.1cm of c1] (d1);
        \diagram* {
            {(a1) -- [graviton] (c1),},
            {(c1) -- [dashed] (b1),},
            {(a2) -- [dotted] (a3),},
            {(b2) -- [dotted] (b3),},
            {(c1) -- [dashed] (c2),}
        };
    \end{feynman}
    \end{tikzpicture}}
    \,
    + {\begin{tikzpicture}[baseline=(d1.mid)]
    \begin{feynman}
        \vertex (a1);
        \vertex[right=1.2cm of a1] (a2);
        \vertex[left=0.6cm of a1] (a3);
        \vertex[right=0.6cm of a1] (a4);
        \vertex[below=1.2cm of a1] (b1);
        \vertex[right=1.2cm of b1] (b2);
        \vertex[left=0.6cm of b1] (b3);
        \vertex[right=0.6 of b1] (b4);
        \vertex[below=0.6cm of a1] (c1);
        \vertex[right=1.2cm of c1] (c2);
        \vertex[below=0.1cm of c1] (d1);
        \diagram* {
            {(a1) -- [dashed] (c1),},
            {(c1) -- [graviton] (b1),},
            {(a2) -- [dotted] (a3),},
            {(b2) -- [dotted] (b3),},
            {(c1) -- [dashed] (c2),}
        };
    \end{feynman}
    \end{tikzpicture}}\nonumber 
    \\
    &+{\begin{tikzpicture}[baseline=(d1.mid)]
    \begin{feynman}
        \vertex (a1);
        \vertex[right=1.2cm of a1] (a2);
        \vertex[left=0.6cm of a1] (a3);
        \vertex[right=0.6cm of a1] (a4);
        \vertex[below=1.2cm of a1] (b1);
        \vertex[right=1.2cm of b1] (b2);
        \vertex[left=0.6cm of b1] (b3);
        \vertex[right=0.6 of b1] (b4);
        \vertex[below=0.6cm of a1] (c1);
        \vertex[right=1.2cm of c1] (c2);
        \vertex[below=0.1cm of c1] (d1);
        \diagram* {
            {(a1) -- [dashed] (b1),},
            {(a1) -- [fermion] (a4),},
            {(a2) -- [dotted] (a3),},
            {(b2) -- [dotted] (b3),},
            {(a4) -- [dashed] (c2),}
        };
    \end{feynman}
    \end{tikzpicture}}
    \,+
    {\begin{tikzpicture}[baseline=(d1.mid)]
    \begin{feynman}
        \vertex (a1);
        \vertex[right=1.2cm of a1] (a2);
        \vertex[left=0.6cm of a1] (a3);
        \vertex[right=0.6cm of a1] (a4);
        \vertex[below=1.2cm of a1] (b1);
        \vertex[right=1.2cm of b1] (b2);
        \vertex[left=0.6cm of b1] (b3);
        \vertex[right=0.6 of b1] (b4);
        \vertex[below=0.6cm of a1] (c1);
        \vertex[right=1.2cm of c1] (c2);
        \vertex[below=0.1cm of c1] (d1);
        \diagram* {
            {(a1) -- [dashed] (b1),},
            {(b1) -- [fermion] (b4),},
            {(a2) -- [dotted] (a3),},
            {(b2) -- [dotted] (b3),},
            {(b4) -- [dashed] (c2),}
        };
    \end{feynman}
    \end{tikzpicture}}\,+
    {\begin{tikzpicture}[baseline=(d1.mid)]
    \begin{feynman}
        \vertex (a1);
        \vertex[right=1.2cm of a1] (a2);
        \vertex[left=0.6cm of a1] (a3);
        \vertex[right=0.6cm of a1] (a4);
        \vertex[below=1.2cm of a1] (b1);
        \vertex[right=1.2cm of b1] (b2);
        \vertex[left=0.6cm of b1] (b3);
        \vertex[right=0.6 of b1] (b4);
        \vertex[below=0.6cm of a1] (c1);
        \vertex[right=1.2cm of c1] (c2);
        \vertex[below=0.1cm of c1] (d1);
        \diagram* {
            {(a1) -- [dashed] (b1),},
            {(a2) -- [dotted] (a3),},
            {(b2) -- [dotted] (b3),},
            {(a1) -- [dashed] (c2),}
        };
    \end{feynman}
    \end{tikzpicture}}
    \,
    + {\begin{tikzpicture}[baseline=(d1.mid)]
    \begin{feynman}
        \vertex (a1);
        \vertex[right=1.2cm of a1] (a2);
        \vertex[left=0.6cm of a1] (a3);
        \vertex[right=0.6cm of a1] (a4);
        \vertex[below=1.2cm of a1] (b1);
        \vertex[right=1.2cm of b1] (b2);
        \vertex[left=0.6cm of b1] (b3);
        \vertex[right=0.6 of b1] (b4);
        \vertex[below=0.6cm of a1] (c1);
        \vertex[right=1.2cm of c1] (c2);
        \vertex[below=0.1cm of c1] (d1);
        \diagram* {
            {(a1) -- [dashed] (b1),},
            {(a2) -- [dotted] (a3),},
            {(b2) -- [dotted] (b3),},
            {(b1) -- [dashed] (c2),}
        };
    \end{feynman}
    \end{tikzpicture}}
\end{align}

Tedious but straightforward calculations give us:
\begin{subequations}
\begin{equation}
    {\begin{tikzpicture}[baseline=(d1.mid)]
    \begin{feynman}
        \vertex (a1);
        \vertex[right=1.2cm of a1] (a2);
        \vertex[left=0.6cm of a1] (a3);
        \vertex[right=0.6cm of a1] (a4);
        \vertex[below=1.2cm of a1] (b1);
        \vertex[right=1.2cm of b1] (b2);
        \vertex[left=0.6cm of b1] (b3);
        \vertex[right=0.6 of b1] (b4);
        \vertex[below=0.6cm of a1] (c1);
        \vertex[right=1.2cm of c1] (c2);
        \vertex[below=0.1cm of c1] (d1);
        \diagram* {
            {(b1) -- [graviton, momentum = $q$] (a1),},
            {(a1) -- [fermion] (a4),},
            {(a2) -- [dotted] (a3),},
            {(b2) -- [dotted] (b3),},
            {(a4) -- [dashed] (c2),}
        };
        \node[below=0.0 of c2]{{\footnotesize \(\Phi(-k)\)}};
    \end{feynman}
    \end{tikzpicture}} = \frac{i\kappa_D^2 m_1 m_2 a}{q^2 k_{v_1}^2}\left\{2\gamma k_{v_1} k_{v_2} - \frac{2k_{v_1}^2}{D-2} - \left[\gamma^2-\frac{1}{D-2}\right](q\cdot k)\right\},
    \label{eq:Mh1-final}
    \end{equation}
    \begin{equation}
    {\begin{tikzpicture}[baseline=(d1.mid)]
    \begin{feynman}
        \vertex (a1);
        \vertex[right=1.2cm of a1] (a2);
        \vertex[left=0.6cm of a1] (a3);
        \vertex[right=0.6cm of a1] (a4);
        \vertex[below=1.2cm of a1] (b1);
        \vertex[right=1.2cm of b1] (b2);
        \vertex[left=0.6cm of b1] (b3);
        \vertex[right=0.6 of b1] (b4);
        \vertex[below=0.8cm of a1] (c1);
        \vertex[right=1.2cm of c1] (c2);
        \vertex[below=0.1cm of c1] (d1);
        \diagram* {
            {(b1) -- [graviton, momentum = $r$] (a1),},
            {(b1) -- [fermion] (b4),},
            {(a2) -- [dotted] (a3),},
            {(b2) -- [dotted] (b3),},
            {(b4) -- [dashed] (c2),}
        };
        \node[above=0.0 of c2]{{\footnotesize \(\Phi(-k)\)}};
    \end{feynman}
    \end{tikzpicture}} = \frac{i\kappa_D^2 m_1m_2a}{r^2k_{v_2}^2}\left\{2\gamma k_{v_1}k_{v_2} -\frac{2k_{v_2}^2}{D-2}+\left[\gamma^2 -\frac{1}{D-2}\right](q\cdot k)\right\},
    \label{eq:Mh2-final}
    \end{equation}
    \begin{equation}
    {\begin{tikzpicture}[baseline=(d1.mid)]
    \begin{feynman}
        \vertex (a1);
        \vertex[right=0.6cm of a1] (a2);
        \vertex[left=0.6cm of a1] (a3);
        \vertex[right=0.6cm of a1] (a4);
        \vertex[below=1.2cm of a1] (b1);
        \vertex[right=0.6cm of b1] (b2);
        \vertex[left=0.6cm of b1] (b3);
        \vertex[right=0.6 of b1] (b4);
        \vertex[below=0.6cm of a1] (c1);
        \vertex[right=0.6cm of c1] (c2);
        \vertex[below=0.1cm of c1] (d1);
        \diagram* {
            {(c1) -- [graviton, momentum = $r$] (a1),},
            {(b1) -- [dashed, momentum = $q$] (c1),},
            {(a2) -- [dotted] (a3),},
            {(b2) -- [dotted] (b3),},
            {(c1) -- [dashed] (c2),}
        };
        \node[right=0.0 of c2]{{\footnotesize \(\Phi(-k)\)}};
    \end{feynman}
    \end{tikzpicture}} = -\frac{2i\kappa_D^2m_1m_2a}{q^2r^2}k_{v_1}^2,
    \label{eq:Mbulk1}
    \end{equation}
    \begin{equation}
    {\begin{tikzpicture}[baseline=(d1.mid)]
    \begin{feynman}
        \vertex (a1);
        \vertex[right=0.6cm of a1] (a2);
        \vertex[left=0.6cm of a1] (a3);
        \vertex[right=0.6cm of a1] (a4);
        \vertex[below=1.2cm of a1] (b1);
        \vertex[right=0.6cm of b1] (b2);
        \vertex[left=0.6cm of b1] (b3);
        \vertex[right=0.6 of b1] (b4);
        \vertex[below=0.6cm of a1] (c1);
        \vertex[right=0.6cm of c1] (c2);
        \vertex[below=0.1cm of c1] (d1);
        \diagram* {
            {(c1) -- [dashed, momentum = $r$] (a1),},
            {(b1) -- [graviton, momentum = $q$] (c1),},
            {(a2) -- [dotted] (a3),},
            {(b2) -- [dotted] (b3),},
            {(c1) -- [dashed] (c2),}
        };
        \node[right=0.0 of c2]{{\footnotesize \(\Phi(-k)\)}};
    \end{feynman}
    \end{tikzpicture}} = -\frac{2i \kappa_D^2a\,m_1m_2}{q^2r^2}k_{v_2}^2,
    \label{eq:Mbulk2}
    \end{equation}
    \begin{equation}
    {\begin{tikzpicture}[baseline=(d1.mid)]
    \begin{feynman}
        \vertex (a1);
        \vertex[right=1.2cm of a1] (a2);
        \vertex[left=0.6cm of a1] (a3);
        \vertex[right=0.6cm of a1] (a4);
        \vertex[below=1.2cm of a1] (b1);
        \vertex[right=1.2cm of b1] (b2);
        \vertex[left=0.6cm of b1] (b3);
        \vertex[right=0.6 of b1] (b4);
        \vertex[below=0.6cm of a1] (c1);
        \vertex[right=1.2cm of c1] (c2);
        \vertex[below=0.1cm of c1] (d1);
        \diagram* {
            {(b1) -- [dashed, momentum=$q$] (a1),},
            {(a1) -- [fermion] (a4),},
            {(a2) -- [dotted] (a3),},
            {(b2) -- [dotted] (b3),},
            {(a4) -- [dashed] (c2),}
        };
        \node[below=0.0 of c2]{{\footnotesize \(\Phi(-k)\)}};
    \end{feynman}
    \end{tikzpicture}} = -\frac{i\kappa_D^2a^3(D-2)m_1m_2}{4q^2k_{v_1}^2} (q\cdot k), \label{eq:Mphi1-dir}
    \end{equation}
    \begin{equation}
    {\begin{tikzpicture}[baseline=(d1.mid)]
    \begin{feynman}
        \vertex (a1);
        \vertex[right=1.2cm of a1] (a2);
        \vertex[left=0.6cm of a1] (a3);
        \vertex[right=0.6cm of a1] (a4);
        \vertex[below=1.2cm of a1] (b1);
        \vertex[right=1.2cm of b1] (b2);
        \vertex[left=0.6cm of b1] (b3);
        \vertex[right=0.6 of b1] (b4);
        \vertex[below=0.8cm of a1] (c1);
        \vertex[right=1.2cm of c1] (c2);
        \vertex[below=0.1cm of c1] (d1);
        \diagram* {
            {(b1) -- [dashed, momentum = $r$] (a1),},
            {(b1) -- [fermion] (b4),},
            {(a2) -- [dotted] (a3),},
            {(b2) -- [dotted] (b3),},
            {(b4) -- [dashed] (c2),}
        };
        \node[above=0.0 of c2]{{\footnotesize \(\Phi(-k)\)}};
    \end{feynman}
    \end{tikzpicture}} = +\frac{i\kappa_D^2(D-2)m_1m_2a^3}{4r^2k_{v_2}^2}(q\cdot k),
    \label{eq:Mphi2-dir}
    \end{equation}
    \begin{equation}
    {\begin{tikzpicture}[baseline=(d1.mid)]
    \begin{feynman}
        \vertex (a1);
        \vertex[right=1.2cm of a1] (a2);
        \vertex[left=0.6cm of a1] (a3);
        \vertex[right=0.6cm of a1] (a4);
        \vertex[below=1.2cm of a1] (b1);
        \vertex[right=1.2cm of b1] (b2);
        \vertex[left=0.6cm of b1] (b3);
        \vertex[right=0.6 of b1] (b4);
        \vertex[below=0.6cm of a1] (c1);
        \vertex[right=1.2cm of c1] (c2);
        \vertex[below=0.1cm of c1] (d1);
        \diagram* {
            {(b1) -- [dashed, momentum = $q$] (a1),},
            {(a2) -- [dotted] (a3),},
            {(b2) -- [dotted] (b3),},
            {(a1) -- [dashed] (c2),}
        };
        \node[below=0.0 of c2]{{\footnotesize \(\Phi(-k)\)}};
    \end{feynman}
    \end{tikzpicture}} = \frac{i \kappa_D^2a^3(D-2)m_1m_2}{2q^2},
    \label{eq:Msg1}
    \end{equation}
    \begin{equation}
    {\begin{tikzpicture}[baseline=(d1.mid)]
    \begin{feynman}
        \vertex (a1);
        \vertex[right=1.2cm of a1] (a2);
        \vertex[left=0.6cm of a1] (a3);
        \vertex[right=0.6cm of a1] (a4);
        \vertex[below=1.2cm of a1] (b1);
        \vertex[right=1.2cm of b1] (b2);
        \vertex[left=0.6cm of b1] (b3);
        \vertex[right=0.6 of b1] (b4);
        \vertex[below=0.8cm of a1] (c1);
        \vertex[right=1.2cm of c1] (c2);
        \vertex[below=0.1cm of c1] (d1);
        \diagram* {
            {(b1) -- [dashed, momentum = $r$] (a1),},
            {(a2) -- [dotted] (a3),},
            {(b2) -- [dotted] (b3),},
            {(b1) -- [dashed] (c2),}
        };
        \node[above=0.0 of c2]{{\footnotesize \(\Phi(-k)\)}};
    \end{feynman}
    \end{tikzpicture}} = \frac{ i \kappa_D^2a^3(D-2)\,m_1m_2}{2r^2}.
    \label{eq:Msg2}
\end{equation}
\end{subequations}
It is important to note that the Feynman rules used to derive the above expression assume that the kinetic term in the dilaton field action is: $-\frac{1}{2\kappa_2^2}\frac{4}{D-2}\int \sqrt{-G}\,(\partial \Phi)^2$. To compare the resulting amplitude with the string theory result, we need to first rescale the dilaton field, $\Phi\to \tilde{\Phi}$, so that it has the canonically normalized kinetic term, $-\frac{1}{2}\int\sqrt{-G}\,(\partial \tilde{\Phi})^2$. This amounts to rescaling the amplitude as:
\begin{equation}\label{Adil_normalization}
    A^{\rm dilaton} \to A'^{\rm dilaton} = \frac{\kappa_D\sqrt{D-2}}{2}A^{\rm dilaton}.
\end{equation}

Summing the diagrams above, reintroducing and performing the integrals over the $\delta$-functions, and rescaling the resulting amplitude by $\frac{\kappa_D\sqrt{D-2}}{2}$, we obtain:
\begin{align}
    -iA_{\text{NS-NS}}^{\text{dilaton}} = &\,\frac{\kappa_D^3m_1m_2}{4\sqrt{\gamma^2-1}}\frac{D-4}{\sqrt{D-2}}\int \frac{d^{D-2} q}{(2\pi)^{D-2}}e^{iq\cdot b}\nonumber 
    \\
    &\times \bigg[4\bigg(\frac{k_{v_1}^2+k_{v_2}^2}{q^2r^2}+\frac{\gamma(\gamma k_{v_1}-k_{v_2})}{k_{v_1}q^2}+\frac{\gamma(\gamma k_{v_2}-k_{v_1})}{k_{v_2}r^2}\bigg)\;\nonumber
    \\[0.2cm]
    &\qquad +\left(\frac{D-6}{4}+\gamma^2\right)\bigg(\frac{r^2}{q^2k_{v_1}^2}+\frac{q^2}{r^2k_{v_2}^2}-\frac{4}{q^2}-\frac{4}{r^2} - \frac{1}{k_{v_1}^2} - \frac{1}{k_{v_2}^2}\bigg)\bigg].
\label{dilaNSNS}
\end{align}
This amplitude agrees with the string theory result \eqref{dil_amp_field_theory_lim} up to the terms proportional to $\delta(b)$.

\subsubsection{R-R sector}
The RR-sector contribution is given by the following diagrams:
\begin{equation}
    A_{\text{RR}}^{\text{dilaton}} = {\begin{tikzpicture}[baseline=(d1.mid)]
    \begin{feynman}
        \vertex (a1);
        \vertex[right=1.2cm of a1] (a2);
        \vertex[left=0.6cm of a1] (a3);
        \vertex[right=0.6cm of a1] (a4);
        \vertex[below=1.2cm of a1] (b1);
        \vertex[right=1.2cm of b1] (b2);
        \vertex[left=0.6cm of b1] (b3);
        \vertex[right=0.6 of b1] (b4);
        \vertex[below=0.6cm of a1] (c1);
        \vertex[right=1.2cm of c1] (c2);
        \vertex[below=0.1cm of c1] (d1);
        \diagram* {
            {(a1) -- [photon] (b1),},
            {(a1) -- [fermion] (a4),},
            {(a2) -- [dotted] (a3),},
            {(b2) -- [dotted] (b3),},
            {(a4) -- [dashed] (c2),}
        };
    \end{feynman}
    \end{tikzpicture}}
    \,+
    {\begin{tikzpicture}[baseline=(d1.mid)]
    \begin{feynman}
        \vertex (a1);
        \vertex[right=1.2cm of a1] (a2);
        \vertex[left=0.6cm of a1] (a3);
        \vertex[right=0.6cm of a1] (a4);
        \vertex[below=1.2cm of a1] (b1);
        \vertex[right=1.2cm of b1] (b2);
        \vertex[left=0.6cm of b1] (b3);
        \vertex[right=0.6 of b1] (b4);
        \vertex[below=0.6cm of a1] (c1);
        \vertex[right=1.2cm of c1] (c2);
        \vertex[below=0.1cm of c1] (d1);
        \diagram* {
            {(a1) -- [photon] (b1),},
            {(b1) -- [fermion] (b4),},
            {(a2) -- [dotted] (a3),},
            {(b2) -- [dotted] (b3),},
            {(b4) -- [dashed] (c2),}
        };
    \end{feynman}
    \end{tikzpicture}}\,+
    {\begin{tikzpicture}[baseline=(d1.mid)]
    \begin{feynman}
        \vertex (a1);
        \vertex[right=1.2cm of a1] (a2);
        \vertex[left=0.6cm of a1] (a3);
        \vertex[right=0.6cm of a1] (a4);
        \vertex[below=1.2cm of a1] (b1);
        \vertex[right=1.2cm of b1] (b2);
        \vertex[left=0.6cm of b1] (b3);
        \vertex[right=0.6 of b1] (b4);
        \vertex[below=0.6cm of a1] (c1);
        \vertex[right=1.2cm of c1] (c2);
        \vertex[below=0.1cm of c1] (d1);
        \diagram* {
            {(a1) -- [photon] (b1),},
            {(a2) -- [dotted] (a3),},
            {(b2) -- [dotted] (b3),},
            {(c1) -- [dashed] (c2),}
        };
    \end{feynman}
    \end{tikzpicture}}
\end{equation}
We obtain:
\begin{subequations}
\begin{equation}
    {\begin{tikzpicture}[baseline=(d1.mid)]
    \begin{feynman}
        \vertex (a1);
        \vertex[right=1.2cm of a1] (a2);
        \vertex[left=0.6cm of a1] (a3);
        \vertex[right=0.6cm of a1] (a4);
        \vertex[below=1.2cm of a1] (b1);
        \vertex[right=1.2cm of b1] (b2);
        \vertex[left=0.6cm of b1] (b3);
        \vertex[right=0.6 of b1] (b4);
        \vertex[below=0.6cm of a1] (c1);
        \vertex[right=1.2cm of c1] (c2);
        \vertex[below=0.1cm of c1] (d1);
        \diagram* {
            {(b1) -- [photon, momentum = $q$] (a1),},
            {(a1) -- [fermion] (a4),},
            {(a2) -- [dotted] (a3),},
            {(b2) -- [dotted] (b3),},
            {(a4) -- [dashed] (c2),}
        };
        \node[below=0.01cm of c2]{{\footnotesize \(\Phi(-k)\)}};
    \end{feynman}
    \end{tikzpicture}} = -\frac{2 i \kappa_D^2a\,m_1m_2}{k_{v_1}^2q^2}
    \left[k_{v_1}k_{v_2}-\gamma(k\cdot q)\right].
    \label{eq:rrphi-MC1-final}
\end{equation}
\begin{equation}
    {\begin{tikzpicture}[baseline=(d1.mid)]
    \begin{feynman}
        \vertex (a1);
        \vertex[right=1.2cm of a1] (a2);
        \vertex[left=0.6cm of a1] (a3);
        \vertex[right=0.6cm of a1] (a4);
        \vertex[below=1.2cm of a1] (b1);
        \vertex[right=1.2cm of b1] (b2);
        \vertex[left=0.6cm of b1] (b3);
        \vertex[right=0.6 of b1] (b4);
        \vertex[below=0.8cm of a1] (c1);
        \vertex[right=1.2cm of c1] (c2);
        \vertex[below=0.1cm of c1] (d1);
        \diagram* {
            {(b1) -- [photon, momentum=$r$] (a1),},
            {(b1) -- [fermion] (b4),},
            {(a2) -- [dotted] (a3),},
            {(b2) -- [dotted] (b3),},
            {(b4) -- [dashed] (c2),}
        };
        \node[above=0.01cm of c2]{{\footnotesize \(\Phi(-k)\)}};
    \end{feynman}
    \end{tikzpicture}} = -\frac{2 i \kappa_D^2a\,m_1m_2}{k_{v_2}^2r^2}
    \left[k_{v_1}k_{v_2}+\gamma(k\cdot q)\right].
    \label{eq:rrphi-MC2-final}
\end{equation}
\begin{equation}
    {\begin{tikzpicture}[baseline=(d1.mid)]
    \begin{feynman}
        \vertex (a1);
        \vertex[right=0.6cm of a1] (a2);
        \vertex[left=0.6cm of a1] (a3);
        \vertex[right=0.6cm of a1] (a4);
        \vertex[below=1.2cm of a1] (b1);
        \vertex[right=0.6cm of b1] (b2);
        \vertex[left=0.6cm of b1] (b3);
        \vertex[right=0.6 of b1] (b4);
        \vertex[below=0.6cm of a1] (c1);
        \vertex[right=0.6cm of c1] (c2);
        \vertex[below=0.1cm of c1] (d1);
        \diagram* {
            {(b1) -- [photon, momentum = $q$] (c1),},
            {(c1) -- [photon, momentum = $r$] (a1),},
            {(a2) -- [dotted] (a3),},
            {(b2) -- [dotted] (b3),},
            {(c1) -- [dashed] (c2),}
        };
        \node[right=0.01cm of c2]{{\footnotesize \(\Phi(-k)\)}};
    \end{feynman}
    \end{tikzpicture}} = +\frac{4 i \kappa_D^2a\,m_1m_2}{q^2r^2}
    \left[k_{v_1}k_{v_2} - \gamma (r\cdot q)\right].
    \label{eq:rrphi-bulk-final}
    \end{equation}
\end{subequations}
Adding \eqref{eq:rrphi-MC1-final}, \eqref{eq:rrphi-MC2-final} and \eqref{eq:rrphi-bulk-final}, normalizing the amplitude according to \eqref{Adil_normalization}, and reintroducing and performing the integrals over the delta functions $\delta(u_1\cdot r)\delta(u_2\cdot q)$, we obtain:
\begin{align}
    -iA_{\text{R-R}}^{\text{dilaton}} = &\,\frac{\kappa_D^3\,m_1m_2}{\sqrt{\gamma^2 -1}}\frac{D-4}{\sqrt{D-2}}\int \frac{d^{D-2}q}{(2\pi)^{D-2}}e^{iq\cdot b}
    \\
    &\,\times \left[\frac{k_{v_1}k_{v_2}-\gamma(k\cdot q)}{k_{v_1}^2q^2} + \frac{k_{v_1}k_{v_2}+\gamma(k\cdot q)}{k_{v_2}^2r^2} + \frac{\gamma(q^2+r^2)-2k_{v_1}k_{v_2}}{q^2r^2} \right].
    \label{eq:rrphi-sum-kv}
\end{align}
This expression agrees with \eqref{RR_dilaton_emission} up to terms proportional to $\delta(b)$.

\newpage
\section{Conclusions and Outlook}
\label{conclu}
In this paper, we have studied the emission of a massless closed-string state in the scattering of two D0-branes. Using the boundary-state formalism, we computed the corresponding five-point amplitude for graviton and dilaton emission in bosonic string theory and in type II superstring theory. We then performed the $\alpha'\to0$ limit of the string amplitudes and compared the resulting expressions with an independent calculation based on the worldline description of D0-branes coupled to the appropriate low-energy gravitational fields.

For nonzero impact parameter, we find agreement between the string and field-theory calculations. In particular, the comparison includes both contributions in which the emitted massless state couples to a field exchanged between the two D0-branes and bremsstrahlung-type contributions in which it is emitted directly from one of the D0-brane worldlines, thereby extending the analysis of \cite{Hussain:1997hz}. In ten dimensions, the complete graviton and dilaton emission amplitudes vanish in the static limit, as expected from the BPS nature of a system of parallel D0-branes. We stress that our analysis does not establish the same agreement for contributions localized at $b=0$ in impact-parameter space. These terms may probe the short-distance regime in which an EFT containing only the massless closed-string modes is insufficient, but we do not exclude the possibility that with the right order of the worldsheet integrations and the $\alpha'\to0$ limit, the field--theory limit of string amplitude matches the EFT result.

A few extensions of this calculation would be interesting. First, the type IIA field-theory result can be compared directly with the eleven-dimensional five-point gravitational amplitude by specializing Eq.~(5.38) of Ref.~\cite{DiVecchia:2023frv} to the kinematics in which the massive ten-dimensional states carry Kaluza--Klein momentum along the eleventh dimension. Beyond the $\alpha'\to0$ limit, the amplitude also contains exchanges of massive string oscillator states, including higher-spin states. It would be interesting to investigate whether these contributions can be organized consistently with the eleven-dimensional massive supermultiplet structure proposed in \cite{Bars:1995uh}.

A second direction is to repeat the calculation in the open-string channel. Such a representation may be useful both for extending the calculation to higher loops and for studying the short-distance regime in which stretched open strings become light.

To analyze higher orders in the eikonal expansion, where D0-brane recoil can no longer be neglected, the conventional boundary-state description requires refinement. A natural framework for such calculations is the open--closed string field theory treatment of D0-brane recoil developed recently in \cite{Sen:2025xaj}.

\appendix 

\section{Calculation of the bosonic string amplitude}
\label{AppA}

In this appendix, we evaluate the amplitude in Eq.~\eqref{ANN1} using the exponentiated vertex operator in Eq.~\eqref{ANN7}. More explicitly we have to compute the following quantity:
\begin{align}
    &\frac{\kappa_D }{2\pi} \frac{ \alpha'N_1 N_2 T_0^2}{32}\sqrt{1-V_1^2}\sqrt{1-V_2^2}   \nonumber 
    \\
    &\times \langle p=0| {}_a\langle 0| {}_{\tilde{a}}\langle 0| \delta( {\hat{q}}^1-V_1 {\hat{q}}^0)\bigg[\prod_{i=2}^{D-1}\delta ( {\hat{q}}^i -y_1^i )\bigg]  \bigg[\prod_{k=2}^{D-1}e^{\sum_{n=1}^{\infty} a_{n,k}\,{\tilde{a}}_{n,k}} \bigg]e^{-\sum_{n=1}^{\infty} a_n \cdot M (V_1) \cdot {\tilde{a}}_n}\nonumber 
    \\
    & \times  \int d^2 z\, V(z)\tilde{V}(\bar{z}) e^{i \hat{q}\cdot k}e^{\alpha' k\cdot {\hat{p}}\log |z|}e^{\sqrt{\frac{\alpha'}{2}} \hat{p}\cdot (\frac{\epsilon}{z} +\frac{{\tilde{\epsilon}}}{{\bar{z}}})}\int \frac{d^2 w}{|w|^4}\,|w|^{\frac{\alpha'}{2} {\hat{p}^2 + N+{\bar{N}}}}\nonumber 
    \\
    & \times  e^{-\sum_{n=1}^{\infty} a_n^\dagger \cdot M (V_2) \cdot {\tilde{a}}_n^\dagger}\bigg[\prod_{k=2}^{D-1} e^{\sum_{n=1}^{\infty} a_{n, k}^\dagger{\tilde{a}}_{n, k}^\dagger} \bigg] \bigg[\prod_{i=2}^{D-1} \delta (\hat{q}^i-y^i)\bigg] \delta ({\hat{q}}^1 - V_2 {\hat{q}}^0 )  |0\rangle_{{\tilde{a}}} |0\rangle_a|p=0\rangle
    \label{A1}
\end{align}
where
\begin{equation}\label{A1a}
    N= \sum_{n=1}^\infty n a_n^\dagger \cdot a_n,\qquad {\bar{N}}= \sum_{n=1}^\infty n {\tilde{a}}_n^\dagger \cdot {\tilde{a}}_n 
\end{equation}
and
\begin{align}
    V(z) &= e^{-(\sqrt{\frac{\alpha'}{2}}k + \epsilon\,\partial_{z})\cdot \sum_{n=1}^\infty \frac{1}{\sqrt{n}}{a}_n {z}^{-n}} e^{(\sqrt{\frac{\alpha'}{2}}k + \tilde{\epsilon}\,\partial_{{z}})\cdot \sum_{n=1}^\infty \frac{1}{\sqrt{n}}{a}_n^\dagger{z}^{n}},
    \\
    \tilde{V}(\bar{z}) &= e^{-(\sqrt{\frac{\alpha'}{2}}k + \tilde{\epsilon}\,\partial_{\bar{z}})\cdot \sum_{n=1}^\infty \frac{1}{\sqrt{n}}\tilde{a}_n \bar{z}^{-n}} e^{(\sqrt{\frac{\alpha'}{2}}k + \tilde{\epsilon}\,\partial_{\bar{z}})\cdot \sum_{n=1}^\infty \frac{1}{\sqrt{n}}\tilde{a}_n^\dagger\bar{z}^{n}}.
\end{align}
In \eqref{A1}  we can separate the contribution of the zero and nonzero modes. The first one is computed in the first subsection while that of the nonzero modes is computed in the second subsection.

\subsection{Bosonic zero modes}\label{AppA1}
In this appendix, we compute the zero-mode contribution to the matrix element in Eq.~\eqref{A1}. For convenience, we first write this contribution explicitly:
\begin{align}
    & \langle p=0| \int\frac{d^{D-2}r}{(2\pi)^{D-2}} e^{-ir^i( {\hat{q}}^i - y_1^i)} \int \frac{dr_1}{2\pi} e^{-ir_1 ({\hat{q}}^1 -V_1 {\hat{q}}^0)}  \bigg(  e^{i {\hat{q}}\cdot k} e^{\alpha' k \cdot{\hat{p}}\log |z|}e^{\sqrt{\frac{\alpha'}{2}}{\hat{p}}\cdot(\frac{\epsilon}{z} +\frac{{\tilde{\epsilon}}}{{\bar{z}}})}\bigg)   \nonumber 
    \\
    & \times |w|^{\frac{\alpha'}{2} {\hat{p}}^2} \int \frac{d^{D-2}q}{(2\pi)^{D-2}}  e^{i q^i ({\hat{q}}^i- y_2^i)}\int \frac{dq_1}{2\pi}e^{iq_1({\hat{q}}^{(1)}- V_2 {\hat{q}}^{(0)})} |p=0\rangle.
    \label{A2}
\end{align}
The factor in the large parentheses comes from the zero modes of the graviton vertex operator. We first consider the contribution from the transverse directions $i=2,3,\ldots,D-1$. It is given by
\begin{align}
    &\int\frac{d^{D-2}r}{(2\pi)^{D-2}}   \int \frac{d^{D-2}q}{(2\pi)^{D-2}}  e^{\alpha' k_\perp\cdot q_\perp\log |z|}e^{\sqrt{\frac{\alpha'}{2}}{q}\cdot (\frac{\epsilon_\perp}{z} +\frac{{\tilde{\epsilon}_\perp}}{{\bar{z}}})}|w|^{\frac{\alpha'}{2} q^2}\nonumber 
    \\
    &\qquad\qquad\qquad\qquad\qquad\qquad\times\langle p=0| e^{i(k_\perp-r_\perp+q_\perp)\cdot \hat{q}_\perp}|p=0\rangle e^{i r_i y_1^i-i q_i y_2^i} \nonumber 
    \\[0.3cm]
    & = \int\frac{d^{D-2}r}{(2\pi)^{D-2}}   \int \frac{d^{D-2}q}{(2\pi)^{D-2}}  e^{\alpha' k_\perp\cdot q_\perp\log |z|}e^{\sqrt{\frac{\alpha'}{2}}q_\perp \cdot (\frac{\epsilon_\perp}{z} +\frac{{\tilde{\epsilon}_\perp}}{{\bar{z}}})}|w|^{\frac{\alpha'}{2} q^2_\perp}\nonumber 
    \\
    &\qquad\qquad\qquad\qquad\qquad\qquad\times(2\pi)^{D-2} \delta^{D-2} (k_\perp+q_\perp-r_\perp)e^{ik_iy_1^i} e^{iq(y_1^i-y_2^i)} \nonumber 
    \\[0.3cm]
    &=\int \frac{d^{D-2}q}{(2\pi)^{D-2}} e^{iq b}  e^{\alpha' k_\perp q_\perp\log |z|}e^{\sqrt{\frac{\alpha'}{2}}{q_\perp}(\frac{\epsilon_\perp}{z} +\frac{{\tilde{\epsilon}_\perp}}{{\bar{z}}})} e^{\frac{\alpha'}{2} q^2_\perp \log |w|}  \label{A3}
\end{align}
where we have introduced the subscript $\perp$ to distinguish the transverse components from the longitudinal directions $0$ and $1$. We have also chosen $y_1=0$ and defined the impact parameter as $b=-y_2$.

The contribution from the longitudinal directions $0$ and $1$ is
\begin{align}
    & \sqrt{1-V_1^2} \sqrt{1-V_2^2} \int \frac{dr_1}{2\pi} \int \frac{dq_1}{2\pi}\times \nonumber 
    \\
    &\quad\;\times \langle p=0| e^{-i r_1( {\hat{q}}^{(1)} -{\hat{q}}^0 V_1) } e^{i k^1 {\hat{q}}^1 -i k^0 {\hat{q}}^{(0)}}e^{\alpha' k\cdot \hat{p}\log |z|}e^{\sqrt{\frac{\alpha'}{2}}{\hat{p}}\cdot (\frac{\epsilon}{z} +\frac{{\tilde{\epsilon}}}{{\bar{z}}})} |w|^{\frac{\alpha'}{2} {\hat{p}}^2}e^{iq_1({\hat{q}}^{(1)}- V_2 {\hat{q}}^{(0)})} |p=0\rangle\nonumber 
    \\[0.3cm]
    &= \sqrt{1-V_1^2} \sqrt{1-V_2^2} \int \frac{dr_1}{2\pi} \int \frac{dq_1}{2\pi} |w|^{\frac{\alpha'}{2} {q_1}^2(1-V_2^2)} \nonumber 
    \\
    & \quad\;\times \left(e^{\alpha' (k^1-k^0 V_2) {q_1}\log |z|}e^{\sqrt{\frac{\alpha'}{2}}{q_1}(\frac{\epsilon^1}{z} +\frac{{\tilde{\epsilon}}^1}{{\bar{z}}})} e^{\sqrt{\frac{\alpha'}{2}}{q_1 V_2}(\frac{\epsilon^0}{z}  +\frac{{\tilde{\epsilon}}^0}{{\bar{z}}})} \right) \nonumber
    \\
    & \quad\;\times\langle p=0| e^{i (-r_1+k^1) {\hat{q}}^{(1)} } e^{-i (k^0+ V_2 q_1- V_1 r_1) {\hat{q}}^{(0)}} e^{iq_1({\hat{q}}^{(1)}- V_2 {\hat{q}}^{(0)})} |p=0\rangle  \label{A4}
\end{align}
The last matrix element can be evaluated explicitly, giving
\begin{align}
    & \langle p=0| e^{i {\hat{q}}^{(1)} (k^1+q^1 - r^1)} e^{-i {\hat{q}}^{(0)}(k^0-V_1 r^1 +V_2 q^1)}|p=0\rangle \nonumber 
    \\
    & = \frac{(2\pi)^2}{|V_1-V_2|} \delta\left(\frac{k^0-k^1 V_2}{V_1-V_2} - r^1\right)\delta \left(\frac{k^0- k^1V_1}{V_1-V_2}-q_1\right) \label{A5}
\end{align}
Inserting  this result into Eq.~\eqref{A4}, we find
\begin{align}
    & \frac{\sqrt{1-V_1^2} \sqrt{1-V_2^2} }{|V_1-V_2|}|w|^{\frac{\alpha'}{2}\left( {\frac{k^0-V_1 k^1 }{V_1-V_2}}\right)^2(1-V_2^2)} \nonumber 
    \\
    & \times\bigg(e^{\frac{\alpha'}{2}  (k^1-k^0 V_2) {\frac{k^0-V_1 k^1 }{V_1-V_2}}\log |z|^2}  e^{\sqrt{\frac{\alpha'}{2}}{\frac{k^0-V_1 k^1 }{V_1-V_2}}(\frac{\epsilon^1}{z} +\frac{{\tilde{\epsilon}}^1}{{\bar{z}}})} e^{-\sqrt{\frac{\alpha'}{2}}{\frac{k^0-V_1 k^1 }{V_1-V_2}V_2}(\frac{\epsilon^0}{z}  +\frac{{\tilde{\epsilon}}^0}{{\bar{z}}})} \bigg) \label{A6}
\end{align}

From Eq.~\eqref{A5}, we obtain the relations
\begin{equation}\label{RF5}
    r_\perp= q_\perp+k_\perp,\quad r_1=\frac{k^0-V_2 k^1 }{V_1-V_2},\quad q^1 = \frac{k^0-V_1 k^1 }{V_1-V_2}.
\end{equation} 
We now introduce the rapidities $v_1$ and $v_2$ of the two D0-branes,
\begin{equation}
    V_1= \tanh v_1,\quad V_2= \tanh v_2
\end{equation}
so that $r_1$ and $q_1$ can be written as
\begin{align}
    & \label{RF0} r^1=\frac{k^0 \cosh v_2 -k^1 \sinh v_2}{\sinh(v_1-v_2)}\cosh v_1,
    \\
    &q^1= \frac{k^0\cosh v_1 -k^1 \sinh v_1}{\sinh(v_1-v_2)}\cosh v_2. \label{RF1}
\end{align}
Their time components are correspondingly
\begin{align}
    & \label{RF4}r^0= \frac{k^0 \cosh v_2 -k^1 \sinh v_2}{\sinh(v_1-v_2)}\sinh v_1 = r^1 \tanh v_1,
    \\
    &q^0=\frac{k^0\cosh v_1 -k^1 \sinh v_1}{\sinh(v_1-v_2)}\sinh v_2 = q^1 \tanh v_2.\label{RF2}.
\end{align}
It is straightforward to check that $r^0=k^0+q^0$ and $r^1=k^1+q^1$. We can therefore define the Lorentz vector $r^\mu=q^\mu+k^\mu$. With these definitions, Eq.~\eqref{A6} takes the simpler form
\begin{align}
    & \frac{1}{|\sinh (v_1-v_2)|}e^{\alpha' k_\parallel q_\parallel \log |z|+ \frac{\alpha'}{2}q^2_\parallel  \log |w| + q_\parallel(\frac{\epsilon}{z}+\frac{{\tilde{\epsilon}} }{{\bar{z}}})} \label{HR5}
\end{align}
where the subscript $\parallel$ denotes the components along the longitudinal directions $0$ and $1$.

Combining the longitudinal contribution in Eq.~\eqref{HR5} with the transverse contribution in Eq.~\eqref{A3}, we obtain the full zero-mode contribution:
\begin{align}
    & \frac{1}{|\sinh v_{12}|} \int \frac{d^{D-2}q}{(2\pi)^{D-2}}  e^{iq\cdot b}  e^{\alpha' k\cdot q\log |z|}e^{\sqrt{\frac{\alpha'}{2}}q\cdot (\frac{\epsilon}{z} +\frac{{\tilde{\epsilon}}}{{\bar{z}}})} e^{\frac{\alpha'}{2} q^2 \log |w|}. \label{BZMX3}
\end{align}
Here we have introduced the shorthand notation $v_{12}=v_1-v_2$.

\subsection{Bosonic nonzero modes}
\label{AppA2}

In this appendix, we compute the contribution of the bosonic non--zero modes to the matrix element \eqref{A1}. In order to avoid infinities due to the fact that, in Lorentzian space, the time component of the oscillators have negative norm, we shall be working in Euclidean space, and then analytically continue the final results to Lorentzian signature.

In Euclidean signature the boosted boundary state is:
\begin{equation}
    \prod_{n=1}^\infty e^{-a_n^{||}\cdot M(v) SM(v) \cdot \tilde{a}_n^{||}} e^{a_n^\perp \cdot \tilde{a}_n^\perp}\ket{0}_a\ket{0}_{\tilde{a}},
\end{equation}
where
\begin{equation}\label{lorentz_boost_matrix}
    M(v) = \begin{pmatrix}
        \cos v & \sin v \\ -\sin v  & \cos v 
    \end{pmatrix},\quad S = \begin{pmatrix}
        1 & 0 \\ 0 & -1
    \end{pmatrix},
\end{equation}
and by $a_n^{||}$ we denote annihilation operators in directions $0,1$, i.e., $a_n^{||} = (a_n^0, a_n^1)$, and superscript $\perp$ corresponds to transverse directions $2,3,\dots,D-1$, $a_n^{\perp} = (a_n^2,\dots,a_n^{D-1})$.

\subsubsection{Unboosted directions}
Let us first focus on the contribution from directions $2,3,\dots, D-1$. To simplify the notation, in this section we shall suppress the superscript $\perp$. The transverse contributions are given by the following matrix element:
\begin{align}\label{transverse_bosonic_nonzero}
\bra{0}e^{\sum_{n=1}^\infty a_n\cdot  \tilde{a}_n} V(z)\tilde{V}(\bar{z})e^{\sum_{n=1}^{\infty}n(a_n^\dagger a_n + \tilde{a}_n^\dagger\tilde{a}_n)}e^{\sum_{n=1}^\infty a_{n}^\dagger \cdot \tilde{a}_n^\dagger}\ket{0},
\end{align}
with vertex operators:
\begin{align}
    V(z) =\;: e^{-(\sqrt{\frac{\alpha'}{2}}k+\epsilon \partial_z)\sum_{n=1}^\infty \frac{1}{\sqrt{n}}(a_n z^{-n} - a_{n}^\dagger z^n)}: \,= \prod_{n=1}^\infty e^{B_n a_n^\dagger }e^{C_n a_n},
    \\
    \tilde{V}(\bar{z}) =\;: e^{-(\sqrt{\frac{\alpha'}{2}}k+{\tilde{\epsilon}}\bar{\partial}_{\bar{z}})\sum_{n=1}^\infty \frac{1}{\sqrt{n}}(\tilde{a}_n \bar{z}^{-n} - \tilde{a}_{n}^\dagger \bar{z}^n)}: \,= \prod_{n=1}^\infty e^{\tilde{B}_n \tilde{a}_n^\dagger }e^{\tilde{C}_n \tilde{a}_n}.
\end{align}
where
\begin{align}
    B_n = \frac{1}{\sqrt{n}}Az^n,\quad \tilde{B}_{n} = \frac{1}{\sqrt{n}}\tilde{A}\bar{z}^n,\quad C_n = -\frac{1}{\sqrt{n}}A z^{-n},\quad \tilde{C}_n = -\frac{1}{\sqrt{n}}\tilde{A}\bar{z}^{-n},\label{BnCn_defs}
\end{align}
and $A = \sqrt{\frac{\alpha'}{2}}k + \epsilon \partial_z$ and $\tilde{A} = \sqrt{\frac{\alpha'}{2}}k + \tilde{\epsilon}\partial_{\bar{z}}$. We can write \eqref{transverse_bosonic_nonzero} more compactly as:
\begin{equation}
    \prod_{m=1}^\infty \langle 0 | e^{{a}_m \cdot {\tilde{a}}_m} e^{B_m a_m^\dagger} e^{\tilde{B}_m \tilde{a}_m^\dagger} e^{C_m a_m} e^{\tilde{C}_m \tilde{a}_m}e^{|w|^{2m}{a}_m^\dagger \cdot{\tilde{a}}_m^\dagger} |0\rangle.
\end{equation}
Inserting the resolution of identity in terms of coherent states:
\begin{equation}
    \mathds{1} = \prod_{n=1}^\infty\int d^{2}f_n d^{2}\tilde{f}_n e^{-|f_n|^2 - |\tilde{f}_n|^2} |f_n \tilde{f}_n \rangle \langle f_n \tilde{f}_n|,
\end{equation}
we obtain:
\begin{align}
    \nonumber &\quad\;\prod_{n=1}^\infty \int d^2f_n d^2\tilde{f}_n e^{-|f_n|^2 - |\tilde{f}_n|^2} \langle 0 | e^{a_n \cdot \tilde{a}_n} e^{B_n a_n^\dagger + \tilde{B}_n \tilde{a}_n^\dagger} |f_n \tilde{f}_n \rangle \langle f_n \tilde{f}_n| e^{C_n a_n + \tilde{C}_n \tilde{a}_n} e^{|w|^{2n} a_n^\dagger \tilde{a}_n^\dagger} |0\rangle 
    \\
    \nonumber &= \prod_{n=1}^\infty \int d^2f_n d^2\tilde{f}_n e^{-|f_n|^2 - |\tilde{f}_n|^2} e^{(f_n+B_n)\cdot(\tilde{f}_n+\tilde{B}_n)} e^{|w|^{2n}(f_n^*+C_n^*)\cdot(\tilde{f}_n^*+\tilde{C}_n^*)} 
    \\
    \nonumber &= \prod_{n=1}^\infty \int d^2f_n d^2\tilde{f}_n \exp\left[ - \begin{pmatrix} f_n^* & \tilde{f}_n \end{pmatrix} \begin{pmatrix} \mathds{1} & -|w|^{2n} \\ - \mathds{1} & \mathds{1} \end{pmatrix} \begin{pmatrix} f_n \\ \tilde{f}_n^* \end{pmatrix}\right]\times 
    \\
    &\qquad\qquad\quad\qquad\;\times \exp\left[\begin{pmatrix} f_n^* & \tilde{f}_n \end{pmatrix} \begin{pmatrix} B_n \\ \tilde{C}_n^* \end{pmatrix} + \begin{pmatrix} \tilde{B}_n & C_n^* \end{pmatrix} \begin{pmatrix} f_n \\ \tilde{f}_n^* \end{pmatrix} + B_n\tilde{B}_n + |w|^{2n}C_n^*\tilde{C}_n^* \right].
\end{align}
Using the Gaussian integral formula 
\begin{equation}
    \int d^{4D}F \exp(-F^\dagger M F + F^\dagger A + B^\dagger F) = (\det M)^{-1} e^{B^\dagger M^{-1}A}
\end{equation}
with $F = \begin{pmatrix} f_n \\ \tilde{f}_n^*\end{pmatrix}$ yields
\begin{align}\label{transverse_bosonic_nonzero_1}
    \nonumber \prod_{m=1}^\infty& (1-|w|^{2m})^{2-D} e^{B_m\cdot\tilde{B}_m + C_m\cdot\tilde{C}_m}\times
    \\
    &\times \exp\left[ \frac{|w|^{2m}}{1-|w|^{2m}} (\tilde{B}_m\cdot\tilde{C}_m^* + |w|^{2m}C_m^*\cdot\tilde{C}_m^* + \tilde{B}_m\cdot B_m + C_m^*\cdot B_m) \right].
\end{align}
Recalling the definitions of the various quantities in  \eqref{BnCn_defs}, we can calculate the sums  in the above expressions:
\begin{subequations}
\begin{align}
    \sum_{m=1}^\infty \tilde{B}_m \cdot B_m &= -A\cdot\tilde{A} \log(1-|z|^2),
    \\
    \sum_{m=1}^\infty \tilde{C}_m \cdot C_m |w|^{2m} &= -A\cdot\tilde{A} \log\left(1-\left|\frac{w}{z}\right|^2\right),
    \\
    \sum_{m=1}^\infty \frac{|w|^{4m}}{1-|w|^{2m}} \tilde{C}_m \cdot C_m &= -A\cdot\tilde{A} \sum_{n=1}^\infty \log(1-|\frac{w}{z}|^{2}|w|^{2n}),
    \\
    \sum_{m=1}^\infty \frac{|w|^{2m}}{1-|w|^{2m}} \tilde{B}_m \cdot B_m &= -A\cdot\tilde{A} \sum_{n=1}^\infty\log(1-|z|^2|w|^{2n}),
    \\
    \sum_{m=1}^\infty \frac{|w|^{2m}}{1-|w|^{2m}} C_m^* \cdot B_m &=  \frac{\alpha'}{2}\sum_{n=1}^\infty k_{\perp}^2 \log(1-|w|^{2n}),
    \\
    \sum_{m=1}^\infty \frac{|w|^{2m}}{1-|w|^{2m}} \tilde{B}_m \cdot \tilde{C}_m^* &= \frac{\alpha'}{2}\sum_{n=1}^\infty k_{\perp}^2 \log(1-|w|^{2n}).
\end{align}
\end{subequations}
With these results we can rewrite \eqref{transverse_bosonic_nonzero_1} as:
\begin{align}
&\left(\prod_{n=1}^\infty (1- |w|^{2n})^{-(D-2) + \alpha'k_{\perp}\cdot k_{\perp}}\right)\left(\prod_{n=0}^\infty (1 - |z|^2 |w|^{2n})^{-A_\perp \cdot A_{\perp}}(1-|\frac{w}{z}|^2|w|^{2n})^{-A_{\perp}\cdot A_{\perp}}\right).
\label{A16}
\end{align}

\subsubsection{Boosted directions}
In  this  subsection we will compute  the contribution from directions $0,1$, which is given by the following matrix element,
\begin{equation}
\label{bosonic_longitudinal_nonzero}
    \prod_{n=1}^\infty \langle 0 | e^{-\tilde{a}_n^{||} M(v_1)^{T} S M(v_1) a_n^{||}} e^{B_n a_n^{||}{}^\dagger} e^{\tilde{B}_n \tilde{a}^{||}_n{}^\dagger} e^{C_n a_n^{||}} e^{\tilde{C}_n \tilde{a}_n^{||}} w^{n a_n^{||}{}^\dagger \cdot a_n^{||}} \bar{w}^{n \tilde{a}_n^{||}{}^\dagger \cdot \tilde{a}_n^{||}} e^{-a_n^{||}{}^{\dagger} M(v_2)^{T} S M(v_2) \tilde{a}_n^{||}{}^\dagger} |0\rangle
\end{equation}
where $a_n^{||} = \begin{pmatrix}a_n^0 \\ a_n^{1}\end{pmatrix}$. To simplify our notation, we will drop the superscript $||$ throughout the rest of this section. Vectors $B_n$ and $C_n$ are defined as in equation \eqref{BnCn_defs}, but now we will redefine them to have only two longitudinal components, 0 and 1, instead of the previous range of $2,3,\dots, D-1$. 

In the calculation below we often use the fact that matrices
\begin{equation}
    N(v) = -M(v)^{T} S M(v) = \begin{pmatrix}
        -\cos(2v) &  -\sin(2v)
        \\
        -\sin(2v) & \cos(2v)
    \end{pmatrix}
\end{equation}
satisfy 
\begin{equation}\label{Nmatrix_identity}
    N(v_1)N(v_2) = M(2v_{21}),
\end{equation}
where  $v_{12} = v_1 - v_2$, and $M(v)$ is the Lorentz-boost matrix, given by Eq.  \eqref{lorentz_boost_matrix}.
Proceeding as in the previous case, by inserting a resolution of identity we recast \eqref{bosonic_longitudinal_nonzero} as:
\begin{align}
    \nonumber &\prod_{n=1}^\infty \int d^{2}f_n d^{2}\tilde{f}_n e^{-|f_n|^2-|\tilde{f}_n|^2}\langle 0 | e^{\tilde{a}_n \cdot N(v_1) \cdot a_n} e^{B_n a_n^\dagger + \tilde{B}_n \tilde{a}_n^\dagger} |f_n, \tilde{f}_n\rangle
    \\  
    \nonumber &\qquad\qquad\quad\qquad\qquad\quad\times\langle f_n, \tilde{f}_n | e^{C_n a_n + \tilde{C}_n \tilde{a}_n} e^{|w|^{2n} a_n^\dagger \cdot N(v_2) \cdot \tilde{a}_n^\dagger} |0\rangle 
    \\[0.3cm]
    \nonumber &= \prod_{n=1}^\infty \int d^{2}f_n d^{2}\tilde{f}_n e^{-|f_n|^2-|\tilde{f}_n|^2} e^{(\tilde{f}_n+\tilde{B}_n)N(v_1)(f_n+B_n)} e^{|w|^{2n}(\tilde{f}_n^*+\tilde{C}_n^*)N(v_2)(f_n^*+C_n^*)} 
    \\[0.3cm]
    \nonumber &= \prod_{n=1}^\infty \int d^{2}f_n d^{2}\tilde{f}_n \exp\left\{ -\begin{pmatrix} f_n^* \\ \tilde{f}_n \end{pmatrix}^T \begin{pmatrix} \mathds{1} & -|w|^{2n}N(v_2) \\ -N(v_1) & \mathds{1} \end{pmatrix} \begin{pmatrix} f_n \\ \tilde{f}_n^* \end{pmatrix} \right\}
    \\
    \nonumber &\quad\times\exp\left\{ \begin{pmatrix} f_n^* \\ \tilde{f}_n \end{pmatrix}^T \begin{pmatrix} |w|^{2n}N(v_2)\tilde{C}_n^* \\ B_n \end{pmatrix} + \begin{pmatrix} \tilde{B}_n N(v_1) & |w|^{2n} C_n^* N(v_2) \end{pmatrix} \begin{pmatrix} f_n \\ \tilde{f}_n^* \end{pmatrix} \right\} 
    \\
    &\quad\times\exp\left\{ \tilde{B}_n N(v_1) B_n + |w|^{2n} C_n^* N(v_2) \tilde{C}_n^* \right\}\label{aux1111}
\end{align}
and computing  the Gaussian integrals, we obtain:
\begin{align}
    \prod_{n=1}^\infty &\det(\mathds{1}-|w|^{2n}M(2v_{21}))^{-1} \exp\left\{\tilde{B}_n N(v_1) B_n + |w|^{2n} C_n^* N(v_2) \tilde{C}_n^*\right\} \nonumber 
    \\
    \times&\exp\left\{ |w|^{2n} \tilde{B}_n N(v_1) N(v_2) (\mathds{1}-|w|^{2n}N(v_1)N(v_2))^{-1} |w|^{2n} N(v_1) N(v_2) \tilde{C}_n^* \right. \nonumber \\
    &\qquad + |w|^{4n} \tilde{B}_n N(v_1) N(v_2) (\mathds{1}-|w|^{2n}N(v_1)N(v_2))^{-1} N(v_1) B_n \nonumber \\
    &\qquad + |w|^{2n} C_n^* N(v_2) (\mathds{1}-|w|^{2n}N(v_1)N(v_2))^{-1} N(v_1)N(v_2) |w|^{2n} \tilde{C}_n^* \nonumber \\
    &\qquad \left. + |w|^{2n} C_n^* N(v_2) (\mathds{1}-|w|^{2n}N(v_1)N(v_2))^{-1} N(v_1) B_n \right\}
\end{align}
Using \eqref{BnCn_defs}, we can calculate the sums in the exponents in the above expression:
\begin{subequations}
\begin{align}
    \sum_{n=1}^\infty \tilde{B}_n N(v_1) B_n &= -\tilde{A}N(v_1)A \log(1-|z|^2) 
    \\
    \sum_{n=1}^\infty \tilde{C}_n N(v_2) C_n &= -\tilde{A}N(v_2)A \log\left(1-\left|\frac{w}{z}\right|^2\right) 
    \\
    \nonumber \sum_{n=1}^\infty |w|^{2n} C_n^* N(v_2)&(\mathds{1}-|w|^{2n}N(v_1)N(v_2))^{-1} N(v_1)N(v_2) |w|^{2n} \tilde{C}_n^*
    \\
    &= -\sum_{m=1}^\infty \left[ A \cdot N(v_2+mv_{21}) \cdot \tilde{A} \right] \log(1-|\frac{w}{z}|^2|w|^{2m})
    \\
    \nonumber \sum_{n=1}^\infty |w|^{2n} \tilde{B}_n N(v_1)N(v_2)&(\mathds{1}-|w|^{2n}N(v_1)N(v_2))^{-1}N(v_1) B_n 
    \\
    &= -\sum_{m=1}^{\infty} [\tilde{A}\cdot N(v_1+mv_{12})\cdot A] \log(1-|z|^2|w|^{2m}).
\end{align}
\end{subequations}
Let us now consider the sum $\sum_{n=1}^\infty |w|^{2n} \tilde{B}_n N(v_1) N(v_2) \tilde{C}_n^*$. Writing explicitly the definition of $B_n$ and $C_n$ vectors, and using multiplication identity \eqref{Nmatrix_identity} for the $N(v)$-matrices, we obtain:
\begin{align}
    \nonumber \sum_{n=1}^{\infty}|w|^{2n}\tilde{B}_{n}N(v_1) N(v_2) \tilde{C}^*_n =&\;-\sum_{n=1}^{\infty}\frac{|w|^{2n}}{n}\left(\sqrt{\tfrac{\alpha'}{2}}k + \frac{n}{\bar{z}}\tilde{\epsilon}\right)\cdot M(2v_{21})\cdot \left(\sqrt{\tfrac{\alpha'}{2}}k - \frac{n}{\bar{z}}\tilde{\epsilon}\right)
    \\
    \nonumber =&\;\frac{\alpha'}{2}[k\cdot M(2v_{21})\cdot k]\log(1 - |w|^{2})
    \\
    \nonumber &+ \sqrt{\frac{\alpha'}{2}}\frac{|w|^2/\bar{z}}{1 -|w|^2} \left[k\cdot M(2v_{21})\cdot \tilde{\epsilon} - \tilde{\epsilon}\cdot M(2v_{21})\cdot k\right] 
    \\
    &+ \frac{\tilde{\epsilon}\cdot M(2v_{21})\cdot \tilde{\epsilon}}{\bar{z}^2}\frac{|w|^2}{(1-|w|^2)^2}\label{confusing_eq}
\end{align}
Clearly, the last term does not contribute to the graviton scattering, since it is quadratic in $\tilde{\epsilon}$. We will set this term to zero. In this sense, we identify:
\begin{align}
    \nonumber \sum_{n=1}^{\infty} |w|^{2n}\tilde{B}_{n}N(v_1) N(v_2) \tilde{C}^*_n =\;&\frac{\alpha'}{2}[k\cdot M(2v_{21})\cdot k]\log(1 - |w|^2)
    \\
    &+ \sqrt{\frac{\alpha'}{2}}\frac{1}{\bar{z}}\frac{4|w|^2}{1 - |w|^2} \sin(v_{12})\cos(v_{12})(k^1 \tilde{\epsilon}^0 - k^0 \tilde{\epsilon}^1).
\end{align}
Similarly,
\begin{align}
    \nonumber \sum_{n=1}^\infty |w|^{4n} \tilde{B}_n N(v_1)N(v_2)&(\mathds{1}-|w|^{2n}N(v_1)N(v_2))^{-1}N(v_1)N(v_2) \tilde{C}_n^*
    \\
    \nonumber \equiv&\;\frac{\alpha'}{2}\sum_{m=2}^\infty [k\cdot M(2mv_{21})\cdot k]\log(1 - |w|^{2m})
    \\
    &+ \sqrt{\frac{\alpha'}{2}}\frac{1}{\bar{z}} \sum_{m=2}^\infty\frac{4|w|^{2m}}{1 - |w|^{2m}} \sin(mv_{12})\cos(mv_{12})(k^1 \tilde{\epsilon}^0 - k^0 \tilde{\epsilon}^1),
\end{align}
and
\begin{align}
    \nonumber \sum_{n=1}^\infty |w|^{2n} C_n^* N(v_2)&(\mathds{1}-|w|^{2n}N(v_1)N(v_2))^{-1}N(v_1) B_n
    \\
    \nonumber =&\; \frac{\alpha'}{2}\sum_{m=1}^\infty \left[ k \cdot M(2mv_{12}) \cdot k \right] \log(1-|w|^{2m})
    \\
    &+ \sqrt{\frac{\alpha'}{2}}\frac{1}{\bar{z}}\sum_{m=1}^\infty \frac{4|w|^{2m}}{1 - |w|^{2m}} \sin(mv_{12})\cos(mv_{12})(k^1 \epsilon^0 - k^0 \epsilon^1),
\end{align}
Collecting the above results, we obtain:
\begin{align}
    &\prod_{n=1}^\infty (1-2|w|^{2n}\cos(2v_{21})+|w|^{4n})^{-1}(1-|w|^{2n})^{\frac{\alpha'}{2}k\cdot M(2nv_{12})\cdot k +\frac{\alpha'}{2}k\cdot M(2nv_{21})\cdot k}\nonumber 
    \\
    \nonumber &\times \prod_{n=0}^\infty (1-|z|^2|w|^{2n})^{-\tilde{A}N(v_1+nv_{12})A} \left(1-\left|\frac{w}{z}\right|^2|w|^{2n}\right)^{-\tilde{A}N(v_2+nv_{21})A}
    \\
    &\times\prod_{n=1}^\infty \exp\left\{\sqrt{\frac{\alpha'}{2}}\frac{4|w|^{2n}}{1-|w|^{2n}}\sin(nv_{12})\cos(n v_{12})\left[\frac{1}{z}(k^1 \epsilon^0 - k^0 \epsilon^1) + \frac{1}{\bar{z}}(k^1 \tilde{\epsilon}^0 - k^0 \tilde{\epsilon}^1)\right]\right\}
\end{align}
It turns out that it is very useful  to introduce the following notation. For any Lorentz vector $X^\mu$, we define
\begin{equation}
X_v = \cos(v) X^0 + \sin(v) X^1.
\end{equation}
With this convention, we can recast various matrix products as:
\begin{subequations}
\begin{align}
    &\sin(n v_{12})(k^1 \epsilon^0 - k^{0}\epsilon^1) = \epsilon_{nv_{1}}k_{nv_2} - \epsilon_{nv_{2}}k_{nv_1},
    \\[0.3cm]
    &\tilde{A} N(v_{1} + n v_{12})A = -2\tilde{A}_{v_1 + n v_{12}}A_{v_1 + nv_{12}} + \tilde{A}^0 A^0 + \tilde{A}^1 A^1,
    \\[0.3cm]
    &k\cdot [M(2nv_{12}) + M(2nv_{21})]\cdot k = (k^0)^2 + (k^1)^2-4\sin^2(n v_{12})((k^0)^2 + (k^1)^2).
\end{align}
\end{subequations}

Joining together contributions from transverse and longitudinal directions, we conclude that the total contribution of bosonic nonzero modes is:
\begin{align}
    \nonumber &\prod_{n=1}^\infty (1 - 2|w|^{2n}\cos(2v_{21}) + |w|^{4n})^{-1}(1 - |w|^{2n})^{2 - D -2\alpha'\sin^2(nv_{12})(k_0^2 + k_1^2) + \alpha' k^2}
    \\
    \nonumber &\times \prod_{n=0}^\infty (1-|z|^2|w|^{2n})^{2\tilde{A}_{v_1 + nv_{12}}A_{v_1 + nv_{12}}-{\tilde{A}}\cdot {A}}(1-|\frac{w}{z}|^2|w|^{2n})^{2\tilde{A}_{v_2 + nv_{21}}A_{v_2 + nv_{21}}-{\tilde{A}}\cdot{A}}
    \\
    &\times \prod_{n=1}^\infty e^{\sqrt{\frac{\alpha'}{2}}\frac{4|w|^{2n}}{1 - |w|^{2n}} \cos(n v_{12})\left[\frac{1}{z}(\epsilon_{nv_{1}}k_{nv_{2}} - \epsilon_{nv_{2}}k_{nv_{1}}) + \frac{1}{\bar{z}}(\tilde{\epsilon}_{nv_{1}}k_{nv_{2}} - \tilde{\epsilon}_{nv_{2}}k_{nv_{1}})\right]}.
\end{align}
Analytic continuation of the above result to Lorentzian signature amounts to replacing $k^0\rightarrow -ik^0,$ $\sin(v_1)\rightarrow i\sinh(v_1)$, $\cos(v_1)\rightarrow \cosh(v_1)$, so that $\cos(v) A^0 + \sin (v)A^1\rightarrow -i(\cosh(v_1)A^0 -\sinh(v_1)A^1) = -iA_{v}$. Therefore, in Lorentzian signature, the  bosonic nonzero modes contribute:
\begin{align}
    \nonumber &\prod_{n=1}^\infty (1 - 2|w|^{2n}\cosh(2v_{21}) + |w|^{4n})^{-1}(1 - |w|^{2n})^{4 - D - 2\alpha'\sinh^2(nv_{12})(k_0^2 - k_1^2)}
    \\
    \nonumber &\times \prod_{n=0}^\infty (1-|z|^2|w|^{2n})^{-2\tilde{A}_{v_1 + nv_{12}}A_{v_1 + nv_{12}}-{\tilde{A}}\cdot {A}}(1-|\frac{w}{z}|^2|w|^{2n})^{-2\tilde{A}_{v_2 + nv_{21}}A_{v_2 + nv_{21}}-{\tilde{A}}\cdot{A}}
    \\
    &\times \prod_{n=1}^\infty e^{\sqrt{\frac{\alpha'}{2}}\frac{4|w|^{2n}}{1 - |w|^{2n}} \cosh(n v_{12})\left[\frac{1}{z}(\epsilon_{nv_{1}}k_{nv_{2}} - \epsilon_{nv_{2}}k_{nv_{1}}) + \frac{1}{\bar{z}}(\tilde{\epsilon}_{nv_{1}}k_{nv_{2}} - \tilde{\epsilon}_{nv_{2}}k_{nv_{1}})\right]}.\label{RSVE4}
\end{align}
Here we have included the ghost contribution, which changes the power of the partition-function factor $(1-|w|^{2n})$ from 2 to 4, and assumed that the emitted massless particle is on-shell, so that $k^2 = 0$.

\section{Calculation of the contribution of NS-NS fermions}
\label{AppB}

The matrix element for NS-NS fermions arising from \eqref{EHK6} reads:
\begin{align}\label{NS_NS_modes_full}
    \prod_{r=1/2}^\infty \langle 0 | e^{-i\eta_1 \tilde{b}_r^{||} N(v_1) b_r^{||} -i\eta_1 \tilde{b}_r^{\perp} \cdot b_r^{\perp}} e^{D_r\cdot b_r^\dagger} e^{\tilde{D}_r\cdot\tilde{b}_r^\dagger} e^{E_r\cdot b_r} e^{\tilde{E}_r\cdot \tilde{b}_r} e^{i\eta_2 |w|^{2r} b_r^{||}{}^\dagger N(v_2) \tilde{b}_r^{||}{}^\dagger+i\eta_2 |w|^{2r} b_r^{\perp}{}^\dagger \cdot \tilde{b}_r^{\perp}{}^\dagger} |0\rangle,
\end{align}
where  
\begin{equation}
    D_r = -R\,z^{r-1/2},\quad E_r = -R\,z^{-r-1/2},\quad \tilde{D}_r = -\tilde{R}\,\bar{z}^{r-1/2},\quad \tilde{E}_r = -\tilde{R}\,\bar{z}^{-r-1/2},
\end{equation}
and
\begin{equation}
    R =  \sqrt{\tfrac{\alpha'}{2}}\,k\theta + \epsilon \phi,\quad \tilde{R} = \sqrt{\tfrac{\alpha'}{2}}\,k\tilde{\theta} + \tilde{\epsilon}\tilde{\phi}.
\end{equation}

\subsection{Boosted directions}\label{B1}
The part of \eqref{NS_NS_modes_full} corresponding to longitudinal directions is:
\begin{align}\label{NS_NS_modes_longitudinal}
    \prod_{r=1/2}^\infty \langle 0 | e^{-i\eta_1 \tilde{b}_r^{||} N(v_1) b_r^{||}} e^{D_r^{||}\cdot b_r^{||}{}^\dagger} e^{\tilde{D}_r^{||}\cdot\tilde{b}_r^{||}{}^\dagger} e^{E_r^{||}\cdot b_r^{||}{}} e^{\tilde{E}_r^{||}\cdot \tilde{b}_r^{||}{}} e^{i\eta_2 |w|^{2r} b_r^{||}{}^\dagger N(v_2) \tilde{b}_r^{||}{}^\dagger} |0\rangle.
\end{align}
From now on, we will suppress the superscript $||$. Let us introduce fermionic coherent states: $|f_r\rangle = e^{b_r^\dagger f_r}|0\rangle$, where $f_r$ is a Grassmann variable, and $\langle f_r| = \langle 0|e^{f_r^* b_r}$. They satisfy:
\begin{subequations}
\begin{align}
    b_r|f_r\rangle &= f_r|f_r\rangle,\quad \langle f_r|b_r^\dagger = -\langle f_r|f_r^*,\quad\langle f_r|f_r\rangle = e^{-f_r^*f_r}.
\end{align}
\end{subequations}
Inserting the resolution of identity:
\begin{equation}
    \mathds{1} = \prod_{r=1/2}^\infty \left( \int df_r df_r^* \int d\tilde{f}_r d\tilde{f}_r^*\; e^{f_r^*f_r} e^{\tilde{f}_r^*\tilde{f}_r} |f_r, \tilde{f}_r\rangle \langle f_r, \tilde{f}_r| \right),
\end{equation}
we recast \eqref{NS_NS_modes_longitudinal} as:
\begin{align}
    &\prod_{r=1/2}^\infty \int d^2f_r d^2\tilde{f}_r\, e^{f_r^*f_r} e^{\tilde{f}_r^*\tilde{f}_r} \langle 0| e^{-i\eta_1 \tilde{b}_r N(v_1) b_r} e^{D_r b_r^\dagger} e^{\tilde{D}_r \tilde{b}_r^\dagger} |f_r, \tilde{f}_r\rangle \nonumber 
    \\
    &\qquad \qquad \qquad \qquad \qquad \;\; \times \langle f_r, \tilde{f}_r| e^{E_r b_r} e^{\tilde{E}_r \tilde{b}_r} e^{i\eta_2 |w|^{2r} b_r^\dagger N(v_2) \tilde{b}_r^\dagger} |0\rangle \nonumber 
    \\[0.2cm]
    &= \prod_{r=1/2}^\infty \int d^2f_r d^2\tilde{f}_r\, e^{f_r^*f_r} e^{\tilde{f}_r^*\tilde{f}_r} \langle 0| e^{-i\eta_1 \tilde{b}_r N(v_1) b_r} |f_r-D_r, \tilde{f}_r-\tilde{D}_r\rangle \nonumber 
    \\
    &\qquad \qquad \qquad \qquad \qquad\;\;\times \langle f_r-E_r, \tilde{f}_r-\tilde{E}_r| e^{i\eta_2|w|^{2r}b_r^\dagger N(v_2)\tilde{b}_r^\dagger} |0\rangle \nonumber 
    \\[0.2cm]
    &= \prod_{r=1/2}^\infty \int d^2f_r d^2\tilde{f}_r\, e^{f_r^*f_r - \tilde{f}_r\tilde{f}_r^*} e^{i\eta_1(f_r-D_r)N(v_1)(\tilde{f}_r-\tilde{D}_r)} e^{i\eta_2|w|^{2r}(f_r^*-E_r)N(v_2)(\tilde{f}_r^*-\tilde{E}_r)} \nonumber 
    \\[0.2cm]
    &= \prod_{r=1/2}^\infty \int d^2f_r d^2\tilde{f}_r\, \exp\left\{i\eta_1 D_r N(v_1)\tilde{D}_r + i\eta_2|w|^{2r}E_r N(v_2)\tilde{E}_r \right\} \nonumber 
    \\
    &\quad \times \exp\left\{ \begin{pmatrix} f_r^* & -\tilde{f}_r \end{pmatrix} \begin{pmatrix} \mathds{1} & i\eta_2|w|^{2r}N(v_2) \\ i\eta_1 N(v_1) & \mathds{1} \end{pmatrix} \begin{pmatrix} f_r \\ \tilde{f}_r^* \end{pmatrix} \right\}\nonumber 
    \\
    &\quad \times \exp\left\{(i\eta_1\tilde{D}_r N(v_1), -i\eta_2|w|^{2r}E_r N(v_2)) \begin{pmatrix} f_r \\ \tilde{f}_r^* \end{pmatrix} \right\}\nonumber
    \\
    &\quad \times \exp\left\{ \begin{pmatrix} f_r^* & -\tilde{f}_r \end{pmatrix} \begin{pmatrix} -i\eta_2|w|^{2r}N(v_2)\tilde{E}_r \\ -i\eta_1 N(v_1)D_r \end{pmatrix}\right\}.
\end{align}
Calculating the Grassmann Gaussian integrals, we obtain:
\begin{align} \label{eq:star}
    \nonumber \prod_{r=1/2}^\infty &\det \begin{pmatrix} \mathds{1} & i\eta_2|w|^{2r}N(v_2) \\ i\eta_1 N(v_1) & \mathds{1} \end{pmatrix}\times 
    \\
    &\times \exp\Big\{i\eta_1 D_r N(v_1)\tilde{D}_r + i\eta_2|w|^{2r} E_r N(v_2)\tilde{E}_r -\eta_1\eta_2|w|^{2r} \tilde{D}_r N(v_1)N(v_2) E_r \nonumber 
    \\
    &\qquad\qquad +|w|^{4r} \tilde{D}_r N(v_1)N(v_2)(\mathds{1}+\eta_1\eta_2|w|^{2r}N(v_1)N(v_2))^{-1}N(v_1)N(v_2)E_r \nonumber 
    \\
    &\qquad\qquad +i\eta_2|w|^{2r} \tilde{D}_r N(v_1)N(v_2)(\mathds{1}+\eta_1\eta_2|w|^{2r}N(v_1)N(v_2))^{-1}N(v_1)D_r \nonumber 
    \\
    &\qquad\qquad -i\eta_1|w|^{4r} E_r N(v_2)(\mathds{1}+\eta_1\eta_2|w|^{2r}N(v_1)N(v_2))^{-1}N(v_1)N(v_2)E_r \nonumber 
    \\
    &\qquad\qquad +\eta_1\eta_2|w|^{2r} E_r N(v_2)(\mathds{1}+\eta_1\eta_2|w|^{2r}N(v_1)N(v_2))^{-1}N(v_1)D_r \Big\}.
\end{align}
Using the standard formula for determinant of a block matrix, we obtain:
\begin{align}
    \nonumber \det\begin{pmatrix}
        \mathds{1} & i\eta_2 |w|^{2r}N(v_2)
        \\
        i\eta_1 N(v_1) & \mathds{1}
    \end{pmatrix} &= \det(\mathds{1} + \eta_1 \eta_2 |w|^{2r}N(v_1)N(v_2))
    \\
    &= 1 + 2\eta_1 \eta_2 |w|^{2r} \cos(2v_{21}) + |w|^{4r}.
\end{align}
Calculating the sums in the exponent from the above expression,
\begin{subequations}
\begin{align}
    \sum_{r=1/2}^\infty i\eta_1 D_r N(v_1) \tilde{D}_r &= -i\eta_1 \tilde{R}N(v_1)R \frac{1}{1-|z|^2}
    \\
    \sum_{r=1/2}^\infty i\eta_2|w|^{2r}E_r N(v_2)\tilde{E}_r & = -i\eta_2\frac{|w|}{|z|^2} \tilde{R}N(v_2)R \frac{1}{1-|w/z|^2} 
    \\
    \sum_{r=1/2}^\infty -\eta_1\eta_2|w|^{2r}\tilde{D}_r N(v_1)N(v_2)\tilde{E}_r &= -\frac{\eta_1\eta_2}{\bar{z}}\tilde{R}M(2v_{21})R \frac{|w|}{1-|w|^2}
    \\
    \nonumber \sum_{r=1/2}^\infty |w|^{4r} \tilde{D}_r N(v_1)N(v_2)&(\mathds{1} +\eta_1\eta_2|w|^{2r}N(v_1)N(v_2))^{-1}N(v_1)N(v_2)\tilde{E}_r 
    \\
    &= \frac{1}{\bar{z}}\sum_{n=2}^\infty \frac{(-\eta_1\eta_2|w|)^n}{1-|w|^{2n}}\tilde{R}M(2nv_{21})\tilde{R},
    \\
    \nonumber \sum_{r=1/2}^\infty \eta_1\eta_2|w|^{2r} E_r N(v_2)&(\mathds{1} +\eta_1\eta_2|w|^{2r}N(v_1)N(v_2))^{-1}N(v_1)D_r 
    \\
    &= \frac{1}{z}\sum_{n=1}^\infty \frac{(-\eta_1\eta_2|w|)^n}{1-|w|^{2n}} R M(2nv_{12})R 
    \\
    \nonumber \sum_{r=1/2}^\infty i\eta_2|w|^{2r}\tilde{D}_r N(v_1)N(v_2)&(\mathds{1} +\eta_1\eta_2|w|^{2r}N(v_1)N(v_2))^{-1}N(v_1)D_r
    \\
    &= -i\eta_1 \sum_{n=1}^\infty \tilde{R}N(v_1+nv_{12})R \frac{(-\eta_1\eta_2|w|)^n}{1-|z|^2|w|^{2n}} 
    \\
    \nonumber \sum_{r=1/2}^\infty -i\eta_1|w|^{4r} E_r N(v_2)&(\mathds{1} +\eta_1\eta_2|w|^{2r}N(v_1)N(v_2))^{-1}N(v_1)N(v_2)\tilde{E}_r
    \\
    &= -i\eta_2\frac{|w|}{|z|^2}\sum_{n=1}^\infty \tilde{R}N(v_2+nv_{21})R \frac{(-\eta_1\eta_2|w|)^n}{1-|\frac{w}{z}|^2|w|^{2n}}
\end{align}
\end{subequations}
we obtain:
\begin{align}\label{NS_NS_longitudinal2}
    &\prod_{r=1/2}^\infty (1+2\eta_1\eta_2|w|^{2r}\cos(2v_{21})+|w|^{4r}) \\
    &\quad \times \prod_{n=0}^\infty e^{-i\eta_1 \tilde{R}N(v_1+nv_{12})R \frac{(-\eta_1\eta_2|w|)^n}{1-|z|^2|w|^{2n}}} e^{-i\eta_2 \frac{|w|}{|z|^2} \tilde{R}N(v_2+nv_{21})R \frac{(-\eta_1\eta_2|w|)^n}{1-|w/z|^2|w|^{2n}}} \nonumber \\
    &\quad \times \prod_{n=1}^\infty \exp\left\{ \frac{(-\eta_1\eta_2|w|)^n}{1-|w|^{2n}}[\bar{z}^{-1}\tilde{R}M(2nv_{21})\tilde{R} + z^{-1}R M(2nv_{12})R]\right\} \nonumber
\end{align}
The argument of the last exponent can be simplified by explicitly multiplying the matrices. One can easily check that:
\begin{equation}
    R\,M(2nv_{12})\,R = -4\sqrt{\frac{\alpha'}{2}}\cos(nv_{12})
    (k_{nv_1}\epsilon_{nv_2}-k_{nv_2}\epsilon_{nv_1})\theta\phi.
\end{equation}

\subsection{Transverse directions}
The calculation of the transverse part of \eqref{NS_NS_modes_full}, given by
\begin{align}\label{NS_NS_modes_transverse}
    \prod_{r=1/2}^\infty \langle 0 | e^{-i\eta_1 \tilde{b}_r^{\perp} \cdot b_r^{\perp}} e^{D_r^\perp\cdot b^\perp_r{}^\dagger} e^{\tilde{D}^\perp_r\cdot\tilde{b}^\perp_r{}^\dagger} e^{E_r^\perp\cdot b_r^\perp} e^{\tilde{E}_r^\perp\cdot \tilde{b}_r^\perp} e^{i\eta_2 |w|^{2r} b_r^{\perp}{}^\dagger \cdot \tilde{b}_r^{\perp}{}^\dagger} |0\rangle,
\end{align}
proceeds in the same way as for directions $0,1$. The only difference is that we need to replace the $2\times 2$ matrices $N(v_1)$, $N(v_2)$ with $(D-2)$-dimensional identity matrices $\mathds{1}$. This gives us:
\begin{equation}\label{NS_NS_transverse2}
    \left(\prod_{r=1/2}^\infty (1+\eta_1\eta_2|w|^{2r})^{D-2}\right) \left(\prod_{n=0}^\infty e^{-i\eta_1\tilde{R}^\perp\cdot R^\perp\frac{(-\eta_1\eta_2|w|)^n}{1-|z|^2|w|^{2n}}} e^{-i\eta_2\frac{|w|}{|z|^2}\tilde{R}^\perp\cdot R^{\perp}\frac{(-\eta_1\eta_2|w|)^n}{1-|w/z|^2|w|^{2n}}} \right)
\end{equation}
Combining \eqref{NS_NS_longitudinal2} and \eqref{NS_NS_transverse2}, using the relation
\begin{equation}
    {\tilde{R}}_\perp R_\perp +{\tilde{R}}N(v)R= -2 {\tilde{R}}_{v}R_{v}+{\tilde{R}}^0 R^0+{\tilde{R}}^1 R^1+{\tilde{R}}_\perp R_\perp\end{equation} 
and continuing both expressions to Lorentzian signature, we obtain:
\begin{align}\label{NSfermions_result}
    \nonumber &\prod_{r=1/2}^\infty (1+2\eta_1\eta_2|w|^{2r}\cosh(2v_{21}) + |w|^{4r})(1+\eta_1\eta_2|w|^{2r})^{D - 2}
    \\
    \nonumber &\times \prod_{n=0}^\infty e^{-i\eta_1(\tilde{R}\cdot R + 2\tilde{R}_{v_1 + nv_{12}}R_{v_1 + nv_{12}})\frac{(-\eta_1 \eta_2 |w|)^n}{1 - |z|^2|w|^{2n}}} e^{-\frac{i\eta_2|w|}{|z|^2}(\tilde{R}\cdot R + 2\tilde{R}_{v_2 + nv_{21}}R_{v_2 + nv_{21}})\frac{(-\eta_1 \eta_2 |w|)^n}{1 - |\frac{w}{z}|^2|w|^{2n}}}
    \\
    &\times \prod_{n=1}^\infty
    e^{-4\cosh(nv_{12})\sqrt{\frac{\alpha'}{2}}
    \frac{(-\eta_1 \eta_2|w|)^n}{1 - |w|^{2n}}
    \left[\frac{1}{z}(k_{nv_1}\epsilon_{nv_2}-k_{nv_2}\epsilon_{nv_1})\,
    \theta\phi + \frac{1}{\bar z}(k_{nv_1}\tilde{\epsilon}_{nv_2}-k_{nv_2}\tilde{\epsilon}_{nv_1})\,
    \tilde{\theta}\tilde{\phi} \right]}.
\end{align}
To account for the NS-NS sector $\beta\gamma$ ghosts, we need to reduce the power of the partition function factor $(1 + \eta_1 \eta_2 |w|^{2r})$ by two, changing it from $D-2$ to $D-4$.

\section{Calculation of the contribution of R-R fermions}
\label{AppC}

\subsection{Nonzero modes}
The contribution of the nonzero fermionic modes in the Ramond-Ramond sector is given by:
\begin{equation}\label{R_R_modes_full}
    \prod_{n=1}^\infty \langle 0 | e^{-i\eta_1 \tilde{d}_n^{||} N(v_1) d_n^{||} -i\eta_1 \tilde{d}_n^{\perp} \cdot d_n^{\perp}} e^{D_n\cdot d_n^\dagger} e^{\tilde{D}_n\cdot\tilde{d}_n^\dagger} e^{E_n\cdot d_n} e^{\tilde{E}_n\cdot \tilde{d}_n} e^{i\eta_2 |w|^{2n} d_n^{||}{}^\dagger N(v_2) \tilde{d}_n^{||}{}^\dagger+i\eta_2 |w|^{2n} d_n^{\perp}{}^\dagger \cdot \tilde{d}_n^{\perp}{}^\dagger} |0\rangle.
\end{equation}
Note that this is the same matrix elements as in the NS-NS sector \eqref{NS_NS_modes_full}, the only difference being that we replace half-integers $r$ with integers $n$ (and we denote fermionic annihilation operators as $d_n$). Following the steps from App.~\ref{AppB}, we rewrite \eqref{R_R_modes_full} as (in Lorentzian signature):
\begin{align}
    \nonumber & \prod_{n=1}^\infty (1+2\eta_1\eta_2|w|^{2n}\cosh(2v_{21}) + |w|^{4n})(1+\eta_1\eta_2|w|^{2n})^{D - 2}
    \\
    \nonumber &\times \prod_{n=0}^\infty e^{-i\eta_1|z|(2\tilde{R}_{v_1 + nv_{12}}R_{v_1 + nv_{12}} + \tilde{R}\cdot R)\frac{(-\eta_1 \eta_2 |w|^2)^n}{1 - |z|^2|w|^{2n}}} e^{- \frac{i\eta_2|w|^2}{|z|^3}(2\tilde{R}_{v_2 + nv_{21}}R_{v_2 + nv_{21}} + \tilde{R}\cdot R)\frac{(-\eta_1 \eta_2 |w|^2)^n}{1 - |\frac{w}{z}|^2|w|^{2n}}}
    \\
    &\times \prod_{n=1}^\infty
    e^{-4\cosh(nv_{12})\sqrt{\frac{\alpha'}{2}}
    \frac{(-\eta_1 \eta_2|w|^2)^n}{1 - |w|^{2n}}
    \left[\frac{1}{z}(k_{nv_1}\epsilon_{nv_2}-k_{nv_2}\epsilon_{nv_1})\,
    \theta\phi + \frac{1}{\bar z}(k_{nv_1}\tilde{\epsilon}_{nv_2}-k_{nv_2}\tilde{\epsilon}_{nv_1})\,
    \tilde{\theta}\tilde{\phi} \right]}\label{RR_nonzero_modes}
\end{align}
To account for the contribution of the $\beta\gamma$ ghosts, we change the power of the partition function  $(1 + \eta_1 \eta_2 |w|^{2r})$ from $D-2$ to $D-4$.

\subsection{Zero modes}
In this part we will work in Lorentzian signature in order to use standard conventions for $\Gamma$-matrices and charge conjugation,
\begin{equation}
    \{\Gamma^\mu,\Gamma^\nu\} = 2\eta^{\mu\nu},\quad (\Gamma^\mu)^T = - C\Gamma^\mu C^{-1},
\end{equation}
where $C$ is the charge-conjugation matrix, $C_{AB} = -C_{BA}$, where indices $A,B$ run from $1$ to $32$. Our goal is to calculate the following matrix element:
\begin{equation}\label{RR_zero_mode}
    (\mathcal{N}_{\eta_1})_{AB}(\mathcal{M}_{\eta_2})_{CD}\bra{A}\bra{\tilde{B}}e^{-z^{-1/2}R\cdot d_0}e^{-\bar{z}^{-1/2}\tilde{R}\cdot \tilde{d}_0}\ket{C}\ket{\tilde{D}},
\end{equation}
with Ramond-Ramond zero modes $d_{0}^\mu$, $\tilde{d}_0^\mu$ acting on the states $\ket{A}\ket{\tilde{B}}$ as
\begin{equation}
    d_0^\mu \ket{A}\ket{\tilde{B}}=\frac{1}{\sqrt{2}}(\Gamma^\mu)_{AC}\ket{C}\ket{\tilde{B}},\quad \tilde{d}_0^\mu \ket{A}\ket{\tilde{B}}=\frac{1}{\sqrt{2}}(\Gamma^{11})_{AC}(\Gamma^\mu)_{BD}\ket{C}\ket{\tilde{D}}. 
\end{equation}
The states $\ket{A}{\ket{\tilde{B}}}$ are normalized to
\begin{equation}
    \langle A|\langle\tilde{B}| |C\rangle|\tilde{D}\rangle = (C^{-1})_{AC}(C^{-1})_{BD}.
\end{equation}
The matrices $\mathcal{N}_{\eta_1}$, $\mathcal{M}_{\eta_2}$ are given by~\footnote{See Eqs. 7.235 and 7.240 of \cite{DiVecchia:1999mal} upon redefinition $\eta_1\to -\eta_1$.}:
\begin{align}
    \mathcal{N}_{\eta_1} = C(\cosh(v_1)\Gamma^0-\sinh(v_1)\Gamma^1)\frac{1-i\eta_1\Gamma_{11}}{1+i\eta_1},
    \\
    \mathcal{M}_{\eta_2} = C(\cosh(v_2) \Gamma^0 - \sinh(v_2)\Gamma^1)\frac{1+i\eta_2 \Gamma_{11}}{1 + i\eta_2}.
\end{align}
Taylor-expanding  the exponents in \eqref{RR_zero_mode}, we obtain:
\begin{align}
\nonumber (\mathcal{N}_{\eta_1})_{AB}(\mathcal{M}_{\eta_2})_{CD}\bra{A}\bra{\tilde{B}}\Big[1 &+ \frac{1}{|z|}(R\cdot d_0)(\tilde{R}\cdot \tilde{d}_0) + \frac{1}{2z}(R\cdot d_0)^2
\\
&+ \frac{1}{2\bar{z}}(\tilde{R}\cdot\tilde{d}_0)^2 + \frac{1}{4|z|^2}(R\cdot d_0)^2 (\tilde{R}\cdot \tilde{d}_0)^2\Big]\ket{C}\ket{\tilde{D}}.
\end{align}
We have five terms to calculate. The first term from the above expression is:
\begin{align}
    \nonumber (\mathcal{N}_{\eta_1})_{AB}(\mathcal{M}_{\eta_2})_{CD}\bra{A}\bra{\tilde{B}}\ket{C}\ket{\tilde{D}} &= (C^{-1})_{AC}(C^{-1})_{BD}(\mathcal{N}_{\eta_1})_{AB}(\mathcal{M}_{\eta_2})_{CD} 
    \\
    \nonumber &= \text{Tr}\{\mathcal{N}_{\eta_1}^TC^{-1}\mathcal{M}_{\eta_2}(C^{-1})^T\}
    \\
    &= -\text{Tr}\{\mathcal{N}_{\eta_1}^TC^{-1}\mathcal{M}_{\eta_2}C^{-1}\}.\label{first_one}
\end{align}
The second term gives us:
\begin{align}
    \nonumber &(\mathcal{N}_{\eta_1})_{AB}(\mathcal{M}_{\eta_2})_{CD}\bra{A}\bra{\tilde{B}}(R\cdot d_0)(\tilde{R}\cdot \tilde{d}_0)\ket{C}\ket{\tilde{D}} 
    \\
    \nonumber &= \frac{1}{\sqrt{2}}R_{\mu}\tilde{R}_{\nu}\bra{A}\bra{\tilde{B}}d_0^\mu (\Gamma^{11})_{CE}(\Gamma^\nu)_{DF}\ket{E}\ket{\tilde{F}}(\mathcal{N}_{\eta_1})_{AB}(\mathcal{M}_{\eta_2})_{CD}
    \\
    \nonumber &= \frac{1}{2}R_{\mu}\tilde{R}_{\nu}(C^{-1})_{AG}(C^{-1})_{BH}(\Gamma^{\mu})_{EG}\delta_{FH}(\Gamma^{11})_{CE}(\Gamma^\nu)_{DF}(\mathcal{N}_{\eta_1})_{AB}(\mathcal{M}_{\eta_2})_{CD}
    \\
    \nonumber &= \frac{1}{2}R_{\mu}\tilde{R}_{\nu}\text{Tr}\{\mathcal{N}_{\eta_1}^TC^{-1}(\Gamma^{\mu})^T(\Gamma_{11})^T\mathcal{M}_{\eta_2}\Gamma^\nu (C^{-1})^T\}
    \\
    &= -\frac{1}{2}R_{\mu}\tilde{R}_{\nu}\text{Tr}\{\mathcal{N}_{\eta_1}^T\Gamma^{\mu}\Gamma_{11}C^{-1}\mathcal{M}_{\eta_2}\Gamma^\nu C^{-1}\}.\label{second_term}
\end{align}
From the third one, we obtain:
\begin{align}
    \nonumber &\bra{A}\bra{\tilde{B}}(R\cdot d_0)^2 \ket{C}\ket{\tilde{D}}(\mathcal{N}_{\eta_1})_{AB}(\mathcal{M}_{\eta_2})_{CD}=
    \\
    \nonumber &= (\mathcal{N}_{\eta_1})_{AB}(\mathcal{M}_{\eta_2})_{CD}R_\mu R_\nu \cdot \frac{1}{2}\bra{A}\bra{\tilde{B}}(\Gamma^\mu)_{EF}(\Gamma^\nu)_{CE}\ket{F}\ket{\tilde{D}}
    \\
    \nonumber &= (\mathcal{N}_{\eta_1})_{AB}(\mathcal{M}_{\eta_2})_{CD} \cdot \frac{1}{2}R_\mu R_\nu (\Gamma^\mu)_{EF}(\Gamma^\nu)_{CE} (C^{-1})_{AF}(C^{-1})_{BD}
    \\
    \nonumber &= \frac{1}{2}R_\mu R_\nu\text{Tr}\{\mathcal{N}_{\eta_1}^TC^{-1}(\Gamma^\mu)^{T}(\Gamma^\nu)^T\mathcal{M}_{\eta_2}(C^{-1})^T\}
    \\
    &= -\frac{1}{2}R_\mu R_\nu\text{Tr}\{\mathcal{N}_{\eta_1}^T\Gamma^\mu\Gamma^\nu C^{-1}\mathcal{M}_{\eta_2}C^{-1}\}\label{third_term}
\end{align}
The fourth one can be recast as:
\begin{align}
    \nonumber &(\mathcal{N}_{\eta_1})_{AB}(\mathcal{M}_{\eta_2})_{CD}\bra{A}\bra{\tilde{B}}(\tilde{R}\cdot \tilde{d}_0)^2 \ket{C}\ket{\tilde{D}}
    \\
    \nonumber &= \frac{1}{2}\tilde{R}_{\mu} \tilde{R}_{\nu} (C^{-1})_{AG}(C^{-1})_{BH}(\Gamma_{11})_{CE}(\Gamma_{11})_{EG}(\Gamma^{\mu})_{FH}(\Gamma^{\nu})_{DF}(\mathcal{N}_{\eta_1})_{AB}(\mathcal{M}_{\eta_2})_{CD}
    \\
    \nonumber &= \frac{1}{2}\tilde{R}_{\mu}\tilde{R}_{\nu}\text{Tr}\{\mathcal{N}_{\eta_1}^TC^{-1}(\Gamma_{11})^T(\Gamma_{11})^T\mathcal{M}_{\eta_2}\Gamma^\nu\Gamma^\mu(C^{-1})^T\}
    \\
    &= -\frac{1}{2}\tilde{R}_{\mu}\tilde{R}_{\nu}\text{Tr}\{\mathcal{N}_{\eta_1}^TC^{-1}\mathcal{M}_{\eta_2}\Gamma^\nu\Gamma^\mu C^{-1}\}.\label{fourth_term}
\end{align}
The fifth one is:
\begin{align}
    \nonumber &(\mathcal{N}_{\eta_1})_{AB}(\mathcal{M}_{\eta_2})_{CD}\bra{A}\bra{\tilde{B}}(R\cdot d_0)^2(\tilde{R}\cdot \tilde{d}_0)^2 \ket{C}\ket{\tilde{D}}
    \nonumber 
    \\
    & = -\frac{1}{4}R_\mu R_{\nu}\tilde{R}_{\rho}\tilde{R}_{\sigma} \text{Tr}\{\mathcal{N}_{\eta_1}^T\Gamma^\mu \Gamma^\nu C^{-1}\mathcal{M}_{\eta_1}\Gamma^\rho \Gamma^\sigma C^{-1}\}\label{fifth_term}
\end{align}
Since
\begin{equation}
    \mathcal{N}_{\eta}^T = C \frac{1 + i\eta \Gamma_{11}}{1 + i\eta}(c_v \Gamma^0 - s_v \Gamma^1)
\end{equation}
we can rewrite \eqref{first_one} as:
\begin{align}
    \nonumber \text{Tr}\{\mathcal{N}_{\eta_1}^T C^{-1}\mathcal{M}_{\eta_2}C^{-1}\} &= \text{Tr}\Big\{\frac{1 + i\eta_1 \Gamma_{11}}{1 + i\eta_1}(c_{v_1}\Gamma^0 - s_{v_1}\Gamma^1)(c_{v_2}\Gamma^0 - s_{v_2}\Gamma^1)\frac{1 + i\eta_2\Gamma_{11}}{1 + i\eta_2}\Big\}
    \\
    \nonumber &= \frac{(1 - \eta_1 \eta_2)}{(1 + i\eta_1)(1 + i\eta_2)}\text{Tr}\{(c_{v_1}\Gamma^0 - s_{v_1}\Gamma^1)(c_{v_2}\Gamma^0 - s_{v_2}\Gamma^1)\} 
    \\
    \nonumber &= \frac{1}{2}(1-\eta_1 \eta_2)\text{Tr}\{(c_{v_1}\Gamma^0 - s_{v_1}\Gamma^1)(c_{v_2}\Gamma^0 - s_{v_2}\Gamma^1)\}
    \\
    &= -16(1-\eta_1\eta_2)\cosh(v_{12}).
\end{align}
The second term, \eqref{second_term}, becomes:
\begin{align}
    \nonumber &-\frac{1}{2}R_{\mu}\tilde{R}_{\nu}\text{Tr}\{\mathcal{N}_{\eta_1}^T\Gamma^{\mu}\Gamma_{11}C^{-1}\mathcal{M}_{\eta_2}\Gamma^\nu C^{-1}\}
    \\
    \nonumber &=-\frac{1}{2}R_\mu \tilde{R}_\nu \text{Tr}\left\{\frac{1 + i\eta_1 \Gamma_{11}}{1 + i\eta_1}(c_{v_1}\Gamma^0 - s_{v_1}\Gamma^1)\Gamma^\mu \Gamma_{11}(c_{v_2}\Gamma^0 - s_{v_2}\Gamma^1)\frac{1 + i\eta_2 \Gamma_{11}}{1 + i\eta_2}\Gamma^\nu\right\}
    \\
    &= -\frac{1}{2}R_{\mu}\tilde{R}_{\nu}\frac{i\eta_1(1-\eta_1 \eta_2)}{(1 + i\eta_1)(1 + i\eta_2)}\text{Tr}\left\{(c_{v_1}\Gamma^0 - s_{v_1}\Gamma^1)\Gamma^\mu(c_{v_2}\Gamma^0 - s_{v_2}\Gamma^1)\Gamma^\nu\right\} \nonumber 
    \\
    & = -\frac{1}{4}R_\mu {\tilde{R}}_\nu i\eta_1(1-\eta_1\eta_2) \{ c_{v_1} c_{v_2} \text{Tr}( \Gamma^0\Gamma^\mu\Gamma^0 \Gamma^\nu )+ s_{v_1}s_{v_2}\text{Tr}(\Gamma^1\Gamma^\mu\Gamma^1 \Gamma^\nu )\nonumber 
    \\
    &\qquad \qquad \qquad \qquad \qquad \quad -c_{v_1}s_{v_2}\text{Tr}(\Gamma^0\Gamma^\mu\Gamma^1 \Gamma^\nu )- c_{v_2} s_{v_1} \text{Tr}(\Gamma^1\Gamma^\mu\Gamma^0 \Gamma^\nu )\}\nonumber 
    \\
    &= -\frac{32}{4}R_\mu {\tilde{R}}_\nu i\eta_1(1 -\eta_1\eta_2) \{ c_{v_1}c_{v_2}(\eta^{\mu \nu}+2\eta^{\mu 0}\eta^{\nu 0})+s_{v_1}s_{v_2}(2\eta^{\mu 1}\eta^{\nu 1}-\eta^{\mu \nu})\nonumber 
    \\
    &\qquad\qquad \qquad \qquad \qquad \qquad -(c_{v_1} s_{v_2}+c_{v_2}s_{v_1})(\eta^{\mu 0}\eta^{\nu 1}+\eta^{\nu0}\eta^{\mu 1}) \}\nonumber 
    \\
    &= - 8R_\mu {\tilde{R}}_\nu i\eta_1(1-\eta_1\eta_2)\{\eta^{\mu \nu} \cosh v_{12} +2 c_{v_1}c_{v_2}\eta^{0\mu}\eta^{0\nu}+2s_{v_1}s_{v_2}\eta^{\mu 1}\eta^{\nu 1}\nonumber 
    \\
    &\qquad\qquad\qquad \qquad\qquad\quad -(c_{v_1} s_{v_2}+c_{v_2}s_{v_1})(\eta^{\mu 0}\eta^{\nu 1}+\eta^{\nu0}\eta^{\mu 1})\}\nonumber 
    \\
    &=-8i\eta_1(1-\eta_1 \eta_2)\{\cosh(v_{12})(\epsilon\cdot \tilde{\epsilon})\phi\tilde{\phi} + R_{v_1}\tilde{R}_{v_2} + R_{v_2}\tilde{R}_{v_1}\}
\end{align}
where
\begin{equation}
    R_v= R^0 c_v-R^1 s_v.
\end{equation}
The third term, \eqref{third_term}, is equal to:
\begin{align}
    \nonumber &-\frac{1}{2}R_\mu R_\nu \text{Tr}\{\mathcal{N}_{\eta_1}^T \Gamma^\mu \Gamma^\nu C^{-1}\mathcal{M}_{\eta_2}C^{-1}\}
    \\
    \nonumber &= -\frac{1}{4}(1 -\eta_1\eta_2)R_\mu R_\nu \text{Tr}\{(c_{v_1}\Gamma^0 - s_{v_1}\Gamma^1)\Gamma^\mu \Gamma^\nu(c_{v_2}\Gamma^0 - s_{v_2}\Gamma^1)\}
    \\
    &= -\frac{1}{4}(1-\eta_1 \eta_2) R_\mu R_\nu \Big[c_{v_1} c_{v_2} \text{Tr}(\Gamma^0 \Gamma^\mu \Gamma^\nu \Gamma^0)+ s_{v_1}s_{v_2}\text{Tr}(\Gamma^1 \Gamma^\mu \Gamma^\nu \Gamma^1) \nonumber \\
    &\qquad \qquad \qquad \qquad \qquad -c_{v_1} s_{v_2}\text{Tr}(\Gamma^0\Gamma^\mu \Gamma^\nu \Gamma^1) -c_{v_2} s_{v_1}\text{Tr}(\Gamma^1 \Gamma^\mu \Gamma^\nu \Gamma^0)\Big]\nonumber 
    \\
    &= -\frac{32}{4}(1-\eta_1 \eta_2)R_\mu R_\nu \Big[-c_{v_1}c_{v_2}\eta^{\mu\nu} + s_{v_1}s_{v_2}\eta^{\mu\nu}\nonumber 
    \\
    &\qquad\qquad\qquad\qquad\qquad \; \,- (c_{v_1} s_{v_2} - c_{v_2}s_{v_1})(\eta^{0\mu}\eta^{\nu 1} - \eta^{0\nu}\eta^{\mu 1})\Big]
\end{align}
Using $R_\mu R_\nu \eta^{\mu\nu} = 0$, we obtain:
\begin{equation}
    -16(1-\eta_1\eta_2) \sinh(v_{12}) \sqrt{\frac{\alpha'}{2}}(k^0 \epsilon^1 - k^1 \epsilon^0)\theta\phi = -16(1-\eta_1\eta_2) \sqrt{\frac{\alpha'}{2}}(k_{v_1} \epsilon_{v_2} - k_{v_2} \epsilon_{v_1})\theta\phi
\end{equation}
The fourth term, \eqref{fourth_term}, is given by:
\begin{align}
    &-\frac{1}{2}\tilde{R}_\mu \tilde{R}_\nu \text{Tr}\{\mathcal{N}_{\eta_1}^T C^{-1}\mathcal{M}_{\eta_2}\Gamma^\nu \Gamma^\mu C^{-1}\}\nonumber 
    \\
    &= -\frac{1}{4}(1-\eta_1 \eta_2)\tilde{R}_\mu \tilde{R}_\nu \text{Tr}\{(c_{v_1}\Gamma^0 - s_{v_1}\Gamma^1)(c_{v_2}\Gamma^0 - s_{v_2}\Gamma^{1})\Gamma^\nu \Gamma^\mu\}\nonumber 
    \\
    &= \frac{1}{4}(1-\eta_1\eta_2)\tilde{R}_\mu \tilde{R}_\nu (c_{v_1}s_{v_2}\text{Tr}\{\Gamma^0\Gamma^1\Gamma^\nu\Gamma^\mu\} + s_{v_1}c_{v_2}\text{Tr}\{\Gamma^1 \Gamma^0\Gamma^\nu\Gamma^\mu\})\nonumber 
    \\
    &= \frac{32}{4}(1-\eta_1\eta_2)\tilde{R}_\mu \tilde{R}_\nu (c_{v_1}s_{v_2} - s_{v_1}c_{v_2})(\eta^{0\mu}\eta^{1\nu} - \eta^{0\nu}\eta^{1\mu})\nonumber 
    \\
    &= - 16(1-\eta_1\eta_2)\sinh(v_{12})\sqrt{\frac{\alpha'}{2}}(k^0\tilde{\epsilon}^1 - k^1 \tilde{\epsilon}^0)\tilde{\theta}\tilde{\phi}\nonumber 
    \\
    &= - 16(1-\eta_1\eta_2)\sqrt{\frac{\alpha'}{2}}(k_{v_1}\tilde{\epsilon}_{v_2} - k_{v_2} \tilde{\epsilon}_{v_1})\tilde{\theta}\tilde{\phi}.
\end{align}
Finally, the fifth term, \eqref{fifth_term}, can be simplified as:
\begin{align}
    \nonumber &-\frac{1}{4}R_\mu R_\nu \tilde{R}_\rho \tilde{R}_\sigma \text{Tr}\{\mathcal{N}_{\eta_1}^T\Gamma^\mu \Gamma^\nu C^{-1}\mathcal{M}_{\eta_2}\Gamma^\rho \Gamma^\sigma C^{-1}\}
    \\
    \nonumber &= -\frac{\alpha'}{16}(1-\eta_1\eta_2)\theta\phi\tilde{\theta}\tilde{\phi}(k_\mu \epsilon_\nu - k_\nu \epsilon_\mu)(k_\rho \tilde{\epsilon}_\sigma - k_\sigma\tilde{\epsilon}_\rho)\times 
    \\
    \nonumber &\quad \times \text{Tr}\{(c_{v_1}\Gamma^0 - s_{v_1}\Gamma^1)\Gamma^\mu \Gamma^\nu(c_{v_2}\Gamma^0 - s_{v_2}\Gamma^1)\Gamma^\rho\Gamma^\sigma\}
    \\
    \nonumber &= -\frac{\alpha'}{16}(1 -\eta_1\eta_2)\theta\phi\tilde{\theta}\tilde{\phi}(k_{\mu_2} \epsilon_{\mu_3} - k_{\mu_3} \epsilon_{\mu_2})(k_{\mu_5} \tilde{\epsilon}_{\mu_6} - k_{\mu_6}\tilde{\epsilon}_{\mu_5})
    \\
    &\quad \times u_{1\mu_1}u_{2\mu_4}\text{Tr}\{\Gamma^{\mu_1}\Gamma^{\mu_2}\Gamma^{\mu_3}\Gamma^{\mu_4}\Gamma^{\mu_5}\Gamma^{\mu_6}\},
\end{align}
where $u_{1\mu}$ and $u_{2\mu}$ are the D-brane 4-velocities. Using the identity
\begin{align*}
\text{Tr}(\Gamma^{\mu_1}\Gamma^{\mu_2}\Gamma^{\mu_3}\Gamma^{\mu_4}\Gamma^{\mu_5}\Gamma^{\mu_6}) &= 32 \Big[ \eta^{\mu_1\mu_2}(\eta^{\mu_3\mu_4}\eta^{\mu_5\mu_6} - \eta^{\mu_3\mu_5}\eta^{\mu_4\mu_6} + \eta^{\mu_3\mu_6}\eta^{\mu_4\mu_5}) \\
&\quad - \eta^{\mu_1\mu_3}(\eta^{\mu_2\mu_4}\eta^{\mu_5\mu_6} - \eta^{\mu_2\mu_5}\eta^{\mu_4\mu_6} + \eta^{\mu_2\mu_6}\eta^{\mu_4\mu_5}) \\
&\quad + \eta^{\mu_1\mu_4}(\eta^{\mu_2\mu_3}\eta^{\mu_5\mu_6} - \eta^{\mu_2\mu_5}\eta^{\mu_3\mu_6} + \eta^{\mu_2\mu_6}\eta^{\mu_3\mu_5}) \\
&\quad - \eta^{\mu_1\mu_5}(\eta^{\mu_2\mu_3}\eta^{\mu_4\mu_6} - \eta^{\mu_2\mu_4}\eta^{\mu_3\mu_6} + \eta^{\mu_2\mu_6}\eta^{\mu_3\mu_4}) \\
&\quad + \eta^{\mu_1\mu_6}(\eta^{\mu_2\mu_3}\eta^{\mu_4\mu_5} - \eta^{\mu_2\mu_4}\eta^{\mu_3\mu_5} + \eta^{\mu_2\mu_5}\eta^{\mu_3\mu_4}) \Big]
\end{align*}
we obtain:
\begin{align}
    -\frac{32}{16}\alpha' (1-\eta_1 \eta_2)\theta\phi \tilde{\theta}\tilde{\phi}\cdot 8(\tilde{\epsilon}\cdot \epsilon)k_{v_1}\,.
\end{align}
Collecting the above results, we can write the full RR zero-mode matrix element as:
\begin{align}
    \nonumber 16(\eta_1 \eta_2-1)\Big[&\cosh(v_{12}) + \frac{i\eta_1}{2 |z|}(\cosh(v_{12})(\epsilon\cdot \tilde{\epsilon})\phi\tilde{\phi} + R_{v_1}\tilde{R}_{v_2} + R_{v_2}\tilde{R}_{v_1})
    \\
    \nonumber& +\frac{1}{2}\sqrt{\frac{\alpha'}{2}}\left(\frac{1}{z}(k_{v_1} \epsilon_{v_2} - k_{v_2} \epsilon_{v_1})\theta\phi + \frac{1}{\bar{z}}(k_{v_1} \tilde{\epsilon}_{v_2} - k_{v_2} \tilde{\epsilon}_{v_1})\tilde{\theta}\tilde{\phi}\right)
    \\
    & + \frac{\alpha'}{4|z|^2} k_{v_1}k_{v_2}(\tilde{\epsilon}\cdot \epsilon)\theta\phi\tilde{\theta}\tilde{\phi}\Big]
\end{align}
Importantly, the above result must be multiplied by an additional factor of $1/2$ arising from the zero modes of RR-sector of the $\beta\gamma$-ghosts \cite{Yost:1988jm,Billo:1998vr}.

\section{Calculation of the field--theory limits}
\label{AppD}

In this appendix, we derive the field-theory limit in both the bosonic string and the superstring. The bosonic-string case is discussed in subsection~\ref{AppD1}, while the NS-NS and R-R sectors of the superstring are treated in subsections~\ref{AppD2} and \ref{AppD3}, respectively.

\subsection{Bosonic string}\label{AppD1}
In this Appendix we compute the field--theory limit for the bosonic string starting from the expression given in \eqref{CBC5}. 

As explained in the main text, the field-theory limit corresponds to the small-$|w|$ expansion of the integrand in Eq.~\eqref{CBC5}. There is, however, an important subtlety. Since the \(z\)-integration ranges from \(|z|=|w|\) to \(|z|=1\), both \(|w/z|\) and \(|z|\) must be kept finite in this expansion. Keeping only the leading terms in this small-\(|w|\) expansion of the integrand in Eq.~\eqref{CBC5}, we obtain
\begin{align}
    A_{5} &= \frac{\kappa_D}{2\pi}\,\frac{\alpha' N_1N_2T_0^2}{32\lvert\sinh(v_{12})\rvert}\, \int \frac{\mathrm{d}^{D-2}q}{(2\pi)^{D-2}} e^{i  q\cdot b}\int_0^1 \frac{\mathrm{d}^2w}{\lvert w\rvert^2}\int_{\lvert w\rvert}^{1}\mathrm{d}^2z\nonumber 
    \\
    &\quad \times \lvert w\rvert^{\frac{\alpha'}{2}q^2}\lvert z\rvert^{\alpha' k\cdot q} e^{\sqrt{\frac{\alpha'}{2}}\,q\cdot\left(\frac{\epsilon}{z}+\frac{\tilde{\epsilon}}{\bar z}\right)}(1-\lvert z\rvert^2)^{-\alpha' k_{v_1}^2} \left(1-\left\lvert \tfrac{w}{z}\right\rvert^2\right)^{-\alpha' k_{v_2}^2}\nonumber
    \\
    &\quad \times \left(\frac{1}{|w|^2} +D-4+2\cosh(2v_{12})\right)\nonumber
    \\
    &\quad \times (1-\lvert z\rvert^2)^{-(2\tilde{\epsilon}_{v_1}\epsilon_{v_1}+\tilde{\epsilon}\cdot\epsilon)\partial_z\partial_{\bar z}-\sqrt{2\alpha'}\,k_{v_1}\epsilon_{v_1}\partial_z- \sqrt{2\alpha'}\,k_{v_1}\tilde{\epsilon}_{v_1}\partial_{\bar z}} \nonumber
    \\
    &\quad \times \left(1-\left\lvert \tfrac{w}{z}\right\rvert^2\right)^{-(2\tilde{\epsilon}_{v_2}\epsilon_{v_2}+\tilde{\epsilon}\cdot\epsilon)\partial_z\partial_{\bar z}-\sqrt{2\alpha'}\,k_{v_2}\epsilon_{v_2}\partial_z- \sqrt{2\alpha'}\,k_{v_2}\tilde{\epsilon}_{v_2}\partial_{\bar z}} \nonumber
    \\
    &\quad \times (1-\lvert z\rvert^2\lvert w\rvert^2)^{-(2\tilde{\epsilon}_{v_1+v_{12}}\epsilon_{v_1+v_{12}}+\tilde{\epsilon}\cdot\epsilon)\partial_z\partial_{\bar z}-\alpha' k_{v_1+v_{12}}^2} \nonumber
    \\
    &\quad \times\nonumber(1-\lvert z\rvert^2\lvert w\rvert^2)^{-\sqrt{2\alpha'}\,k_{v_1+v_{12}}\epsilon_{v_1+v_{12}}\partial_z- \sqrt{2\alpha'}\,k_{v_1+v_{12}}\tilde{\epsilon}_{v_1+v_{12}}\partial_{\bar z}}
    \\
    &\quad \times \left(1-\left\lvert \tfrac{w}{z}\right\rvert^2\lvert w\rvert^2\right)^{-(2\tilde{\epsilon}_{v_2+v_{21}}\epsilon_{v_2+v_{21}}+\tilde{\epsilon}\cdot\epsilon)\partial_z\partial_{\bar z}-\alpha' k_{v_2+v_{21}}^2} \nonumber
    \\
    &\quad \times \nonumber \left(1-\left\lvert \tfrac{w}{z}\right\rvert^2\lvert w\rvert^2\right)^{-\sqrt{2\alpha'}\,k_{v_2+v_{21}}\epsilon_{v_2+v_{21}}\partial_z- \sqrt{2\alpha'}\,k_{v_2+v_{21}}\tilde{\epsilon}_{v_2+v_{21}}\partial_{\bar z}}
    \\
    &\quad \times e^{4\sqrt{\frac{\alpha'}{2}}\lvert w\rvert^2\cosh(v_{12}) \left[\frac1z(\epsilon_{v_1}k_{v_2}-\epsilon_{v_2}k_{v_1})+\frac1{\bar{z}}(\tilde{\epsilon}_{v_1}k_{v_2}-\tilde{\epsilon}_{v_2}k_{v_1})\right]}.\label{AppD_eq1}
\end{align}
We split this expression into two contributions, $A_5^{\text{massless}}$ and $A_5^{\text{tachyon}}$, corresponding respectively to the massless and tachyonic exchanges in the partition function. The term proportional to $D-4+2\cosh(2v_{12})$ in the third line gives the contribution of the massless   states, while the term proportional to $1/|w|^2$ gives the tachyonic contribution,
\begin{equation}
    A_{5} = A_{5}^{\text{massless}} + A_{5}^{\text{tachyon}}.
\end{equation}
The tachyonic contribution is
\begin{align}
    A_{5}^{\rm tachyon} &= \frac{\kappa_D}{2\pi}\,\frac{\alpha' N_1N_2T_0^2}{32\lvert\sinh(v_{12})\rvert}\, \int \frac{\mathrm{d}^{D-2}q}{(2\pi)^{D-2}} e^{i  q\cdot b}\int_0^1 \frac{\mathrm{d}^2w}{\lvert w\rvert^4}\int_{\lvert w\rvert}^{1}\mathrm{d}^2z\nonumber 
    \\
    &\quad \times \lvert w\rvert^{\frac{\alpha'}{2}q^2}\lvert z\rvert^{\alpha' k\cdot q} e^{\sqrt{\frac{\alpha'}{2}}\,q\cdot\left(\frac{\epsilon}{z}+\frac{\tilde{\epsilon}}{\bar z}\right)}(1-\lvert z\rvert^2)^{-\alpha' k_{v_1}^2} \left(1-\left\lvert \tfrac{w}{z}\right\rvert^2\right)^{-\alpha' k_{v_2}^2}\nonumber
    \\
    &\quad \times (1-\lvert z\rvert^2)^{-(2\tilde{\epsilon}_{v_1}\epsilon_{v_1}+\tilde{\epsilon}\cdot\epsilon)\partial_z\partial_{\bar z}-\sqrt{2\alpha'}\,k_{v_1}\epsilon_{v_1}\partial_z- \sqrt{2\alpha'}\,k_{v_1}\tilde{\epsilon}_{v_1}\partial_{\bar z}} \nonumber
    \\
    &\quad \times \left(1-\left\lvert \tfrac{w}{z}\right\rvert^2\right)^{-(2\tilde{\epsilon}_{v_2}\epsilon_{v_2}+\tilde{\epsilon}\cdot\epsilon)\partial_z\partial_{\bar z}-\sqrt{2\alpha'}\,k_{v_2}\epsilon_{v_2}\partial_z- \sqrt{2\alpha'}\,k_{v_2}\tilde{\epsilon}_{v_2}\partial_{\bar z}} \nonumber
    \\
    &\quad \times (1-\lvert z\rvert^2\lvert w\rvert^2)^{-(2\tilde{\epsilon}_{v_1+v_{12}}\epsilon_{v_1+v_{12}}+\tilde{\epsilon}\cdot\epsilon)\partial_z\partial_{\bar z}-\alpha' k_{v_1+v_{12}}^2} \nonumber
    \\
    &\quad \times\nonumber(1-\lvert z\rvert^2\lvert w\rvert^2)^{-\sqrt{2\alpha'}\,k_{v_1+v_{12}}\epsilon_{v_1+v_{12}}\partial_z- \sqrt{2\alpha'}\,k_{v_1+v_{12}}\tilde{\epsilon}_{v_1+v_{12}}\partial_{\bar z}}
    \\
    &\quad \times \left(1-\left\lvert \tfrac{w}{z}\right\rvert^2\lvert w\rvert^2\right)^{-(2\tilde{\epsilon}_{v_2+v_{21}}\epsilon_{v_2+v_{21}}+\tilde{\epsilon}\cdot\epsilon)\partial_z\partial_{\bar z}-\alpha' k_{v_2+v_{21}}^2} \nonumber
    \\
    &\quad \times \nonumber \left(1-\left\lvert \tfrac{w}{z}\right\rvert^2\lvert w\rvert^2\right)^{-\sqrt{2\alpha'}\,k_{v_2+v_{21}}\epsilon_{v_2+v_{21}}\partial_z- \sqrt{2\alpha'}\,k_{v_2+v_{21}}\tilde{\epsilon}_{v_2+v_{21}}\partial_{\bar z}}
    \\
    &\quad \times e^{4\sqrt{\frac{\alpha'}{2}}\lvert w\rvert^2\cosh(v_{12}) \left[\frac1z(\epsilon_{v_1}k_{v_2}-\epsilon_{v_2}k_{v_1})+\frac1{\bar{z}}(\tilde{\epsilon}_{v_1}k_{v_2}-\tilde{\epsilon}_{v_2}k_{v_1})\right]}.
\end{align}
Since the massless contribution  is finite in the small-$|w|$ expansion, we may discard all terms in the integrand of Eq.~\eqref{AppD_eq1} that vanish as $|w|\to0$, while keeping $|w/z|$ and $|z|$ fixed. The resulting massless contribution is therefore
\begin{align}
    A_{5}^{\mathrm{massless}} &= \frac{\kappa_D}{2\pi}\,\frac{\alpha' N_1N_2T_0^2}{32\lvert\sinh(v_{12})\rvert}\, \int \frac{\mathrm{d}^{D-2}q}{(2\pi)^{D-2}} e^{i  q\cdot b}\int_0^1 \frac{\mathrm{d}^2w}{\lvert w\rvert^2}\int_{\lvert w\rvert}^{1}\mathrm{d}^2z\nonumber 
    \\
    &\quad \times (D-4+2\cosh(2v_{12})) \lvert w\rvert^{\frac{\alpha'}{2}q^2}\lvert z\rvert^{\alpha' k\cdot q} (1-\lvert z\rvert^2)^{-\alpha' k_{v_1}^2}\left(1-\left\lvert \tfrac{w}{z}\right\rvert^2\right)^{-\alpha' k_{v_2}^2} \nonumber
    \\
    &\quad \times (1-\lvert z\rvert^2)^{-(2\tilde{\epsilon}_{v_1}\epsilon_{v_1}+\tilde{\epsilon}\cdot\epsilon)\partial_z\partial_{\bar z}} \left(1-\left\lvert \tfrac{w}{z}\right\rvert^2\right)^{-(2\tilde{\epsilon}_{v_2}\epsilon_{v_2}+\tilde{\epsilon}\cdot\epsilon)\partial_z\partial_{\bar z}} \nonumber
    \\
    &\quad \times (1-\lvert z\rvert^2)^{-\sqrt{2\alpha'}k_{v_1}\epsilon_{v_1}\partial_z} \left(1-\left\lvert \tfrac{w}{z}\right\rvert^2\right)^{-\sqrt{2\alpha'}k_{v_2}\epsilon_{v_2}\partial_z} e^{\sqrt{\frac{\alpha'}{2}}\,q\cdot\epsilon/z}\nonumber
    \\
    &\quad \times (1-\lvert z\rvert^2)^{-\sqrt{2\alpha'}k_{v_1}\tilde{\epsilon}_{v_1}\partial_{\bar z}} \left(1-\left\lvert \tfrac{w}{z}\right\rvert^2\right)^{-\sqrt{2\alpha'}k_{v_2}\tilde{\epsilon}_{v_2}\partial_{\bar z}} e^{\sqrt{\frac{\alpha'}{2}}\,q\cdot\tilde{\epsilon}/\bar z} .
\end{align}
We now extract the terms bilinear in the polarizations. For the massless part, this gives
\begin{align}
    A_{5}^{\mathrm{massless}} &= \frac{\kappa_D}{2\pi}\,\frac{\alpha' N_1N_2T_0^2}{32\lvert\sinh(v_{12})\rvert}\int \frac{\mathrm{d}^{D-2}q}{(2\pi)^{D-2}} e^{i  q\cdot b} \int_0^1 \frac{\mathrm{d}^2w}{\lvert w\rvert^2} \int_{\lvert w\rvert}^{1}\mathrm{d}^2z \nonumber
    \\
    &\quad\times (D-4+2\cosh(2v_{12}))\lvert w\rvert^{\frac{\alpha'}{2}q^2}\lvert z\rvert^{\alpha' k\cdot q} (1-\lvert z\rvert^2)^{-\alpha' k_{v_1}^2}(1-\lvert\tfrac{w}{z}\rvert^2)^{-\alpha' k_{v_2}^2} \nonumber
    \\
    &\quad\times\Big\{\left(\sqrt{\tfrac{\alpha'}{2}}\,\tfrac{q\cdot\epsilon}{z} -\sqrt{2\alpha'}k_{v_1}\epsilon_{v_1}\partial_z\log(1-\lvert z\rvert^2) -\sqrt{2\alpha'}k_{v_2}\epsilon_{v_2}\partial_z \log(1-\lvert\tfrac{w}{z}\rvert^2)\right)\nonumber 
    \\
    &\qquad\;\;\; \times \left(\sqrt{\tfrac{\alpha'}{2}}\tfrac{q\cdot\tilde{\epsilon}}{\bar z} -\sqrt{2\alpha'}k_{v_1}\tilde{\epsilon}_{v_1}\partial_{\bar z}\log(1-\lvert z\rvert^2)-\sqrt{2\alpha'}k_{v_2}\tilde{\epsilon}_{v_2}\partial_{\bar z}\log(1-\lvert\tfrac{w}{z}\rvert^2)\right)\nonumber 
    \\
    &\qquad \;\;\;-(2\tilde{\epsilon}_{v_1}\epsilon_{v_1} +\tilde{\epsilon}\cdot\epsilon) \partial_z \partial_{\bar z} \log(1-\lvert z\rvert^2)\nonumber
    \\
    &\qquad \;\;\; -(2\tilde{\epsilon}_{v_2}\epsilon_{v_2}+\tilde{\epsilon}\cdot\epsilon) \partial_z\partial_{\bar z}\log(1-\lvert\tfrac{w}{z}\rvert^2)\Big\}.
\end{align}

Similarly, the tachyonic part becomes
\begin{align}
    A_{5}^{\mathrm{tachyon}} =&\; \frac{\kappa_D}{2\pi}\,\frac{\alpha' N_1N_2T_0^2}{32\lvert\sinh(v_{12})\rvert}\, \int \frac{\mathrm{d}^{D-2}q}{(2\pi)^{D-2}} e^{i  q\cdot b}\int_0^1 \frac{\mathrm{d}^2w}{\lvert w\rvert^4}\int_{\lvert w\rvert}^{1}\mathrm{d}^2z\nonumber 
    \\
     &\times \lvert w\rvert^{\frac{\alpha'}{2}q^2}\lvert z\rvert^{\alpha' k\cdot q} (1-\lvert z\rvert^2)^{-\alpha' k_{v_1}^2}\left(1-\left\lvert \tfrac{w}{z}\right\rvert^2\right)^{-\alpha' k_{v_2}^2} \nonumber
    \\
    &\times(1-\lvert z\rvert^2\lvert w\rvert^2)^{-\alpha' k_{v_1+v_{12}}^2} \left(1-\left\lvert \tfrac{w}{z}\right\rvert^2\lvert w\rvert^2\right)^{-\alpha' k_{v_2+v_{21}}^2} \nonumber
    \\
    &\times\Big\{-(2\tilde{\epsilon}_{v_1}\epsilon_{v_1}+\tilde{\epsilon}\cdot\epsilon)\partial_z\partial_{\bar z}\log(1-\lvert z\rvert^2) \nonumber 
    \\
    &\qquad -(2\tilde{\epsilon}_{v_2}\epsilon_{v_2}+\tilde{\epsilon}\cdot\epsilon)\partial_z\partial_{\bar z}\log\!\left(1-\left\lvert \tfrac{w}{z}\right\rvert^2\right) \nonumber
    \\
    &\qquad-(2\tilde{\epsilon}_{v_1+v_{12}}\epsilon_{v_1+v_{12}}+\tilde{\epsilon}\cdot\epsilon)\partial_z\partial_{\bar z}\log(1-\lvert z\rvert^2\lvert w\rvert^2) \nonumber
    \\
    &\qquad-(2\tilde{\epsilon}_{v_2+v_{21}}\epsilon_{v_2+v_{21}}+\tilde{\epsilon}\cdot\epsilon)\partial_z\partial_{\bar z}\log\!\left(1-\left\lvert \tfrac{w}{z}\right\rvert^2\lvert w\rvert^2\right) \nonumber
    \\
    &\qquad+\bigg[\sqrt{\tfrac{\alpha'}{2}}[\tfrac{q\cdot\epsilon}{z}+\tfrac{4}{z}\lvert w\rvert^2\cosh(v_{12})(\epsilon_{v_1}k_{v_2}-\epsilon_{v_2}k_{v_1})] \nonumber 
    \\
    &\qquad\qquad  -\sqrt{2\alpha'}k_{v_1}\epsilon_{v_1}\partial_z\log(1-\lvert z\rvert^2) -\sqrt{2\alpha'}k_{v_2}\epsilon_{v_2}\partial_z\log\!\left(1-\left\lvert \tfrac{w}{z}\right\rvert^2\right)\nonumber 
    \\
    &\qquad\qquad -\sqrt{2\alpha'}k_{v_1+v_{12}}\epsilon_{v_1+v_{12}}\partial_z\log(1-\lvert z\rvert^2\lvert w\rvert^2)
    \nonumber
    \\
    &\qquad\qquad -\sqrt{2\alpha'}k_{v_2+v_{21}}\epsilon_{v_2+v_{21}}\partial_z\log\!\left(1-\left\lvert \tfrac{w}{z}\right\rvert^2\lvert w\rvert^2\right)\bigg] \nonumber
    \\
    &\qquad\times\bigg[\sqrt{\tfrac{\alpha'}{2}}[\tfrac{q\cdot\tilde\epsilon}{\bar{z}}+\tfrac{4}{\bar{z}}\lvert w\rvert^2\cosh(v_{12})(\tilde\epsilon_{v_1}k_{v_2}-\tilde\epsilon_{v_2}k_{v_1})] \nonumber 
    \\
    &\qquad\qquad  -\sqrt{2\alpha'}k_{v_1}\tilde\epsilon_{v_1}\partial_{\bar{z}}\log(1-\lvert z\rvert^2) -\sqrt{2\alpha'}k_{v_2}\epsilon_{v_2}\partial_{\bar{z}}\log\!\left(1-\left\lvert \tfrac{w}{z}\right\rvert^2\right)\nonumber 
    \\
    &\qquad\qquad -\sqrt{2\alpha'}k_{v_1+v_{12}}\tilde\epsilon_{v_1+v_{12}}\partial_{\bar{z}}\log(1-\lvert z\rvert^2\lvert w\rvert^2)
    \nonumber
    \\
    &\qquad\qquad -\sqrt{2\alpha'}k_{v_2+v_{21}}\tilde\epsilon_{v_2+v_{21}}\partial_{\tilde{z}}\log\!\left(1-\left\lvert \tfrac{w}{z}\right\rvert^2\lvert w\rvert^2\right)\bigg] \Big\}.\label{A_tachyonic_aux1}
\end{align}
By taking the derivatives with respect to $z$ and $\bar{z}$, we can recast $A_5^{\rm massless}$ as:
\begin{align}
    A_{5}^{\mathrm{massless}} &= \frac{\kappa_D}{2\pi}\,\frac{\alpha' N_1N_2T_0^2}{32\lvert\sinh(v_{12})\rvert}\int \frac{\mathrm{d}^{D-2}q}{(2\pi)^{D-2}} e^{i  q\cdot b} \int_0^1 \frac{\mathrm{d}^2w}{\lvert w\rvert^2} \int_{\lvert w\rvert}^{1}\mathrm{d}^2z \nonumber
    \\
    &\quad\times (D-4+2\cosh(2v_{12}))\lvert w\rvert^{\frac{\alpha'}{2}q^2}\lvert z\rvert^{\alpha' k\cdot q} (1-\lvert z\rvert^2)^{-\alpha' k_{v_1}^2}(1-\lvert\tfrac{w}{z}\rvert^2)^{-\alpha' k_{v_2}^2} \nonumber
    \\
    &\quad\times\Big\{\left(\sqrt{\tfrac{\alpha'}{2}}\,\tfrac{q\cdot\epsilon}{z} +\sqrt{2\alpha'}k_{v_1}\epsilon_{v_1}\frac{{\bar{z}}}{1-|z|^2} -\sqrt{2\alpha'}k_{v_2}\epsilon_{v_2} \frac{\frac{1}{z}\frac{|w|^2}{|z|^2} }{1-\frac{|w|^2}{|z|^2}} \right)\nonumber 
    \\
    &\qquad\;\;\; \times \left(\sqrt{\tfrac{\alpha'}{2}}\tfrac{q\cdot\tilde{\epsilon}}{\bar z} + \sqrt{2\alpha'}k_{v_1}\tilde{\epsilon}_{v_1}\frac{z}{1-|z|^2}-\sqrt{2\alpha'}k_{v_2}\tilde{\epsilon}_{v_2} \frac{\frac{1}{{\bar{z}}}\frac{|w|^2}{|z|^2}}{1-\frac{|w|^2}{|z|^2}}\right)\nonumber 
    \\
    &\qquad \;\;\;+(2\tilde{\epsilon}_{v_1}\epsilon_{v_1} +\tilde{\epsilon}\cdot\epsilon) \frac{1}{(1-|z|^2)^2}  +(2\tilde{\epsilon}_{v_2}\epsilon_{v_2}+\tilde{\epsilon}\cdot\epsilon)\frac{\frac{|w|^2}{|z|^2}\frac{1}{|z|^2}}{(1- \frac{|w|^2}{|z|^2}} \Big\}.\label{A_massless_bosonic_aux}
\end{align}
Integrating over phases $\operatorname{arg}(z), \operatorname{arg}(w)$, and changing the radial integration variables from $|z|$ and $|w|$ to:
\begin{align}
    x=|z|^2,\qquad y=\frac{|w|^2}{|z|^2},\qquad  \int_0^1 \frac{d^2w}{|w|^2}\int_{|w|}^1 \frac{d^2 z}{|z|^2} =\pi^2 \int_0^1\frac{dx}{x} \int_0^1 \frac{dy}{y}\label{change}  
\end{align} 
we obtain 
\begin{align}
    A_{5}^{\mathrm{massless}} =&\, \frac{\kappa_D}{2\pi}\,\frac{\alpha' N_1N_2T_0^2 \pi^2}{32\lvert\sinh(v_{12})\rvert}\int \frac{\mathrm{d}^{D-2}q}{(2\pi)^{D-2}} e^{i  q\cdot b} \int_0^1 \frac{dy}{y} y^{\frac{\alpha'}{4} q^2}  (1-y)^{-\alpha'k_{v_2}^2}\int_0^1 \frac{dx}{x}\,x^{\frac{\alpha'}{4} (q+k)^2}(1-x)^{-\alpha' k_{v_1}^2}\nonumber 
    \\
    &\times (D-4+2\cosh(2v_{12}))\bigg\{\bigg(\sqrt{\frac{\alpha'}{2}}q\epsilon +\sqrt{2\alpha'}k_{v_1} \epsilon_{v_1}\frac{x}{1-x}-\sqrt{2\alpha'}k_{v_2} \epsilon_{v_2}\frac{y}{1-y}\bigg) \nonumber 
    \\
    &\times  \bigg(\sqrt{\frac{\alpha'}{2}}q{\tilde{\epsilon}} +\sqrt{2\alpha'}k_{v_1} {\tilde{\epsilon}}_{v_1}\frac{x}{1-x}-\sqrt{2\alpha'}k_{v_2} {\tilde{\epsilon}}_{v_2}\frac{y}{1-y}\bigg) \nonumber 
    \\
    &+ (2{\tilde{\epsilon}}_{v_1}\epsilon_{v_1}+(\epsilon {\tilde{\epsilon}}))\frac{x}{(1-x)^2}+(2{\tilde{\epsilon}}_{v_2}\epsilon_{v_2}+(\epsilon {\tilde{\epsilon}}))\frac{y}{(1-y)^2}\bigg\}.
\end{align}
Evaluating the $x$- and $y$-integrals in the $\alpha'\to0$ limit using Appendix~\ref{AppInte} yields:
\begin{align}
    A_{5}^{\mathrm{massless}} =&\, \frac{\kappa_D}{2\pi}\,\frac{ N_1N_2T_0^2 \pi^2}{32\lvert\sinh(v_{12})\rvert}\int \frac{\mathrm{d}^{D-2}q}{(2\pi)^{D-2}} e^{i  q\cdot b}(D-4+2\cosh(2v_{12})) \nonumber 
    \\
    &\times \bigg\{ \frac{2}{q^2 (k+q)^2}\bigg( (q\cdot \epsilon) -\frac{\epsilon_{v_1}(k+q)^2}{k_{v_1}}+\frac{\epsilon_{v_2}q^2}{k_{v_2}}\bigg)\bigg( (q\cdot {\tilde{\epsilon}}) -\frac{{\tilde{\epsilon}}_{v_1}(k+q)^2}{k_{v_1}}+\frac{{\tilde{\epsilon}}_{v_2}q^2}{k_{v_2}}\bigg) \nonumber 
    \\
    &\qquad +(\epsilon\cdot {\tilde{\epsilon}})\bigg(\frac{(k+q)^2}{q^2 k_{v_1}^2} +\frac{q^2}{(k+q)^2 k_{v_2}^2}-\frac{4}{q^2} -\frac{4}{(k+q)^2}\bigg\}.\label{FinalMassless8}
\end{align}

Let us now go back to the tachyonic contribution \eqref{A_tachyonic_aux1}.  Integrating by parts the double-derivative terms in 4th, 5th, 6th and 7th line, and changing the integration variables as in \eqref{change}, we get{\allowdisplaybreaks
\begin{align}
    A_{5}^{\mathrm{tachyon}} =&\; \frac{\kappa_D}{2\pi}\,\frac{\alpha' N_1N_2T_0^2\pi^2}{32\lvert\sinh(v_{12})\rvert}\, \int \frac{\mathrm{d}^{D-2}q}{(2\pi)^{D-2}} e^{i  q\cdot b}\int_0^1 \frac{dx}{x^2 } \int_0^1 \frac{dy}{y^2} y^{\frac{\alpha'}{4}q^2}  x^{\frac{\alpha'}{4} (k+q)^2}\nonumber 
    \\
    & \times(1-x)^{-\alpha' k_{v_1}^2}\left( 1-y \right)^{-\alpha' k_{v_2}^2} (1-x^2y)^{-\alpha' k_{v_1+v_{12}}^2} 
    \left( 1- xy^2\right)^{-\alpha' k_{v_2+v_{21}}^2} \nonumber
    \\
    &\times\Big\{-\frac{(2\epsilon_{v_1}{\tilde{\epsilon}}_{v_1}+({\tilde{\epsilon}}\cdot \epsilon))x}{1-x}\bigg( \frac{\alpha'}{2} kq +\frac{\alpha' k^2_{v_1}x}{1-x} - \frac{\alpha'k^2_{v_2}y}{1-y}+ \frac{\alpha' k^2_{v_1+v_{12}} x^2y}{1-x^2 y}- \frac{\alpha' k^2_{v_2+v_{21}}xy^2}{1-xy^2} \bigg)\nonumber 
    \\
    &\qquad +\frac{(2\epsilon_{v_2}{\tilde{\epsilon}}_{v_2}+({\tilde{\epsilon}}\cdot\epsilon))y}{1-y}\bigg( \frac{\alpha'}{2} (k\cdot q) +\frac{\alpha' k^2_{v_1}x}{1-x} - \frac{\alpha'k^2_{v_2}y}{1-y}+ \frac{\alpha' k^2_{v_1+v_{12}} x^2y}{1-x^2 y}- \frac{\alpha' k^2_{v_2+v_{21}}xy^2}{1-xy^2} \bigg)\nonumber 
    \\
    &\qquad - (2 {\tilde{\epsilon}}_{v_1+v_{12}}\epsilon_{v_1+v_{12}} +{\tilde{\epsilon}}\cdot \epsilon) \frac{x^2y}{(1-x^2y)}\bigg(\frac{\alpha'}{2}(k\cdot q)+\frac{\alpha'k_{v_1}^2 x}{1-x}- \frac{\alpha'k^2_{v_2}y}{1- y}\bigg)\nonumber 
    \\
    &\qquad  + (2 {\tilde{\epsilon}}_{v_2+v_{21}}\epsilon_{v_2+v_{21}} +{\tilde{\epsilon}}\cdot \epsilon) \frac{xy^2}{(1-xy^2)}\bigg(\frac{\alpha'}{2}(k\cdot q) + \frac{\alpha'k_{v_1}^2 x}{1-x}- \frac{\alpha'k^2_{v_2}y}{1- y}\bigg)\nonumber 
    \\
    &\qquad +\bigg[\sqrt{\tfrac{\alpha'}{2}}[q\cdot\epsilon+4 xy\cosh(v_{12})(\epsilon_{v_1}k_{v_2}-\epsilon_{v_2}k_{v_1})] \nonumber 
    \\
    &\qquad\qquad  +\sqrt{2\alpha'}k_{v_1}\epsilon_{v_1}\frac{{{x}}}{1-x} -\sqrt{2\alpha'}k_{v_2}\epsilon_{v_2}\frac{y}{1- y}\nonumber 
    \\
    &\qquad\qquad +\sqrt{2\alpha'}k_{v_1+v_{12}}\epsilon_{v_1+v_{12}}\frac{x^2y}{1-x^2 y} -\sqrt{2\alpha'}k_{v_2+v_{21}}\epsilon_{v_2+v_{21}}\frac{xy^2}{1- xy^2}\bigg] \nonumber
    \\
    &\qquad\times \bigg[\sqrt{\tfrac{\alpha'}{2}}[q\cdot\tilde\epsilon+4 xy \cosh(v_{12})(\tilde\epsilon_{v_1}k_{v_2}-\tilde\epsilon_{v_2}k_{v_1})] \nonumber 
    \\
    &\qquad\qquad  +\sqrt{2\alpha'}k_{v_1}\tilde\epsilon_{v_1} \frac{x}{1-x} -\sqrt{2\alpha'}k_{v_2}{\tilde{\epsilon}}_{v_2}\frac{y}{1-y}\nonumber 
    \\
    &\qquad\qquad +\sqrt{2\alpha'}k_{v_1+v_{12}}\tilde\epsilon_{v_1+v_{12}}\frac{x^2y}{1-x^2 y}  -\sqrt{2\alpha'}k_{v_2+v_{21}}\tilde\epsilon_{v_2+v_{21}}\frac{xy^2}{1-xy^2}\bigg] \Big\} \, .
    \label{UVZ4}
\end{align}}
It is easy to see that the terms proportional to $k\cdot q$ in lines 3,4,5,6 and the four terms involving powers of order $x^4 y^2, x^3 y^3, y^4 x^2$ do not contribute in the field--theory limit. Since we are interested in the moduli space region where $|w|$ is small, we may approximate $1-x^2 y= 1-x|w|^2\simeq 1$ and $1-x y^2= 1-y|w|^2\simeq 1$.  From the remaining terms we get
\begin{align}
    A_{5}^{\rm tachyon} =&\, \frac{\kappa_D}{2\pi}\,\frac{\alpha' N_1N_2T_0^2\pi^2}{32\lvert\sinh(v_{12})\rvert}\, \int \frac{\mathrm{d}^{D-2}q}{(2\pi)^{D-2}} e^{i  q\cdot b}\int_0^1 \frac{dx}{x^2 } \int_0^1 \frac{dy}{y^2} y^{\frac{\alpha'}{4}q^2}  x^{\frac{\alpha'}{4} (k+q)^2}\nonumber 
    \\
    & \times(1-x)^{-\alpha' k_{v_1}^2}\left( 1-y \right)^{-\alpha' k_{v_2}^2} \nonumber
    \\
    &\times\Big\{ 2k_{v_2+v_{21}}^2 k_{v_1}^2\frac{x^2 y^2}{1-x}\bigg( \frac{\epsilon_{v_1}}{k_{v_1}}-\frac{\epsilon_{v_2+v_{21}}}{k_{v_2+v_{21}}}\bigg)\bigg( \frac{{\tilde{\epsilon}}_{v_1}}{k_{v_1}}-\frac{{\tilde{\epsilon}}_{v_2+v_{21}}}{k_{v_2+v_{21}}}\bigg) \nonumber 
    \\
    &\qquad  +2k^2_{v_1+v_{12}}k_{v_2}^2 \frac{x^2 y^2}{1-y}\bigg( \frac{\epsilon_{v_2}}{k_{v_2}}-\frac{\epsilon_{v_1+v_{12}}}{k_{v_1+v_{12}}}\bigg)\bigg( \frac{{\tilde{\epsilon}}_{v_2}}{k_{v_2}}-\frac{{\tilde{\epsilon}}_{v_1+v_{12}}}{k_{v_1+v_{12}}}\bigg)\nonumber 
    \\
    &\qquad  -2k_{v_1}^2 k_{v_1+v_{12}}^2 \frac{x^3y}{1-x} \bigg( \frac{\epsilon_{v_1}}{k_{v_1}}-\frac{\epsilon_{v_1+v_{12}}}{k_{v_1+v_{12}}}\bigg)\bigg( \frac{{\tilde{\epsilon}}_{v_1}}{k_{v_1}}-\frac{{\tilde{\epsilon}}_{v_1+v_{12}}}{k_{v_1+v_{12}}}\bigg)\nonumber 
    \\
    &\qquad  -2 k_{v_2}^2 k_{v_2+v_{21}}^2\frac{xy^3}{1-y}\bigg( \frac{\epsilon_{v_2}}{k_{v_2}}-\frac{\epsilon_{v_2+v_{21}}}{k_{v_2+v_{21}}}\bigg)\bigg( \frac{{\tilde{\epsilon}}_{v_2}}{k_{v_2}}-\frac{{\tilde{\epsilon}}_{v_2+v_{21}}}{k_{v_2+v_{21}}}\bigg)\nonumber 
    \\
    &\qquad + \frac{2xy}{(1-x)(1-y)}k_{v_1}^2 k_{v_2}^2 \bigg(\frac{\epsilon_{v_1}}{k_{v_1}}-\frac{\epsilon_{v_2}}{k_{v_2}}\bigg)\bigg(\frac{{\tilde{\epsilon}}_{v_1}}{k_{v_1}}-\frac{{\tilde{\epsilon}}_{v_2}}{k_{v_2}}\bigg)\nonumber 
    \\
    &\qquad + (\epsilon\cdot {\tilde{\epsilon}}) \bigg[ \frac{(k_{v_1}^2+k_{v_2}^2)xy}{(1-x)(1-y)}-\frac{x^3 y(k^2_{v_{1}+v_{12}}+k_{v_1}^2)}{1-x} - \frac{xy^3 (k^2_{v_2+v_{21}}+k^2_{v_2})}{1-y}\nonumber 
    \\
    &\qquad\qquad\qquad  + \frac{x^2y^2 (k^2_{v_2 +v_{21}}+k^2_{v_1})}{1-x}+ \frac{x^2 y^2(k^2_{v_1+ v_{12}}+k^2_{v_2})}{1-y} \bigg]  \nonumber 
    \\
    &\qquad +2xy \cosh(v_{12})k_{v_1} k_{v_2}\bigg[(q\cdot {\tilde{\epsilon}})\bigg(\frac{\epsilon_{v_1}}{k_{v_1}}-\frac{\epsilon_{v_2}}{k_{v_2}}\bigg)+ (q\cdot \epsilon)\bigg(\frac{{\tilde{\epsilon}}_{v_1}}{k_{v_1}}-\frac{{\tilde{\epsilon}}_{v_2}}{k_{v_2}}\bigg)\bigg] \nonumber 
    \\
    &\qquad + \frac{4x^2y}{1-x}\cosh(v_{12}) k_{v_1}^3 k_{v_2} \bigg[ \frac{{\tilde{\epsilon}}_{v_1}}{k_{v_1}}\bigg(\frac{\epsilon_{v_1}}{k_{v_1}}-\frac{\epsilon_{v_2}}{k_{v_2}}\bigg)+ \frac{\epsilon_{v_1}}{k_{v_1}}\bigg(\frac{{\tilde{\epsilon}}_{v_1}}{k_{v_1}}-\frac{{\tilde{\epsilon}}_{v_2}}{k_{v_2}}\bigg)\bigg] \nonumber 
    \\
    &\qquad + \frac{4xy^2}{1-y}\cosh(v_{12}) k_{v_1} k_{v_2}^3 \bigg[ \frac{{\tilde{\epsilon}}_{v_2}}{k_{v_2}}\bigg(\frac{\epsilon_{v_1}}{k_{v_1}}-\frac{\epsilon_{v_2}}{k_{v_2}}\bigg)+ \frac{\epsilon_{v_2}}{k_{v_2}}\bigg(\frac{{\tilde{\epsilon}}_{v_1}}{k_{v_1}}-\frac{{\tilde{\epsilon}}_{v_2}}{k_{v_2}}\bigg)\bigg] \bigg\}.
    \label{HGJ7}
\end{align}
Upon calculating the remaining terms using the integrals from App.~\ref{AppInte} and dropping the terms proportional to $\delta(b)$, we obtain:
\begin{align}
    A_{5}^{\mathrm{tachyon}} =&\, \frac{\kappa_D}{2\pi}\,\frac{(\alpha')^2 N_1N_2T_0^2\pi^2}{32\lvert\sinh(v_{12})\rvert}\, \int \frac{\mathrm{d}^{D-2}q}{(2\pi)^{D-2}} e^{i  q\cdot b}\nonumber 
    \\
    &\times\Big\{2 k_{v_1+v_{12}}^2  \bigg( \frac{\epsilon_{v_1}}{k_{v_1}}-\frac{\epsilon_{v_1+v_{12}}}{k_{v_1+v_{12}}}\bigg)\bigg( \frac{{\tilde{\epsilon}}_{v_1}}{k_{v_1}}-\frac{{\tilde{\epsilon}}_{v_1+v_{12}}}{k_{v_1+v_{12}}}\bigg) \frac{4}{q^2}\nonumber 
    \\
    &\qquad +2k_{v_2+v_{21}}\bigg( \frac{\epsilon_{v_2}}{k_{v_2}}-\frac{\epsilon_{v_2+v_{21}}}{k_{v_2+v_{21}}}\bigg)\bigg( \frac{{\tilde{\epsilon}}_{v_2}}{k_{v_2}}-\frac{{\tilde{\epsilon}}_{v_2+v_{21}}}{k_{v_2+v_{21}}}\bigg)\frac{4}{(q+k)^2} \nonumber 
    \\
    &\qquad +k_{v_1}^2 k_{v_2}^2 \bigg(\frac{\epsilon_{v_1}}{k_{v_1}}-\frac{\epsilon_{v_2}}{k_{v_2}}\bigg)\bigg(\frac{{\tilde{\epsilon}}_{v_1}}{k_{v_1}}-\frac{{\tilde{\epsilon}}_{v_2}}{k_{v_2}}\bigg)\bigg(-\frac{8}{(q+k)^2 k^2_{v_2}} -\frac{8}{q^2 k^2_{v_1}}+\frac{32}{q^2 (q+k)^2} \bigg)\nonumber
    \\
    &\qquad + (\epsilon\cdot{\tilde{\epsilon}})\bigg[ \frac{16(k^2_{v_1}+k^2_{v_2})}{q^2 (k+q)^2} + \frac{4}{q^2}\frac{k^2_{v_1+v_{12}}-k^2_{v_2}}{k^2_{v_1}} +\frac{4}{(q+k)^2}\frac{k^2_{v_2+v_{21}} -k^2_{v_1}}{k^2_{v_2}}\bigg] \nonumber 
    \\
    &\qquad +\frac{32 \cosh(v_{12})k_{v_1} k_{v_2}}{q^2 (k+q)^2}\bigg[ (q\cdot \epsilon)\bigg(\frac{{\tilde{\epsilon}}_{v_1}}{k_{v_1}}- \frac{{\tilde{\epsilon}}_{v_2}}{k_{v_2}}\bigg)+ (q\cdot {\tilde{\epsilon}})\bigg( \frac{{{\epsilon}}_{v_1}}{k_{v_1}}- \frac{{{\epsilon}}_{v_2}}{k_{v_2}}\bigg)\bigg] \nonumber 
    \\
    &\qquad - \frac{16 \cosh(v_{12})k_{v_1} k_{v_2}}{q^2}\bigg[ \bigg(\frac{{{\epsilon}}_{v_1}}{k_{v_1}}- \frac{{{\epsilon}}_{v_2}}{k_{v_2}}\bigg) \frac{{\tilde{\epsilon}}_{v_1}}{k_{v_1}}+\bigg(\frac{ {\tilde{\epsilon}}_{v_1}}{k_{v_1}}- \frac{{\tilde{\epsilon}}_{v_2}}{k_{v_2}}\bigg) \frac{{{\epsilon}}_{v_1}}{k_{v_1}}\bigg] \nonumber 
    \\
    &\qquad  + \frac{16 \cosh(v_{12}) k_{v_1} k_{v_2}}{(k+q)^2}\bigg[\bigg( \frac{{{\epsilon}}_{v_1}}{k_{v_1}}- \frac{{{\epsilon}}_{v_2}}{k_{v_2}}\bigg) \frac{{\tilde{\epsilon}}_{v_2}}{k_{v_2}}+\bigg( \frac{ {\tilde{\epsilon}}_{v_1}}{k_{v_1}}- \frac{{\tilde{\epsilon}}_{v_2}}{k_{v_2}}\bigg) \frac{{{\epsilon}}_{v_2}}{k_{v_2}}\bigg] \bigg\}
    \label{HGJ8}
\end{align}
Using identities:
\begin{align}
    &k_{v_1}^2 k_{v_1+v_{12}}^2  \bigg( \frac{\epsilon_{v_1}}{k_{v_1}}-\frac{\epsilon_{v_1+v_{12}}}{k_{v_1+v_{12}}}\bigg)\bigg( \frac{{\tilde{\epsilon}}_{v_1}}{k_{v_1}}-\frac{{\tilde{\epsilon}}_{v_1+v_{12}}}{k_{v_1+v_{12}}}\bigg) =  k_{v_1}^2 k_{v_2}^2 \bigg(\frac{\epsilon_{v_1}}{k_{v_1}}-\frac{\epsilon_{v_2}}{k_{v_2}}\bigg)\bigg(\frac{{\tilde{\epsilon}}_{v_1}}{k_{v_1}}-\frac{{\tilde{\epsilon}}_{v_2}}{k_{v_2}}\bigg),
    \\
    &k_{v_2}^2 k_{v_2+v_{21}}^2  \bigg( \frac{\epsilon_{v_2}}{k_{v_2}}-\frac{\epsilon_{v_2+v_{21}}}{k_{v_2+v_{21}}}\bigg)\bigg( \frac{{\tilde{\epsilon}}_{v_2}}{k_{v_2}}-\frac{{\tilde{\epsilon}}_{v_2+v_{21}}}{k_{v_2+v_{21}}}\bigg) =  k_{v_1}^2 k_{v_2}^2 \bigg(\frac{\epsilon_{v_1}}{k_{v_1}}-\frac{\epsilon_{v_2}}{k_{v_2}}\bigg)\bigg(\frac{{\tilde{\epsilon}}_{v_1}}{k_{v_1}}-\frac{{\tilde{\epsilon}}_{v_2}}{k_{v_2}}\bigg),
\end{align}
and noting that the terms in lines 2 and 3 cancel with two of the terms in the fourth line, we get
\begin{align}
    A_{5}^{\mathrm{tachyon}} =&\, \frac{\kappa_D}{2\pi}\,\frac{ N_1N_2T_0^2\pi^2}{32\lvert\sinh(v_{12})\rvert}\, \int \frac{\mathrm{d}^{D-2}q}{(2\pi)^{D-2}} e^{i  q\cdot b}\nonumber 
    \\
    &\times\Bigg\{k_{v_1}^2 k_{v_2}^2 \bigg(\frac{\epsilon_{v_1}}{k_{v_1}}-\frac{\epsilon_{v_2}}{k_{v_2}}\bigg) \bigg(\frac{{\tilde{\epsilon}}_{v_1}}{k_{v_1}}-\frac{{\tilde{\epsilon}}_{v_2}}{k_{v_2}}\bigg)\frac{32}{q^2 (q+k)^2} \nonumber 
    \\
    &\qquad + (\epsilon\cdot {\tilde{\epsilon}})\bigg[ \frac{16(k^2_{v_1}+k^2_{v_2}}{q^2 (k+q)^2} + \frac{4}{q^2}\frac{k^2_{v_1+v_{12}}-k^2_{v_2}}{k^2_{v_1}} +\frac{4}{(q+k)^2}\frac{k^2_{v_2+v_{21}} -k^2_{v_1}}{k^2_{v_2}}\bigg]\nonumber 
    \\
    &\qquad  +\frac{32 \cosh v_{12}k_{v_1} k_{v_2}}{q^2 (k+q)^2}\bigg[ (q\cdot \epsilon)\bigg( \frac{{\tilde{\epsilon}}_{v_1}}{k_{v_1}}- \frac{{\tilde{\epsilon}}_{v_2}}{k_{v_2}}\bigg)+ (q\cdot {\tilde{\epsilon}})\bigg( \frac{{{\epsilon}}_{v_1}}{k_{v_1}}- \frac{{{\epsilon}}_{v_2}}{k_{v_2}}\bigg)\bigg] \nonumber 
    \\
    &\qquad - \frac{16 \cosh v_{12}k_{v_1} k_{v_2}}{ q^2}\bigg[ \bigg( \frac{{{\epsilon}}_{v_1}}{k_{v_1}}- \frac{{{\epsilon}}_{v_2}}{k_{v_2}}\bigg) \frac{{\tilde{\epsilon}}_{v_1}}{k_{v_1}}+\bigg( \frac{ {\tilde{\epsilon}}_{v_1}}{k_{v_1}}- \frac{{\tilde{\epsilon}}_{v_2}}{k_{v_2}}\bigg) \frac{{{\epsilon}}_{v_1}}{k_{v_1}}\bigg] \nonumber 
    \\
    &\qquad  + \frac{16 \cosh v_{12}k_{v_1} k_{v_2}}{(k+q)^2}\bigg[ \bigg( \frac{{{\epsilon}}_{v_1}}{k_{v_1}}- \frac{{{\epsilon}}_{v_2}}{k_{v_2}}\bigg) \frac{{\tilde{\epsilon}}_{v_2}}{k_{v_2}}+\bigg( \frac{ {\tilde{\epsilon}}_{v_1}}{k_{v_1}}- \frac{{\tilde{\epsilon}}_{v_2}}{k_{v_2}}\bigg) \frac{{{\epsilon}}_{v_2}}{k_{v_2}}\bigg] \Bigg\}\, .
    \label{FinalTachyon8}
\end{align}
It can be written in a more compact way as follows
\begin{align}
    A_{5}^{\mathrm{tachyon}} =&\,\frac{\kappa_D}{2\pi}\,\frac{ N_1N_2T_0^2\pi^2}{32\lvert\sinh(v_{12})\rvert}\, \int \frac{d^{D-2}q}{(2\pi)^{D-2}} e^{i  q\cdot b}\nonumber 
    \\
    &\times \Bigg\{\frac{32 P_{12}^\mu P_{12}^\nu-16\gamma (Q_{12}^\mu P_{12}^\nu+ Q_{12}^\nu P_{12}^\mu)}{q^2 (k+q)^2}\nonumber 
    \\
    &\qquad +\eta^{\mu \nu}\bigg( \frac{16(k^2_{v_1}+k^2_{v_2})}{q^2 (k+q)^2} + \frac{4}{q^2}\frac{k^2_{v_1+v_{12}}-k^2_{v_2}}{k^2_{v_1}} +\frac{4}{(q+k)^2}\frac{k^2_{v_2+v_{21}} -k^2_{v_1}}{k^2_{v_2}} \bigg) \Bigg\}
    \label{Ftachyon2}
\end{align}

\subsection{NS-NS sector}\label{AppD2}
This subsection is devoted to the field-theory limit of the NS-NS sector of the superstring. The GSO-projected amplitude is
\begin{equation}\label{NSNSGSO}
    A_{\text{NS}}=\frac14\left[A_{\text{NS}}(++)-A_{\text{NS}}(+-)-A_{\text{NS}}(-+)+A_{\text{NS}}(--)\right],
\end{equation}
where the contributions from the different spin structures are written explicitly in Eq.~\eqref{NSNSFINAL}. In the small-$|w|$ limit, with both $|z|$ and $\left|\frac{w}{z}\right|$ kept finite, these contributions can be approximated as
\begin{align}
    A_{\text{NS}}(\eta_1,\eta_2) &=\frac{\kappa_D}{2\pi}\,\frac{\alpha' N_1N_2T_0^2}{32\lvert\sinh(v_{12})\rvert}\, \int \frac{\mathrm{d}^{D-2}q}{(2\pi)^{D-2}} e^{i  q\cdot b} \int\mathrm{d}\theta\,\mathrm{d}\phi\int\mathrm{d}\tilde\theta\,\mathrm{d}\tilde\phi \nonumber
    \\
    &\quad\times\int_0^1\frac{\mathrm{d}^2w}{\lvert w\rvert^2}\,\lvert w\rvert^{\frac{\alpha'}{2}q^2} \int_{\lvert w\rvert}^{1}\mathrm{d}^2z\, \lvert z\rvert^{\alpha' k\cdot q}\left(\frac1{\lvert w\rvert}+\eta_1\eta_2(D-4+2\cosh(2v_{12}))\right) \nonumber 
    \\
    &\quad \times (1-\lvert z\rvert^2)^{-2\tilde A_{v_1}A_{v_1}-\tilde A\cdot A} \left(1-\left\lvert \tfrac{w}{z}\right\rvert^2\right)^{-2\tilde A_{v_2}A_{v_2}-\tilde A\cdot A}\nonumber
    \\
    &\quad\times e^{4\sqrt{\tfrac{\alpha'}{2}}\eta_1\eta_2\lvert w\rvert\cosh(v_{12})\left[\frac{1}{z}(k_{v_1}\epsilon_{v_2}-k_{v_2}\epsilon_{v_1})\,\theta\phi + \frac{1}{\bar{z}}(k_{v_1}\tilde{\epsilon}_{v_2}-k_{v_2}\tilde{\epsilon}_{v_1})\,\tilde\theta\tilde\phi\right]}
    \nonumber
    \\
    &\quad\times e^{-i \eta_1(2\tilde R_{v_1}R_{v_1}+\tilde R\cdot R)\frac1{1-\lvert z\rvert^2}} e^{-i \eta_2\frac{\lvert w\rvert}{\lvert z\rvert^2}(2\tilde R_{v_2}R_{v_2}+\tilde R\cdot R)\frac1{1-\lvert w/z\rvert^2}} \nonumber
    \\
    &\quad\times e^{i \eta_2\lvert w\rvert(2\tilde R_{v_1+v_{12}}R_{v_1+v_{12}}+\tilde R\cdot R)}e^{i \eta_1\frac{\lvert w\rvert^2}{\lvert z\rvert^2}(2\tilde R_{v_2+v_{21}}R_{v_2+v_{21}}+\tilde R\cdot R)}.\label{ANSeta1eta2_field_thry}
\end{align}
In Eq.~\eqref{ANSeta1eta2_field_thry}, the third and fourth lines contain Grassmann-valued terms proportional to $\theta\phi$, $\tilde\theta\tilde\phi$, or $\tilde\theta\tilde\phi\theta\phi$. By contrast, the fifth and sixth lines contain terms proportional to $\theta\tilde\phi$, $\tilde\theta\phi$, $\theta\tilde\theta$, $\phi\tilde\phi$, and $\tilde\theta\tilde\phi\theta\phi$. We can therefore write the contributions from lines 3 and 4, denoted by $A_{3,4}$, and from lines 5 and 6, denoted by $A_{5,6}$, in the schematic form
\begin{align}
    A_{3,4} &= a_0 + a_1 \theta\phi + a_2\tilde{\theta}\tilde{\phi} + a_3 \tilde{\theta}\tilde{\phi} + a_4\tilde{\theta}\tilde{\phi}\theta\phi,\label{A34}
    \\
    A_{5,6} &= b_0 + b_1\theta\tilde{\theta} + b_2 \tilde{\theta}\phi + b_3 \tilde{\phi}\theta + b_4\tilde{\phi}\phi + b_5 \tilde{\theta}\tilde{\phi}\theta\phi,\label{A56}
\end{align}
which implies that:
\begin{equation}\label{grassmann_integral}
    \int d^2\theta\,d^2\phi \, A_{3,4}A_{5,6} = a_0 b_5 + a_3 b_0.
\end{equation}
Thus, after the Grassmann integration, the polarization-dependent parts of the matrix elements coming from lines 3,4 and from lines 5,6 of Eq.~\eqref{ANSeta1eta2_field_thry} contribute separately. We may therefore take the field-theory limit of the $A_{3,4}$ and $A_{5,6}$ parts independently.

Let us first extract the terms bilinear in the polarizations from $A_{5,6}$. We focus on the contribution
\begin{align}
    &e^{-i \eta_1(2\tilde R_{v_1}R_{v_1}+\tilde R\cdot R)\frac1{1- \lvert z\rvert^2}} e^{-i \eta_2\frac{\lvert w\rvert}{\lvert z\rvert^2}(2\tilde R_{v_2}R_{v_2}+\tilde R\cdot R)\frac1{1-\lvert \tfrac{w}{z}\rvert^2}} \nonumber 
    \\
    &\times e^{i \eta_2\lvert w\rvert(2\tilde R_{v_1+v_{12}}R_{v_1+v_{12}}+\tilde R\cdot R)} e^{i \eta_1\frac{\lvert w\rvert^2}{\lvert z\rvert^2}(2\tilde R_{v_2+v_{21}}R_{v_2+v_{21}}+\tilde R\cdot R)},\label{lines_56}
\end{align}
Using the identities
\begin{align}
    &\tilde R_{v_1}R_{v_1}+\frac12\tilde R\cdot R =\frac{\alpha'}{2}k_{v_1}^2\tilde\theta\theta+ \sqrt{\frac{\alpha'}{2}}\,\tilde{\epsilon}_{v_1}k_{v_1}\tilde\theta\theta+ \sqrt{\frac{\alpha'}{2}}\, \epsilon_{v_1}k_{v_1}\tilde\theta\theta+ \left(\tilde{\epsilon}_{v_1}\epsilon_{v_1}+\frac{\tilde{\epsilon}\cdot\epsilon}{2}\right)\tilde\theta\theta,\label{RR_identity}
    \\
    &(\tilde R_{v_1}R_{v_1}+\tfrac12\tilde R\cdot R)(\tilde R_{v_2}R_{v_2}+\tfrac12\tilde R\cdot R)\nonumber 
    \\
    &\qquad \qquad \qquad \quad\;\; =-\frac{\alpha'}{2}k_{v_1}^2k_{v_2}^2\left[\left(\frac{\epsilon_{v_1}}{k_{v_1}}-\frac{\epsilon_{v_2}}{k_{v_2}}\right) \left(\frac{\tilde{\epsilon}_{v_1}}{k_{v_1}}-\frac{\tilde{\epsilon}_{v_2}}{k_{v_2}}\right) +\frac{\tilde{\epsilon}\cdot\epsilon}{2}\left(\frac1{k_{v_1}^2}+\frac1{k_{v_2}^2}\right)\right]\theta\phi\tilde\theta\tilde\phi,\label{RRRR_identity}
\end{align}
and extracting from Eq.~\eqref{lines_56} the terms bilinear in $R$ and $\tilde R$, we obtain
\begin{align}
    &-\frac12\frac1{(1-\lvert z\rvert^2)^2}(2\tilde R_{v_1}R_{v_1}+\tilde R\cdot R)^2 \nonumber 
    \\
    &-\frac12\frac{\lvert w\rvert^2}{\lvert z\rvert^4}\frac1{(1-\lvert \tfrac{w}{z}\rvert^2)^2}(2\tilde R_{v_2}R_{v_2}+\tilde R\cdot R)^2 \nonumber
    \\
    &- \eta_1\eta_2\frac{\lvert w\rvert}{\lvert z\rvert^2}\frac1{(1-\lvert z\rvert^2)(1-\lvert \tfrac{w}{z}\rvert^2)}(2\tilde R_{v_1}R_{v_1}+\tilde R\cdot R)(2\tilde R_{v_2}R_{v_2}+\tilde R\cdot R) \nonumber
    \\
    &+\eta_1\eta_2\frac{\lvert w\rvert}{1-\lvert z\rvert^2}(2\tilde R_{v_1}R_{v_1}+\tilde R\cdot R)(2\tilde R_{v_1+v_{12}}R_{v_1+v_{12}}+\tilde R\cdot R) \nonumber
    \\
    &+\eta_1\eta_2\frac{\lvert w\rvert}{\lvert z\rvert^2}\frac1{1-\lvert \tfrac{w}{z}\rvert^2}(2\tilde R_{v_2}R_{v_2}+\tilde R\cdot R)(2\tilde R_{v_2+v_{21}}R_{v_2+v_{21}}+\tilde R\cdot R) \nonumber
    \\
    &+\frac{\lvert w\rvert^2}{\lvert z\rvert^2}\frac1{1-\lvert z\rvert^2}(2\tilde R_{v_1}R_{v_1}+\tilde R\cdot R)(2\tilde R_{v_2+v_{21}}R_{v_2+v_{21}}+\tilde R\cdot R) \nonumber
    \\
    &+\frac{\lvert w\rvert^2}{\lvert z\rvert^2}\frac{1}{1-\lvert \tfrac{w}{z}\rvert^2}(2\tilde R_{v_2}R_{v_2}+\tilde R\cdot R)(2\tilde R_{v_1+v_{12}}R_{v_1+v_{12}}+\tilde R\cdot R)\nonumber 
    \\
    &-\frac{1}{2}|w|^2(2\tilde{R}_{v_1 + v_{12}}R_{v_1 + v_{12}} + \tilde{R}\cdot R)^2\nonumber 
    \\
    &-\frac{1}{2}\frac{|w|^4}{|z|^4}(2\tilde{R}_{v_2 + v_{21}}R_{v_2 + v_{21}} + \tilde{R}\cdot R)^2\nonumber 
    \\
    &+\eta_1 \eta_2 \frac{|w|^3}{|z|^2} (2\tilde{R}_{v_1 + v_{12}}R_{v_1 + v_{12}} + \tilde{R}\cdot R)(2\tilde{R}_{v_2 + v_{21}}R_{v_2 + v_{21}} + \tilde{R}\cdot R). \label{eq:fermion-bilinear-star}
\end{align}
The terms in the last three lines of the above equation are of higher order in the small-$|w|$ expansion and will be neglected from now on.

The GSO projection \eqref{NSNSGSO} selects from $A_{\text{NS}}(\eta_1,\eta_2)$ only the terms proportional to $\eta_1\eta_2$; the terms independent of either $\eta_1$ or $\eta_2$ cancel in the spin-structure sum. The relevant small-$|w|$ expansion of the partition function of NS-fermions is
\begin{align}
    \frac{1}{|w|}\prod_{r=1/2}^\infty (1 + 2\eta_1 \eta_2 \cosh(v_{12})|w|^{2r} + |w|^{4r})(1 + \eta_1 \eta_2 |w|^{2r})^{D-4}\nonumber 
    \\
    = \frac1{\lvert w\rvert}+\eta_1\eta_2(D-4+2\cosh(2v_{12}))+O(\lvert w\rvert).
\end{align}
It follows that the terms in the first and fourth lines of Eq.~\eqref{eq:fermion-bilinear-star} must be multiplied by $\eta_1\eta_2(D-4+2\cosh(2v_{12}))\lvert w\rvert$. The terms in the fourth line are then of higher order in the small-$|w|$ expansion and can be neglected. Therefore, multiplying Eq.~\eqref{eq:fermion-bilinear-star} by the fermionic partition function and performing the GSO projection gives
\begin{align}
    &\frac1{1-\lvert z\rvert^2}(2\tilde R_{v_1}R_{v_1}+\tilde R\cdot R)(2\tilde R_{v_1+v_{12}}R_{v_1+v_{12}}+\tilde R\cdot R)\nonumber  
    \\
    &+\frac{1}{\lvert z\rvert^2}\frac{\lvert \tfrac{w}{z}\rvert^2}{1-\lvert \tfrac{w}{z}\rvert^2}(2\tilde R_{v_2}R_{v_2}+\tilde R\cdot R)(2\tilde R_{v_2+v_{21}}R_{v_2+v_{21}}+\tilde R\cdot R) \nonumber
    \\
    &-\frac{1}{\lvert z\rvert^2}\frac1{(1-\lvert z\rvert^2)(1-\lvert \tfrac{w}{z}\rvert^2)}(2\tilde R_{v_1}R_{v_1}+\tilde R\cdot R)(2\tilde R_{v_2}R_{v_2}+\tilde R\cdot R) \nonumber
    \\
    &-\frac{D-4+2\cosh(2v_{12})}{2\lvert z\rvert^2} \bigg[\frac{\lvert z\rvert^2}{(1-\lvert z\rvert^2)^2}(2\tilde R_{v_1}R_{v_1}+\tilde R\cdot R)^2\nonumber 
    \\
    &\qquad \qquad \qquad \qquad \qquad \quad +\frac{\lvert \tfrac{w}{z}\rvert^2}{(1-\lvert \tfrac{w}{z}\rvert^2)^2}(2\tilde R_{v_2}R_{v_2}+\tilde R\cdot R)^2\bigg].
\end{align}
Using Eq.~\eqref{RRRR_identity}, this expression can be rewritten as
\begin{align}
    &\frac{\alpha'}{\lvert z\rvert^2}\Bigg\{-\frac{2\lvert z\rvert^2}{1-\lvert z\rvert^2}\left[k_{v_1}^2 k_{v_1 + v_{12}}^2\left(\frac{\epsilon_{v_1}}{k_{v_1}}-\frac{\epsilon_{v_1+v_{12}}}{k_{v_1 + v_{12}}}\right)\left(\frac{\tilde \epsilon_{v_1}}{k_{v_1}}-\frac{\tilde \epsilon_{v_1+v_{12}}}{k_{v_1 + v_{12}}}\right)+\frac{\tilde{\epsilon}\cdot\epsilon}{2}(k_{v_1}^2+k_{v_1+v_{12}}^2)\right] \nonumber
    \\
    &\qquad\;\;\,-\frac{2\lvert \tfrac{w}{z}\rvert^2}{1-\lvert \tfrac{w}{z}\rvert^2}\left[k_{v_2}^2 k_{v_2 + v_{21}}^2\left(\frac{\epsilon_{v_2}}{k_{v_2}}-\frac{\epsilon_{v_2+v_{21}}}{k_{v_2 + v_{21}}}\right)\left(\frac{\tilde \epsilon_{v_1}}{k_{v_1}}-\frac{\tilde \epsilon_{v_1+v_{12}}}{k_{v_1 + v_{12}}}\right)+\frac{\tilde{\epsilon}\cdot\epsilon}{2}(k_{v_2}^2+k_{v_2+v_{21}}^2)\right]\nonumber
    \\
    &\qquad\;\;\,+\frac{2}{(1-\lvert z\rvert^2)(1-\lvert \tfrac{w}{z}\rvert^2)} \left[k_{v_1}^2 k_{v_2}^2\left(\frac{\epsilon_{v_1}}{k_{v_1}}-\frac{\epsilon_{v_{2}}}{k_{v_{2}}}\right)\left(\frac{\tilde \epsilon_{v_1}}{k_{v_1}}-\frac{\tilde \epsilon_{v_{2}}}{k_{v_{2}}}\right)+\frac{\tilde{\epsilon}\cdot\epsilon}{2}(k_{v_1}^2+k_{v_{2}}^2)\right]\nonumber
    \\
    &\qquad\;\;\, + (D-4+2\cosh(2v_{12}))\left(\frac{k_{v_1}^2\lvert z\rvert^2}{(1-\lvert z\rvert^2)^2}+\frac{k_{v_2}^2\lvert \tfrac{w}{z}\rvert^2}{(1-\lvert \tfrac{w}{z}\rvert^2)^2}\right)(\tilde{\epsilon}\cdot\epsilon)\Bigg\}\theta\phi\tilde\theta\tilde\phi.
\end{align}
The corresponding contribution to the full amplitude, namely the GSO projection of the $a_0 b_5$ term from Eq.~\eqref{grassmann_integral}, is
\begin{align}
    &\frac{\kappa_D}{2\pi}\frac{\alpha'^2 N_1N_2T_0^2}{32\lvert\sinh(v_{12})\rvert}\int \frac{\mathrm{d}^{D-2}q}{(2\pi)^{D-2}} e^{i  q\cdot b} \int_0^1\frac{\mathrm{d}^2w}{\lvert w\rvert^2}\,\lvert w\rvert^{\frac{\alpha'}{2}q^2} \int_{\lvert w\rvert}^{1}\frac{\mathrm{d}^2z}{\lvert z\rvert^2}\, \lvert z\rvert^{\alpha' k\cdot q} \nonumber
    \\
    &\times\Bigg\{-\frac{2\lvert z\rvert^2}{1-\lvert z\rvert^2}\left[k_{v_1}^2 k_{v_1 + v_{12}}^2\left(\frac{\epsilon_{v_1}}{k_{v_1}}-\frac{\epsilon_{v_1+v_{12}}}{k_{v_1 + v_{12}}}\right)\left(\frac{\tilde \epsilon_{v_1}}{k_{v_1}}-\frac{\tilde \epsilon_{v_1+v_{12}}}{k_{v_1 + v_{12}}}\right)+\frac{\tilde{\epsilon}\cdot\epsilon}{2}(k_{v_1}^2+k_{v_1+v_{12}}^2)\right] \nonumber
    \\
    &\qquad-\frac{2\lvert \tfrac{w}{z}\rvert^2}{1-\lvert \tfrac{w}{z}\rvert^2}\left[k_{v_2}^2 k_{v_2 + v_{21}}^2\left(\frac{\epsilon_{v_2}}{k_{v_2}}-\frac{\epsilon_{v_2+v_{21}}}{k_{v_2 + v_{21}}}\right)\left(\frac{\tilde \epsilon_{v_1}}{k_{v_1}}-\frac{\tilde \epsilon_{v_1+v_{12}}}{k_{v_1 + v_{12}}}\right)+\frac{\tilde{\epsilon}\cdot\epsilon}{2}(k_{v_2}^2+k_{v_2+v_{21}}^2)\right]\nonumber
    \\
    &\qquad+\frac{2}{(1-\lvert z\rvert^2)(1-\lvert \tfrac{w}{z}\rvert^2)} \left[k_{v_1}^2 k_{v_2}^2\left(\frac{\epsilon_{v_1}}{k_{v_1}}-\frac{\epsilon_{v_{2}}}{k_{v_{2}}}\right)\left(\frac{\tilde \epsilon_{v_1}}{k_{v_1}}-\frac{\tilde \epsilon_{v_{2}}}{k_{v_{2}}}\right)+\frac{\tilde{\epsilon}\cdot\epsilon}{2}(k_{v_1}^2+k_{v_{2}}^2)\right]\nonumber
    \\
    &\qquad + (D-4+2\cosh(2v_{12}))\left(\frac{k_{v_1}^2\lvert z\rvert^2}{(1-\lvert z\rvert^2)^2}+\frac{k_{v_2}^2\lvert \tfrac{w}{z}\rvert^2}{(1-\lvert \tfrac{w}{z}\rvert^2)^2}\right)(\tilde{\epsilon}\cdot\epsilon)\Bigg\}.
\end{align}
The terms in the second and third lines of this expression can be simplified using the identities
\begin{align}
    &k_{v_1}\epsilon_{v_1 + v_{12}}-k_{v_1 + v_{12}}\epsilon_{v_1}  = -\sinh(v_{12})(\epsilon^0k^1-\epsilon ^1k^0) = -(k_{v_1}\epsilon_{v_2} -k_{v_2}\epsilon_{v_1}),
    \\
    &k_{v_1}^2 k_{v_1 + v_{12}}^2\left(\frac{\epsilon_{v_1}}{k_{v_1}} - \frac{\epsilon_{v_1 + v_{12}}}{k_{v_1 + v_{12}}}\right)\left(\frac{\tilde{\epsilon}_{v_1}}{k_{v_1}} - \frac{\tilde{\epsilon}_{v_1 + v_{12}}}{k_{v_1 + v_{12}}}\right) 
    =
    k_{v_1}^2 k_{v_{2}}^2\left(\frac{\epsilon_{v_1}}{k_{v_1}} - \frac{\epsilon_{v_2}}{k_{v_2}}\right)\left(\frac{\tilde{\epsilon}_{v_1}}{k_{v_1}} - \frac{\tilde{\epsilon}_{v_2}}{k_{v_{2}}}\right),\label{v1v12_identity}
\end{align}
which follow from
\begin{equation}
    k_{v_1}\epsilon_{v_2}-k_{v_2}\epsilon_{v_1}  =\sinh(v_{12})(\epsilon^0k^1-\epsilon ^1k^0).
\end{equation}
 Moreover, as in the bosonic string case  we change variables to $x=|z|^2$ and $y=\left|\frac{w}{z}\right|^2$. This gives
\begin{align}
    &\frac{\kappa_D}{2\pi}\frac{\pi^2 \alpha'^2 N_1N_2T_0^2}{32\lvert\sinh(v_{12})\rvert}\int \frac{\mathrm{d}^{D-2}q}{(2\pi)^{D-2}} e^{i  q\cdot b} \int_0^1dx\,x^{\frac{\alpha'}{4}(k+q)^2 - 1}(1-x)^{-\alpha' k^2_{v_1}} \int_{0}^{1}dy\, y^{\frac{\alpha'}{4}q^2 - 1}(1-y)^{-\alpha'k^2_{v_2}} \nonumber
    \\
    &\times\Bigg\{-\frac{2x}{1-x}\left[k_{v_1}^2 k_{v_2}^2\left(\frac{\epsilon_{v_1}}{k_{v_1}}-\frac{\epsilon_{v_2}}{k_{v_{2}}}\right)\left(\frac{\tilde \epsilon_{v_1}}{k_{v_1}}-\frac{\tilde \epsilon_{v_{2}}}{k_{v_{2}}}\right)+\frac{\tilde{\epsilon}\cdot\epsilon}{2}(k_{v_1}^2+k_{v_1+v_{12}}^2)\right] \nonumber
    \\
    &\qquad-\frac{2y}{1-y}\left[k_{v_1}^2 k_{v_2}^2\left(\frac{\epsilon_{v_1}}{k_{v_1}}-\frac{\epsilon_{v_2}}{k_{v_2}}\right)\left(\frac{\tilde \epsilon_{v_1}}{k_{v_1}}-\frac{\tilde \epsilon_{v_{2}}}{k_{v_{2}}}\right)+\frac{\tilde{\epsilon}\cdot\epsilon}{2}(k_{v_2}^2+k_{v_2+v_{21}}^2)\right]\nonumber
    \\
    &\qquad+\frac{2}{(1- x)(1-y)} \left[k_{v_1}^2 k_{v_2}^2\left(\frac{\epsilon_{v_1}}{k_{v_1}}-\frac{\epsilon_{v_{2}}}{k_{v_{2}}}\right)\left(\frac{\tilde \epsilon_{v_1}}{k_{v_1}}-\frac{\tilde \epsilon_{v_{2}}}{k_{v_{2}}}\right)+\frac{\tilde{\epsilon}\cdot\epsilon}{2}(k_{v_1}^2+k_{v_{2}}^2)\right]\nonumber
    \\
    &\qquad + (D-4+2\cosh(2v_{12}))\left(\frac{k_{v_1}^2 x }{(1- x )^2}+\frac{k_{v_2}^2y}{(1-y)^2}\right)(\tilde{\epsilon}\cdot\epsilon)\Bigg\}.
\end{align}
From each of the $x$- and $y$-integrals, we need to extract only the terms of order $1/\alpha'$ in the $\alpha'\to0$ limit, since only these terms can cancel the overall $\alpha'^2$ factor multiplying the amplitude. Thus, using the integrals in Appendix~\ref{AppInte}, and denoting $r=k+q$, we obtain:
\begin{align}
    &\frac{\kappa_D}{2\pi}\frac{\pi^2N_1N_2T_0^2}{32\lvert\sinh(v_{12})\rvert}\int \frac{\mathrm{d}^{D-2}q}{(2\pi)^{D-2}} e^{i  q\cdot b} \nonumber 
    \\
    &\times \bigg\{ \frac8{k_{v_1}^2q^2} \left[(k_{v_1}\epsilon_{v_2}-k_{v_2}\epsilon_{v_1})(k_{v_1}\tilde{\epsilon}_{v_2}-k_{v_2}\tilde{\epsilon}_{v_1})+\frac{\tilde{\epsilon}\cdot\epsilon}{2}(k_{v_1}^2+k_{v_1+v_{12}}^2)\right] \nonumber
    \\
    &\qquad+\frac8{k_{v_2}^2r^2} \left[(k_{v_1}\epsilon_{v_2}-k_{v_2}\epsilon_{v_1})(k_{v_1}\tilde{\epsilon}_{v_2}-k_{v_2}\tilde{\epsilon}_{v_1})+\frac{\tilde{\epsilon}\cdot\epsilon}{2}(k_{v_2}^2+k_{v_2+v_{21}}^2)\right] \nonumber
    \\
    &\qquad+\left(\frac{16k_{v_1}^2k_{v_2}^2}{r^2 q^2} - \frac{4k_{v_1}^2}{ r^2} - \frac{4k_{v_2}^2}{ q^2} + 1\right) \bigg[2\left(\frac{\epsilon_{v_1}}{k_{v_1}}-\frac{\epsilon_{v_2}}{k_{v_2}}\right)\left(\frac{\tilde{\epsilon}_{v_1}}{k_{v_1}}-\frac{\tilde{\epsilon}_{v_2}}{k_{v_2}}\right) + (\tilde{\epsilon}\cdot\epsilon)\left(\frac{1}{k_{v_1}^2}+\frac{1}{k_{v_2}^2}\right)\bigg] \bigg\}.
\label{eq:superstring-fermionic-result}
\end{align}
Dropping the terms proportional to $\delta(b)$, we can recast \eqref{eq:superstring-fermionic-result} as:
\begin{align}
    \frac{\kappa_D}{2\pi}\frac{\pi^2N_1N_2T_0^2}{32\lvert\sinh(v_{12})\rvert}\int \frac{\mathrm{d}^{D-2}q}{(2\pi)^{D-2}} e^{i  q\cdot b} \bigg\{\frac4{k_{v_1}^2q^2}(\tilde{\epsilon}\cdot\epsilon)(k_{v_1+v_{12}}^2-k_{v_2}^2) +\frac4{k_{v_2}^2r^2}(\tilde{\epsilon}\cdot\epsilon)(k_{v_2+v_{21}}^2-k_{v_1}^2) \nonumber 
    \\
    +  \frac{32}{q^2 r^2} \left[(k_{v_1}\epsilon_{v_2}-k_{v_2}\epsilon_{v_1})(k_{v_1}\tilde{\epsilon}_{v_2}-k_{v_2}\tilde{\epsilon}_{v_1})+\frac{\tilde{\epsilon}\cdot\epsilon}{2}(k_{v_1}^2+k_{v_2}^2)\right]\bigg\}.\label{A56_final}
\end{align}
Let us now analyze the contribution from the third and fourth lines of Eq.~\eqref{ANSeta1eta2_field_thry}. Before the GSO projection, this contribution is
\begin{align}
    &\frac{\kappa_D}{2\pi}\,\frac{\alpha' N_1N_2T_0^2}{32\lvert\sinh(v_{12})\rvert}\, \int \frac{\mathrm{d}^{D-2}q}{(2\pi)^{D-2}} e^{i  q\cdot b} \int_0^1\frac{\mathrm{d}^2w}{\lvert w\rvert^2} \int_{ \lvert w\rvert}^{1}\mathrm{d}^2z\,\lvert w\rvert^{\frac{\alpha'}{2}q^2} \lvert z\rvert^{\alpha'k\cdot q} \int\mathrm{d}\theta\mathrm{d}\phi\int\mathrm{d}\tilde\theta\mathrm{d}\tilde\phi\nonumber
    \\
    &\times\left(\frac1{\lvert w\rvert}+(D-4+2\cosh(2v_{12}))\eta_1\eta_2\right)\nonumber 
    \\
    &\times (1-\lvert z\rvert^2)^{- \alpha'k_{v_1}^2}\left(1-\left\lvert  \tfrac{w}{z} \right\rvert^2\right)^{- \alpha'k_{v_2}^2} e^{\sqrt{\frac{\alpha'}2}q\cdot\left(\frac{\epsilon}{z}\theta\phi+\frac{\tilde{\epsilon}}{\bar z}\tilde\theta\tilde\phi\right)} \nonumber
    \\
    &\times(1-\lvert z\rvert^2)^{- \sqrt{2\alpha'}\epsilon_{v_1}k_{v_1}\theta\phi\partial_z- \sqrt{2\alpha'}\tilde{\epsilon}_{v_1}k_{v_1}\tilde\theta\tilde\phi\partial_{\bar z}-(2\tilde{\epsilon}_{v_1}\epsilon_{v_1}+\tilde{\epsilon}\cdot\epsilon)\tilde\theta\tilde\phi\theta\phi\partial_z\partial_{\bar z}} \nonumber
    \\
    &\times\left(1-\left\lvert  \tfrac{w}{z} \right\rvert^2\right)^{- \sqrt{2\alpha'}\epsilon_{v_2}k_{v_2}\theta\phi\partial_z- \sqrt{2\alpha'}\tilde{\epsilon}_{v_2}k_{v_2}\tilde\theta\tilde\phi\partial_{\bar z}-(2\tilde{\epsilon}_{v_2}\epsilon_{v_2}+\tilde{\epsilon}\cdot\epsilon)\tilde\theta\tilde\phi\theta\phi\partial_z\partial_{\bar z}} \nonumber
    \\
    &\times e^{4\sqrt{\tfrac{\alpha'}{2}}\eta_1\eta_2 |w| \cosh( v_{12})\left[\frac{1}{z}(k_{v_1}\epsilon_{v_2}-k_{v_2}\epsilon_{v_1})\,\theta\phi + \frac{1}{\bar z}(k_{v_1}\tilde{\epsilon}_{v_2}-k_{v_2}\tilde{\epsilon}_{v_1})\tilde\theta\tilde\phi\right]}.
\end{align}
Performing the Grassmann integrals gives
\begin{align}
    &\frac{\kappa_D}{2\pi}\,\frac{\alpha' N_1N_2T_0^2}{32\lvert\sinh(v_{12})\rvert}\, \int \frac{\mathrm{d}^{D-2}q}{(2\pi)^{D-2}} e^{i  q\cdot b} \int_0^1\frac{\mathrm{d}^2w}{\lvert w\rvert^2}\,\lvert w\rvert^{\frac{\alpha'}{2}q^2}\int_{ \lvert w\rvert}^{1}\mathrm{d}^2z\,  \lvert z\rvert^{\alpha'k\cdot q} \nonumber
    \\
    &\times\left(\frac1{\lvert w\rvert}+(D-4+2\cosh(2v_{12}))\eta_1\eta_2\right)(1-\lvert z\rvert^2)^{-\alpha'k_{v_1}^2} \left(1-\left\lvert  \tfrac{w}{z} \right\rvert^2\right)^{-\alpha'k_{v_2}^2} \nonumber
    \\
    &\times\bigg\{2\alpha'\bigg[U(z,\bar{z})+2\eta_1\eta_2\cosh(v_{12})\frac{\lvert w\rvert}{z} (k_{v_1}\epsilon_{v_2}-k_{v_2}\epsilon_{v_1})\bigg] \nonumber
    \\
    &\qquad\;\;\,\times\bigg[\tilde{U}(z,\bar{z}) +2\eta_1\eta_2\cosh(v_{12})\frac{\lvert w\rvert}{\bar z} (k_{v_1}\tilde{\epsilon}_{v_2}-k_{v_2}\tilde{\epsilon}_{v_1})\bigg] \nonumber
    \\
    &\qquad-(2\tilde{\epsilon}_{v_1}\epsilon_{v_1}+\tilde{\epsilon}\cdot\epsilon) \partial_z\partial_{\bar z}\log(1-\lvert z\rvert^2) -(2\tilde{\epsilon}_{v_2}\epsilon_{v_2}+\tilde{\epsilon}\cdot\epsilon) \partial_z\partial_{\bar z}\log\!\left(1 - \left\lvert  \tfrac{w}{z} \right\rvert^2\right)\Bigg\} .
\end{align}
Here we have introduced the shorthand notation
\begin{subequations}
\begin{align}
    U(z,\bar{z}) &= \frac{q\cdot\epsilon}{2}-\epsilon_{v_1}k_{v_1}\partial_z\log(1-\lvert z\rvert^2) -\epsilon_{v_2}k_{v_2}\partial_z\log\!\left(1- \left\lvert  \tfrac{w}{z} \right\rvert^2\right),
    \\
    \tilde{U}(z,\bar{z}) &= \frac{q\cdot\tilde{\epsilon}}{2}-\tilde{\epsilon}_{v_1}k_{v_1}\partial_{\bar z}\log(1- \lvert z\rvert^2)-\tilde{\epsilon}_{v_2}k_{v_2}\partial_{\bar z}\log\!\left(1- \left\lvert  \tfrac{w}{z} \right\rvert^2\right).
\end{align}
\end{subequations}
Applying the GSO projection, we retain only the terms proportional to $\eta_1\eta_2$, obtaining
\begin{align}
    &\frac{\kappa_D}{2\pi}\,\frac{\alpha' N_1N_2T_0^2}{32\lvert\sinh(v_{12})\rvert}\, \int \frac{\mathrm{d}^{D-2}q}{(2\pi)^{D-2}} e^{i  q\cdot b} \int_0^1\frac{\mathrm{d}^2w}{\lvert w\rvert^2}\lvert w\rvert^{\frac{\alpha'}{2}q^2}\int_{ \lvert w\rvert}^{1}\mathrm{d}^2z\, \lvert z\rvert^{\alpha'k\cdot q}\nonumber 
    \\
    &\times (1-\lvert z\rvert^2)^{- \alpha'k_{v_1}^2}\left(1-\left\lvert  \tfrac{w}{z} \right\rvert^2\right)^{-\alpha'k_{v_2}^2} \nonumber
    \\
    &\times\bigg\{(D-4+2\cosh(2v_{12}))\bigg[2\alpha' U(z,\bar{z}) \tilde{U}(z,\bar{z})-(2\tilde{\epsilon}_{v_1}\epsilon_{v_1}+\tilde{\epsilon}\cdot\epsilon)\partial_z\partial_{\bar z}\log(1-\lvert z\rvert^2) \nonumber
    \\
    &\qquad\qquad\qquad \qquad \qquad \qquad \  -(2\tilde{\epsilon}_{v_2}\epsilon_{v_2}+\tilde{\epsilon}\cdot\epsilon)\partial_z\partial_{\bar z}\log\!\left(1- \left\lvert  \tfrac{w}{z} \right\rvert^2\right)\bigg] \nonumber
    \\
    &\qquad\;+\frac{4\alpha'}{z}\cosh(v_{12})(k_{v_1}\epsilon_{v_2}-k_{v_2}\epsilon_{v_1}) U(z,\bar{z}) + \frac{4\alpha'}{\bar{z}}\cosh(v_{12})(k_{v_1}\tilde{\epsilon}_{v_2}-k_{v_2}\tilde{\epsilon}_{v_1})\tilde{U}(z,\bar{z})\bigg\} .
\end{align}
{\allowdisplaybreaks The terms proportional to $D-4+2\cosh(2v_{12})$ coincide with the $A_{5}^{\text{massless}}$ term in the bosonic string result, Eq.~\eqref{A_massless_bosonic_aux}. As in \ref{AppD1}, we evaluate the derivatives of the logarithms, integrate by parts the terms involving $\partial_z\partial_{\bar z}\log(1-|z|^2)$ and $\partial_z\partial_{\bar z}\log(1-\left|\frac{w}{z}\right|^2)$, and change variables to $x=|z|^2$ and $y=\left|\frac{w}{z}\right|^2$. We obtain
\begin{align}
    &\frac{\kappa_D}{2\pi}\,\frac{\pi^2\alpha'^2 N_1N_2T_0^2}{32\lvert\sinh(v_{12})\rvert}\, \int \frac{\mathrm{d}^{D-2}q}{(2\pi)^{D-2}} e^{i  q\cdot b} \int_0^1\frac{\mathrm{d} x}{x}x^{\frac{\alpha'}{4}(k+q)^2}(1-x)^{-\alpha'k_{v_1}^2} \int_0^1\frac{\mathrm{d} y}{y}y^{\frac{\alpha'}{4}q^2}(1-y)^{-\alpha'k_{v_2}^2} \nonumber
    \\
    &\quad\times\bigg\{4\cosh(v_{12})(k_{v_1}\epsilon_{v_2}-k_{v_2}\epsilon_{v_1}) \left[\frac{q\cdot\tilde{\epsilon}}{2}+\tilde{\epsilon}_{v_1}k_{v_1}\frac{x}{1-x}-\tilde{\epsilon}_{v_2}k_{v_2}\frac{y}{1-y}\right] \nonumber
    \\
    &\qquad\;\;\;+4\cosh(v_{12})(k_{v_1}\tilde{\epsilon}_{v_2}-k_{v_2}\tilde{\epsilon}_{v_1}) \left[\frac{q\cdot\epsilon}{2}+\epsilon_{v_1}k_{v_1}\frac{x}{1-x}- \epsilon_{v_2}k_{v_2}\frac{y}{1-y}\right] \nonumber
    \\
    &\qquad\;\;\;+(D-4+2\cosh(2v_{12})) \bigg[\frac{(q\cdot\epsilon)(q\cdot\tilde{\epsilon})}{2} - \left(\frac{x}{1-x}\right)^2k_{v_1}^2(\tilde{\epsilon}\cdot\epsilon) - \left(\frac{y}{1-y}\right)^2k_{v_2}^2(\tilde{\epsilon}\cdot\epsilon) \nonumber 
    \\
    &\qquad \qquad +\frac{x}{1-x}\Bigl(k_{v_1}\epsilon_{v_1}(q\cdot\tilde{\epsilon})+k_{v_1}\tilde{\epsilon}_{v_1}(q\cdot\epsilon)-(\tilde{\epsilon}_{v_1}\epsilon_{v_1}+\tfrac12\tilde{\epsilon}\cdot\epsilon)(k\cdot q)\Bigr) \nonumber
    \\
    &\qquad\qquad- \frac{y}{1-y}\Bigl(k_{v_2}\epsilon_{v_2}(q\cdot\tilde{\epsilon})+k_{v_2}\tilde{\epsilon}_{v_2}(q\cdot\epsilon)-(\tilde{\epsilon}_{v_2}\epsilon_{v_2}+\tfrac12\tilde{\epsilon}\cdot\epsilon)(k\cdot q)\Bigr) \nonumber
    \\
    &\qquad\qquad+ \frac{xy}{(1-x)(1-y)}\bigg(2k_{v_1}^2 k_{v_2}^2 \left(\frac{\epsilon_{v_1}}{k_{v_1}} - \frac{\epsilon_{v_2}}{k_{v_2}}\right)\left(\frac{\tilde{\epsilon}_{v_1}}{k_{v_1}} - \frac{\tilde{\epsilon}_{v_2}}{k_{v_2}}\right) + (k_{v_1}^2 + k_{v_2}^2)(\tilde{\epsilon}\cdot \epsilon)\bigg)\bigg]\bigg\}.
\end{align}
Extracting from the $x$- and $y$-integrals the contributions proportional to $1/\alpha'$, and neglecting terms localized at $b=0$, we find
\begin{align}
    &\frac{\kappa_D}{2\pi}\frac{\pi^2N_1N_2T_0^2}{32\lvert\sinh(v_{12})\rvert}\int \frac{\mathrm{d}^{D-2}q}{(2\pi)^{D-2}} e^{i  q\cdot b} \nonumber 
    \\
    &\times \bigg\{4\cosh(v_{12})(k_{v_1}\epsilon_{v_2}-k_{v_2}\epsilon_{v_1}) \left[\frac{8(q\cdot\tilde{\epsilon})}{q^2(k+q)^2}-\frac{4\tilde{\epsilon}_{v_1}}{k_{v_1}}\frac1{q^2}+\frac{4\tilde{\epsilon}_{v_2}}{k_{v_2}}\frac1{(k+q)^2}\right] \nonumber
    \\
    &\qquad+4\cosh(v_{12})(k_{v_1}\tilde{\epsilon}_{v_2}-k_{v_2}\tilde{\epsilon}_{v_1}) \left[\frac{8(q\cdot\epsilon)}{q^2(k+q)^2}-\frac{4\epsilon_{v_1}}{k_{v_1}}\frac1{q^2}+\frac{4\epsilon_{v_2}}{k_{v_2}}\frac1{(k+q)^2}\right] \nonumber
    \\
    &\qquad+(D-4+2\cosh(2v_{12}))\bigg[\frac{8(q\cdot\epsilon)(q\cdot\tilde{\epsilon})}{q^2(k+q)^2} -4\tilde{\epsilon}\cdot\epsilon\left(\frac1{q^2}+\frac1{(k+q)^2}\right)\nonumber 
    \\
    &\qquad\quad -\frac4{k_{v_1}^2q^2}\bigl(k_{v_1}\epsilon_{v_1}(q\cdot\tilde{\epsilon})+k_{v_1}\tilde{\epsilon}_{v_1}(q\cdot\epsilon)-(\tilde{\epsilon}_{v_1}\epsilon_{v_1}+\tfrac12\tilde{\epsilon}\cdot\epsilon)(k\cdot q)\bigr) \nonumber
    \\
    &\qquad\quad + \frac4{k_{v_2}^2(k+q)^2}\bigl(k_{v_2}\epsilon_{v_2}(q\cdot\tilde{\epsilon})+k_{v_2}\tilde{\epsilon}_{v_2}(q\cdot\epsilon)-(\tilde{\epsilon}_{v_2}\epsilon_{v_2}+\tfrac12\tilde{\epsilon}\cdot\epsilon)(k\cdot q)\bigr) \bigg]\bigg\}.\label{A34_final}
\end{align}}

Combining Eqs.~\eqref{A56_final} and \eqref{A34_final}, we obtain the field-theory limit of the NS-NS amplitude:
\begin{align}
    A_{\text{NS-NS}} =&\;\frac{\kappa_D}{2\pi}\frac{\pi^2N_1N_2T_0^2}{32\lvert\sinh(v_{12})\rvert}\int \frac{\mathrm{d}^{D-2}q}{(2\pi)^{D-2}} e^{i  q\cdot b}\nonumber 
    \\
    &\times \bigg\{(D-4+2\cosh(2v_{12})) \bigg[\frac{8(q\cdot\epsilon)(q\cdot\tilde{\epsilon})}{q^2(k+q)^2}-4(\tilde{\epsilon}\cdot\epsilon)\left(\frac1{q^2}+\frac1{(k+q)^2}\right)\nonumber
    \\
    &\qquad-\frac4{k_{v_1}^2q^2}\Bigl(k_{v_1}\epsilon_{v_1}(q\cdot\tilde{\epsilon})+k_{v_1}\tilde{\epsilon}_{v_1}(q\cdot\epsilon)-(\tilde{\epsilon}_{v_1}\epsilon_{v_1}+\tfrac12\tilde{\epsilon}\cdot\epsilon)(k\cdot q)\Bigr)\nonumber
    \\
    &\qquad+\frac4{k_{v_2}^2(k+q)^2}\Bigl(k_{v_2}\epsilon_{v_2}(q\cdot\tilde{\epsilon})+k_{v_2}\tilde{\epsilon}_{v_2}(q\cdot\epsilon)-(\tilde{\epsilon}_{v_2}\epsilon_{v_2}+\tfrac12\tilde{\epsilon}\cdot\epsilon)(k\cdot q)\Bigr) \bigg] \nonumber
    \\
    &\qquad+4\cosh(v_{12})(k_{v_1}\epsilon_{v_2}-k_{v_2}\epsilon_{v_1}) \left[\frac{8(q\cdot\tilde{\epsilon})}{q^2(k+q)^2}-\frac{4\tilde{\epsilon}_{v_1}}{k_{v_1}}\frac1{q^2}+\frac{4\tilde{\epsilon}_{v_2}}{k_{v_2}}\frac1{(k+q)^2}\right] \nonumber
    \\
    &\qquad+4\cosh(v_{12})(k_{v_1}\tilde{\epsilon}_{v_2}-k_{v_2}\tilde{\epsilon}_{v_1}) \left[\frac{8(q\cdot\epsilon)}{q^2(k+q)^2}-\frac{4\epsilon_{v_1}}{k_{v_1}}\frac1{q^2}+\frac{4\epsilon_{v_2}}{k_{v_2}}\frac1{(k+q)^2}\right] \nonumber
    \\
    &\qquad+\frac{32}{q^2(k+q)^2} \left[(k_{v_1}\epsilon_{v_2}-k_{v_2}\epsilon_{v_1})(k_{v_1}\tilde{\epsilon}_{v_2}-k_{v_2}\tilde{\epsilon}_{v_1})+\frac{\tilde{\epsilon}\cdot\epsilon}{2}(k_{v_1}^2+k_{v_2}^2)\right] \nonumber
    \\
    &\qquad+\frac4{k_{v_1}^2q^2}(\tilde{\epsilon}\cdot\epsilon)(k_{v_1+v_{12}}^2-k_{v_2}^2) +\frac4{k_{v_2}^2(k+q)^2}(\tilde{\epsilon}\cdot\epsilon)(k_{v_2+v_{21}}^2-k_{v_1}^2) \bigg\}.
\end{align}
This agrees precisely with the bosonic-string result, given by the sum of Eqs.~\eqref{FinalMassless8} and \eqref{FinalTachyon8}. This agreement is expected, since the massless sector of the bosonic string gives rise to NS-NS gravity.

\subsection{R-R sector}
\label{AppD3}
In this subsection, we compute the field-theory limit of the R-R-sector contribution to the superstring scattering amplitude. The GSO-projected amplitude is
\begin{equation}
    A_{\text{R-R}} = \frac{1}{4}[A_{\text{R-R}}(++) + A_{\text{R-R}}(+-) + A_{\text{R-R}}(-+) + A_{\text{R-R}}(--)],\label{RR_sector_GSO_projection}
\end{equation}
with $A_{\text{R-R}}(\eta_1,\eta_2)$ given in Eq.~\eqref{RRFINAL}. In the limit $|w|\to0$, with $|z|$ and $\left|\frac{w}{z}\right|$ kept fixed, the fermionic matrix element, given by the product of \eqref{RR_nonzero_modes} and \eqref{RR_zero_mode}, together with the factor of $1/2$ arising from the zero modes of the $\beta\gamma$ ghosts, reduces to
\begin{align}
    &8(\eta_1 \eta_2 - 1)\,e^{-i\eta_1|z|(2\tilde{R}_{v_1}R_{v_1} + \tilde{R}\cdot R)\frac{1}{1 - |z|^2}} e^{- \frac{i\eta_2|w|^2}{|z|^3}(2\tilde{R}_{v_2}R_{v_2} + \tilde{R}\cdot R)\frac{1}{1 - |\frac{w}{z}|^2}}\nonumber 
    \\
    &\Big[\cosh(v_{12}) + \frac{i\eta_1}{2}(\cosh(v_{12})(\epsilon\cdot \tilde{\epsilon})\phi\tilde{\phi} + R_{v_1}\tilde{R}_{v_2} + R_{v_2}\tilde{R}_{v_1})\nonumber 
    \\
    &\quad +\frac{1}{2}\sqrt{\frac{\alpha'}{2}}\left(\frac{1}{z}(k_{v_1} \epsilon_{v_2} - k_{v_2} \epsilon_{v_1})\theta\phi + \frac{1}{\bar{z}}(k_{v_1} \tilde{\epsilon}_{v_2} - k_{v_2} \tilde{\epsilon}_{v_1})\tilde{\theta}\tilde{\phi}\right) + \frac{\alpha'}{4|z|^2} k_{v_1}k_{v_2}(\tilde{\epsilon}\cdot \epsilon)\theta\phi\tilde{\theta}\tilde{\phi}\Big].
    \label{POL3}
\end{align}
Expanding the exponentials in the first line of Eq.~\eqref{POL3} in the Grassmann variables $R$ and $\tilde R$, we obtain
\begin{align}
    &e^{-i\eta_1|z|(2\tilde{R}_{v_1}R_{v_1} + \tilde{R}\cdot R)\frac{1}{1 - |z|^2}} e^{- \frac{i\eta_2|w|^2}{|z|^3}(2\tilde{R}_{v_2}R_{v_2} + \tilde{R}\cdot R)\frac{1}{1 - |\frac{w}{z}|^2}}
    \\
    &=1 -i\eta_1|z|(2\tilde{R}_{v_1}R_{v_1} + \tilde{R}\cdot R)\frac{1}{1 - |z|^2}-\frac{i\eta_2|w|^2}{|z|^3}(2\tilde{R}_{v_2}R_{v_2} + \tilde{R}\cdot R)\frac{1}{1 - |\frac{w}{z}|^2} \nonumber 
    \\
    & -\frac{1}{2}\bigg(|z|(2\tilde{R}_{v_1}R_{v_1} + \tilde{R}\cdot R)\frac{1}{1 - |z|^2}\bigg)^2 -\frac{1}{2}\bigg(\frac{|w|^2}{|z|^3}(2\tilde{R}_{v_2}R_{v_2} + \tilde{R}\cdot R)\frac{1}{1 - |\frac{w}{z}|^2}\bigg)^2 \nonumber 
    \\
    & -\eta_1 \eta_2 (2\tilde{R}_{v_1}R_{v_1} + \tilde{R}\cdot R)\frac{1}{1 - |z|^2}\frac{|w|^2}{|z|^2}(2\tilde{R}_{v_2}R_{v_2} + \tilde{R}\cdot R)\frac{1}{1 - |\frac{w}{z}|^2}.
\end{align}
After the GSO projection \eqref{RR_sector_GSO_projection}, only the terms independent of $\eta_1$ and $\eta_2$ survive; terms proportional to either $\eta_1$ or $\eta_2$ cancel in the spin-structure sum. Thus Eq.~\eqref{POL3} reduces to
\begin{align}
    &\nonumber -8\cosh(v_{12}) -\frac{2\alpha'}{|z|^2}k_{v_1}k_{v_2}(\tilde{\epsilon}\cdot \epsilon)\theta\phi\tilde{\theta}\tilde{\phi}
    \\
    &\nonumber-\frac{4}{z}\sqrt{\frac{\alpha'}{2}}(k_{v_1}\epsilon_{v_2} - k_{v_2}\epsilon_{v_1})\theta\phi -\frac{4}{\bar{z}}\sqrt{\frac{\alpha'}{2}}(k_{v_1}\tilde{\epsilon}_{v_2} - k_{v_2}\tilde{\epsilon}_{v_1})\tilde{\theta}\tilde{\phi}
    \\
    &\nonumber +\frac{4}{|z|^2}\frac{|z|^2}{1-|z|^2}(\cosh(v_{12})(\epsilon\cdot \tilde{\epsilon})\phi \tilde{\phi} + R_{v_1}\tilde{R}_{v_2} + R_{v_2}\tilde{R}_{v_1})(2\tilde{R}_{v_1}R_{v_1} + \tilde{R}\cdot R)
    \\
    &\nonumber +\frac{4}{|z|^2}\frac{|\frac{w}{z}|^2}{1 - |\frac{w}{z}|^2}(\cosh(v_{12})(\epsilon\cdot \tilde{\epsilon})\phi \tilde{\phi} + R_{v_1}\tilde{R}_{v_2} + R_{v_2}\tilde{R}_{v_1})(2\tilde{R}_{v_2}R_{v_2} + \tilde{R}\cdot R)
    \\
    \nonumber &+ \frac{4\cosh(v_{12})}{|z|^2} \frac{|z|^4}{(1-|z|^2)^2}(2\tilde{R}_{v_1}R_{v_1} + \tilde{R}\cdot R)(2\tilde{R}_{v_1}R_{v_1} + \tilde{R}\cdot R)
    \\
    \nonumber &+ \frac{4\cosh(v_{12})}{|z|^2}\frac{|w/z|^4}{(1-|w/z|^2)^2}(2\tilde{R}_{v_2}R_{v_2} + \tilde{R}\cdot R)(2\tilde{R}_{v_2}R_{v_2} + \tilde{R}\cdot R)
    \\
    &- \frac{8\cosh(v_{12})}{|z|^2}\frac{|w|^2}{(1-|z|^2)(1-|w/z|^2)} (2\tilde{R}_{v_1}R_{v_1} + \tilde{R}\cdot R)(2\tilde{R}_{v_2}R_{v_2} + \tilde{R}\cdot R) \label{aux_eq_RRfield_theory}
\end{align}
Using the identities \cref{RR_identity,RRRR_identity}, we can evaluate the products involving the variables $R$ and $\tilde R$. In particular, we find
\begin{align}
    (\cosh(v_{12})(\epsilon\cdot \tilde{\epsilon})\phi \tilde{\phi} + R_{v_1}\tilde{R}_{v_2} + R_{v_2}\tilde{R}_{v_1})(2\tilde{R}_{v_1}R_{v_1} + \tilde{R}\cdot R) \nonumber 
    \\
    = \alpha'(\tilde{\epsilon}\cdot \epsilon)\Big[k_{v_1}^2 \cosh(v_{12}) + k_{v_1}k_{v_2}\Big]\theta\phi\tilde{\theta}\tilde{\phi}.
\end{align}
With these identities, Eq.~\eqref{aux_eq_RRfield_theory} can be rewritten as
\begin{align}
    &\nonumber -8\cosh(v_{12}) -\frac{2\alpha'}{|z|^2}k_{v_1}k_{v_2}(\tilde{\epsilon}\cdot \epsilon)\theta\phi\tilde{\theta}\tilde{\phi}
    \\
    &\nonumber +\frac{4\alpha'}{|z|^2}\frac{|z|^2}{1-|z|^2}(\tilde{\epsilon}\cdot \epsilon)\Big[k_{v_1}^2 \cosh(v_{12}) + k_{v_1}k_{v_2}\Big]\theta\phi\tilde{\theta}\tilde{\phi}
    \\
    &\nonumber -\frac{4\alpha'}{|z|^2}\frac{|\frac{w}{z}|^2}{1 - |\frac{w}{z}|^2}(\tilde{\epsilon}\cdot \epsilon)\Big[k_{v_2}^2 \cosh(v_{12}) + k_{v_1}k_{v_2}\Big]\theta\phi\tilde{\theta}\tilde{\phi}
    \\
    &\nonumber -\frac{4}{z}\sqrt{\frac{\alpha'}{2}}(k_{v_1}\epsilon_{v_2} - k_{v_2}\epsilon_{v_1})\theta\phi -\frac{4}{\bar{z}}\sqrt{\frac{\alpha'}{2}}(k_{v_1}\tilde{\epsilon}_{v_2} - k_{v_2}\tilde{\epsilon}_{v_1})\tilde{\theta}\tilde{\phi}
    \\
    \nonumber &- \frac{8\alpha'}{|z|^2} \frac{\cosh(v_{12})|z|^4}{(1-|z|^2)^2} k_{v_1}^2 (\tilde{\epsilon}\cdot \epsilon)\theta\phi\tilde{\theta}\tilde{\phi} - \frac{8\alpha'}{|z|^2}\frac{\cosh(v_{12})|\tfrac{w}{z}|^4}{(1-|\tfrac{w}{z}|^2)^2}k_{v_2}^2 (\tilde{\epsilon}\cdot \epsilon)\theta\phi\tilde{\theta}\tilde{\phi}
    \\
    &+ \frac{16\alpha'\cosh(v_{12})|\tfrac{w}{z}|^2}{(1-|z|^2)(1-|\tfrac{w}{z}|^2)} \Big[k_{v_1}^2 k_{v_2}^2\left(\frac{\epsilon_{v_1}}{k_{v_1}} - \frac{\epsilon_{v_2}}{k_{v_2}}\right)\left(\frac{\tilde{\epsilon}_{v_1}}{k_{v_1}} - \frac{\tilde{\epsilon}_{v_2}}{k_{v_2}}\right) +\frac{(\tilde{\epsilon}\cdot \epsilon)}{2}(k_{v_1}^2 + k_{v_2}^2)\Big]\theta\phi\tilde{\theta}\tilde{\phi}.\label{aux_eq_RRfield_theory1}
\end{align}
We now combine this result with the contribution from the bosonic modes in the R-R sector, namely
\begin{align}
    \nonumber &\frac{\kappa_D}{2\pi} \frac{\alpha' N_1 N_2 T_{0}^2}{32|\sinh(v_{12})|}\int \frac{d^{D-2}{q}}{(2\pi)^{D-2}}e^{i{q} \cdot b} \int\frac{d^2 w }{|w|^2}\int_{|w|}^1    d^2 z \int d\theta \int d\phi \int d\tilde{\theta} \int d\tilde{\phi}
    \\
    &\times (1-|z|^2)^{-2\tilde{A}_{v_1}A_{v_1}-{\tilde{A}}\cdot {A}}(1-|\tfrac{w}{z}|^2)^{-2\tilde{A}_{v_2}A_{v_2}-{\tilde{A}}\cdot{A}}.
\end{align}
Note that in the R-R sector we can keep fewer terms in the small-$|w|$ expansion of the bosonic matrix element than in the bosonic string case. This is because the Ramond sector contains no tachyon, and therefore the partition function has no divergent contribution at $|w|\to0$ that could make higher-order terms in $|w|$ finite in the field-theory limit.

More explicitly, the field-theory limit of the R-R amplitude is
\begin{align}
    \nonumber A_{\text{R-R}} = &\frac{\kappa_D}{2\pi} \frac{\alpha' N_1 N_2 T_{0}^2}{32|\sinh(v_{12})|}\int \frac{d^{D-2}{q}}{(2\pi)^{D-2}}e^{i{q} \cdot b} \int\frac{d^2 w }{|w|^2}\int_{|w|}^1    d^2 z \int d\theta \int d\phi \int d\tilde{\theta} \int d\tilde{\phi}
    \\
    \nonumber &\times (1-|z|^2)^{-2\tilde{A}_{v_1}A_{v_1}-{\tilde{A}}\cdot {A}}(1-|\tfrac{w}{z}|^2)^{-2\tilde{A}_{v_2}A_{v_2}-{\tilde{A}}\cdot{A}}
    \\
    \nonumber &\times \bigg\{-8\cosh(v_{12}) -\frac{2\alpha'}{|z|^2}k_{v_1}k_{v_2}(\tilde{\epsilon}\cdot \epsilon)\theta\phi\tilde{\theta}\tilde{\phi}
    \\
    &\nonumber +\frac{4\alpha'}{|z|^2}\frac{|z|^2}{1-|z|^2}(\tilde{\epsilon}\cdot \epsilon)\Big[k_{v_1}^2 \cosh(v_{12}) + k_{v_1}k_{v_2}\Big]\theta\phi\tilde{\theta}\tilde{\phi}
    \\
    &\nonumber -\frac{4\alpha'}{|z|^2}\frac{|\frac{w}{z}|^2}{1 - |\frac{w}{z}|^2}(\tilde{\epsilon}\cdot \epsilon)\Big[k_{v_2}^2 \cosh(v_{12}) + k_{v_1}k_{v_2}\Big]\theta\phi\tilde{\theta}\tilde{\phi}
    \\
    &\nonumber -\frac{4}{z}\sqrt{\frac{\alpha'}{2}}(k_{v_1}\epsilon_{v_2} - k_{v_2}\epsilon_{v_1})\theta\phi -\frac{4}{\bar{z}}\sqrt{\frac{\alpha'}{2}}(k_{v_1}\tilde{\epsilon}_{v_2} - k_{v_2}\tilde{\epsilon}_{v_1})\tilde{\theta}\tilde{\phi}
    \\
    \nonumber &- \frac{8\alpha'}{|z|^2} \frac{\cosh(v_{12})|z|^4}{(1-|z|^2)^2} k_{v_1}^2 (\tilde{\epsilon}\cdot \epsilon)\theta\phi\tilde{\theta}\tilde{\phi} - \frac{8\alpha'}{|z|^2}\frac{\cosh(v_{12})|\tfrac{w}{z}|^4}{(1-|\tfrac{w}{z}|^2)^2}k_{v_2}^2 (\tilde{\epsilon}\cdot \epsilon)\theta\phi\tilde{\theta}\tilde{\phi}
    \\
    &+ \frac{16\alpha'\cosh(v_{12})|\tfrac{w}{z}|^2}{(1-|z|^2)(1-|\tfrac{w}{z}|^2)} \Big[k_{v_1}^2 k_{v_2}^2\left(\frac{\epsilon_{v_1}}{k_{v_1}} - \frac{\epsilon_{v_2}}{k_{v_2}}\right)\left(\frac{\tilde{\epsilon}_{v_1}}{k_{v_1}} - \frac{\tilde{\epsilon}_{v_2}}{k_{v_2}}\right) +\frac{(\tilde{\epsilon}\cdot \epsilon)}{2}(k_{v_1}^2 + k_{v_2}^2)\Big]\theta\phi\tilde{\theta}\tilde{\phi}\bigg\}.\label{RR_amp_field_thry_lim_aux1}
\end{align}
Now we need to perform the integrals over the Grassmann variables. This is immediate for the terms inside the large curly bracket that are already proportional to $\theta\phi\tilde\theta\tilde\phi$.

A second type of contribution comes from multiplying the $\theta\phi\tilde\theta\tilde\phi$ terms in the bosonic part by the Grassmann-independent term $-8\cosh(v_{12})$ from the fermionic part. This contribution is obtained by extracting the terms bilinear in the polarization tensors, exactly as for the $A_{5}^{\text{massless}}$ term in the bosonic-string calculation. The only difference is that the overall factor $D-4+2\cosh(2v_{12})$ appearing in $A_{5}^{\text{massless}}$ is replaced by the Grassmann-independent term in Eq.~\eqref{aux_eq_RRfield_theory1}, namely $-8\cosh(v_{12})$.

The final type of contribution, which is absent in the bosonic string and NS-NS-sector calculations, arises from the terms
\begin{equation}
    -\frac{4}{z}\sqrt{\frac{\alpha'}{2}}(\epsilon_{v_1}k_{v_2} - \epsilon_{v_2}k_{v_1})\theta\phi -\frac{4}{\bar{z}}\sqrt{\frac{\alpha'}{2}}(\epsilon_{v_1}k_{v_2} - \epsilon_{v_2}k_{v_1})\tilde{\theta}\tilde{\phi}.
\end{equation}
These terms combine with the terms in the bosonic contribution that are linear in the polarizations, namely:
\begin{align}
    \nonumber &\frac{\kappa_D}{2\pi} \frac{\alpha' N_1 N_2 T_{0}^2}{32|\sinh(v_{12})|}\int \frac{d^{D-2}{q}}{(2\pi)^{D-2}}e^{i{q} \cdot b} \int\frac{d^2 w }{|w|^2}\int_{|w|}^1    d^2 z \int d\theta \int d\phi \int d\tilde{\theta} \int d\tilde{\phi}
    \\
    \nonumber &\times |\tfrac{w}{z}|^{\alpha'q^2/2}|z|^{\alpha'(q+k)^2/2} (1-|z|^2)^{-\alpha'k_{v_1}^2}(1-|\frac{w}{z}|^2)^{-\alpha'k_{v_2}^2}
    \\
    \nonumber&\times\sqrt{2\alpha'}\Big[\left(\frac{q\cdot \epsilon}{2z}-k_{v_1}\epsilon_{v_1}\partial_z\log(1-|z|^2) -k_{v_2}\epsilon_{v_2}\partial_z\log(1-|\frac{w}{z}|^2)\right)\phi\theta
    \\
    &\qquad\quad  +\left(\frac{q\cdot \tilde{\epsilon}}{2\bar{z}} - k_{v_2}\tilde{\epsilon}_{v_2}\partial_{\bar{z}}\log(1-|\frac{w}{z}|^2) -  k_{v_1}\tilde{\epsilon}_{v_1}\partial_{\bar{z}}\log(1-|z|^2)\right)\tilde{\phi}\tilde{\theta}\Big]
\end{align}
This gives
\begin{align}
    \nonumber &\frac{\kappa_D}{2\pi} \frac{\alpha' N_1 N_2 T_{0}^2}{32|\sinh(v_{12})|}\int \frac{d^{D-2}{q}}{(2\pi)^{D-2}}e^{i{q} \cdot b} \int\frac{d^2 w }{|w|^2}\int_{|w|}^1    \frac{d^2 z}{|z|^2}\times
    \\
    \nonumber &\times |\tfrac{w}{z}|^{\alpha'q^2/2}|z|^{\alpha'(q+k)^2/2} (1-|z|^2)^{-\alpha'k_{v_1}^2}(1-|\tfrac{w}{z}|^2)^{-\alpha'k_{v_2}^2}
    \\
    \nonumber&\times\Big[4\alpha'(\tilde{\epsilon}_{v_1}k_{v_2} - \tilde{\epsilon}_{v_2}k_{v_1})\left(\frac{q\cdot \epsilon}{2} + k_{v_1}\epsilon_{v_1}\frac{|z|^2}{1-|z|^2} - k_{v_2}\epsilon_{v_2}\frac{|\tfrac{w}{z}|^2}{1-|\tfrac{w}{z}|^2}\right)
    \\
    &\quad\;\;+4\alpha'(\epsilon_{v_1}k_{v_2} - \epsilon_{v_2}k_{v_1})\left(\frac{q\cdot \tilde{\epsilon}}{2} + k_{v_1}\tilde{\epsilon}_{v_1}\frac{|z|^2}{1-|z|^2} - k_{v_2}\tilde{\epsilon}_{v_2}\frac{|\tfrac{w}{z}|^2}{1-|\tfrac{w}{z}|^2}\right)\Big]
\end{align}
The remaining steps of the calculation are closely analogous to those in the previous two subsections, and we can rewrite \eqref{RR_amp_field_thry_lim_aux1} as:
\begin{align}
    \nonumber  A_{\text{R-R}}=&\;\frac{\kappa_D}{2\pi} \frac{\pi^2 \alpha'^2 N_1 N_2 T_0^2 }{64|\sinh(v_{12})|} \int_0^1 d x\int_0^1 d y\int\frac{ d^{D-2}{q}}{(2\pi)^{D-2}}\,
    \\
    &\nonumber \times e^{iqb}x^{\frac{\alpha'}{4}(q+k)^2 - 1}y^{\frac{\alpha'}{4}q^2 - 1}(1-x)^{-\alpha' k_{v_1}^2}(1-y)^{-\alpha' k_{v_2}^2}
    \\
    &\nonumber \times \bigg\{-8\cosh(v_{12})\Big[(q\cdot \epsilon)(q\cdot\tilde{\epsilon}) +\frac{2x^2}{(1-x)^2}k_{v_1}^2(\tilde{\epsilon}\cdot\epsilon) + \frac{2y^2}{(1-y)^2}k_{v_2}^2(\tilde{\epsilon}\cdot\epsilon)
    \\
    \nonumber &\qquad +\frac{2x}{1-x}(k_{v_1}\epsilon_{v_1}(q\cdot\tilde{\epsilon}) + k_{v_1}\tilde{\epsilon}_{v_1}(q\cdot\epsilon) - (q\cdot k)\tilde{\epsilon}_{v_1}\epsilon_{v_1} + \frac{1}{2}(q\cdot k)(\tilde{\epsilon}\cdot\epsilon))
    \\
    \nonumber &\qquad -\frac{2y}{1-y}(k_{v_2}\epsilon_{v_2}(q\cdot\tilde{\epsilon}) + k_{v_2}\tilde{\epsilon}_{v_2}(q\cdot\epsilon) - (q\cdot k)\tilde{\epsilon}_{v_2}\epsilon_{v_2} + \frac{1}{2}(q\cdot k)(\tilde{\epsilon}\cdot\epsilon))\Big]
    \\
    &\nonumber -8(\tilde{\epsilon}\cdot \epsilon)\Big[k_{v_1}^2 \cosh(v_{12}) + k_{v_1}k_{v_2}\Big]\frac{x}{1-x} - 8(\tilde{\epsilon}\cdot \epsilon)\Big[k_{v_2}^2 \cosh(v_{12}) + k_{v_1}k_{v_2}\Big]\frac{y}{1 - y}
    \\
    \nonumber & - 4k_{v_1}k_{v_2}(\tilde{\epsilon}\cdot \epsilon) - 16\cosh(v_{12})k_{v_1}^2 (\tilde{\epsilon}\cdot \epsilon)\frac{x^2}{(1-x)^2} - 16 \cosh(v_{12})k_{v_2}^2 (\tilde{\epsilon}\cdot \epsilon)\frac{y^2}{(1-y)^2}
    \\
    \nonumber &+ \frac{32\cosh(v_{12})xy}{(1-x)(1-y)} \Big[k_{v_1}^2 k_{v_2}^2\left(\frac{\epsilon_{v_1}}{k_{v_1}} - \frac{\epsilon_{v_2}}{k_{v_2}}\right)\left(\frac{\tilde{\epsilon}_{v_1}}{k_{v_1}} - \frac{\tilde{\epsilon}_{v_2}}{k_{v_2}}\right) +\frac{(\tilde{\epsilon}\cdot \epsilon)}{2}(k_{v_1}^2 + k_{v_2}^2)\Big]
    \\
    \nonumber & + 8(k_{v_1} \tilde{\epsilon}_{v_2} - k_{v_2} \tilde{\epsilon}_{v_1})\left(\frac{q\cdot \epsilon}{2} + k_{v_1}\epsilon_{v_1}\frac{x}{1-x} - k_{v_2}\epsilon_{v_2}\frac{y}{1-y}\right)
    \\
    &  +8(k_{v_1} \epsilon_{v_2} - k_{v_2} \epsilon_{v_1})\left(\frac{q\cdot \tilde{\epsilon}}{2} + k_{v_1}\tilde{\epsilon}_{v_1}\frac{x}{1-x} - k_{v_2}\tilde{\epsilon}_{v_2}\frac{y}{1-y}\right)
    \bigg\} \label{RRfield_theory}
\end{align}
Evaluating the $x$- and $y$-integrals using the results from Appendix~\ref{AppInte}, we arrive at the final result
\begin{align}
    \nonumber  A_{\text{R-R}}\simeq&\;\frac{\kappa_{D}}{2\pi}\frac{\pi^2 N_1N_2 T_0^2}{64|\sinh(v_{12})|}\cdot 8\int\frac{ d^{D-2}{q}}{(2\pi)^{D-2}}\,e^{iq\cdot b}\times 
    \\
    &\nonumber \times \Bigg\{-16\cosh(v_{12})\Big[(q\cdot \epsilon)(q\cdot\tilde{\epsilon})\frac{1}{q^2(k+q)^2}+\frac{1}{2q^2}(\tilde{\epsilon}\cdot\epsilon) + \frac{1}{2(k+q)^2}(\tilde{\epsilon}\cdot\epsilon)
    \\
    \nonumber &\qquad -\frac{1}{2k_{v_1}^2q^2}(k_{v_1}\epsilon_{v_1}(q\cdot\tilde{\epsilon}) + k_{v_1}\tilde{\epsilon}_{v_1}(q\cdot\epsilon) - (q\cdot k)\tilde{\epsilon}_{v_1}\epsilon_{v_1} + \frac{1}{2}(q\cdot k)(\tilde{\epsilon}\cdot\epsilon))
    \\
    \nonumber &\qquad + \frac{1}{2k_{v_2}^2(k+q)^2}(k_{v_2}\epsilon_{v_2}(q\cdot\tilde{\epsilon}) + k_{v_2}\tilde{\epsilon}_{v_2}(q\cdot\epsilon) - (q\cdot k)\tilde{\epsilon}_{v_2}\epsilon_{v_2} + \frac{1}{2}(q\cdot k)(\tilde{\epsilon}\cdot\epsilon))\Big]
    \\
    &\nonumber +\frac{4(\tilde{\epsilon}\cdot \epsilon)}{q^2}\Big[\cosh(v_{12}) + \frac{k_{v_2}}{k_{v_1}}\Big] + \frac{4(\tilde{\epsilon}\cdot \epsilon)}{(k+q)^2}\Big[\cosh(v_{12}) + \frac{k_{v_1}}{k_{v_2}}\Big]
    \\
    \nonumber & - \frac{8k_{v_1}k_{v_2}(\tilde{\epsilon}\cdot \epsilon)}{q^2(k+q)^2} - \frac{8\cosh(v_{12}) (\tilde{\epsilon}\cdot \epsilon)}{q^2} - \frac{8\cosh(v_{12})(\tilde{\epsilon}\cdot \epsilon)}{(k+q)^2}
    \\
    \nonumber & + (k_{v_1} \tilde{\epsilon}_{v_2} - k_{v_2} \tilde{\epsilon}_{v_1})\left(\frac{8(q\cdot \epsilon)}{q^2(k+q)^2}-\frac{4\epsilon_{v_1}}{k_{v_1}q^2} + \frac{4\epsilon_{v_2}}{k_{v_2}(k+q)^2}\right)
    \\
    &  + (k_{v_1} \epsilon_{v_2} - k_{v_2} \epsilon_{v_1})\left(\frac{8(q\cdot \tilde{\epsilon})}{q^2(k+q)^2} - \frac{4\tilde{\epsilon}_{v_1}}{k_{v_1}q^2} + \frac{4\tilde{\epsilon}_{v_2}}{k_{v_2}(k+q)^2}\right)
    \Bigg\}.
    \label{FINALR-R}
\end{align}

\section{\texorpdfstring{$\alpha'\to 0$}~~limit of a few integrals}
\label{AppInte}

In this Appendix we compute the field--theory limit of the various integrals that appear in Appendix \ref{AppD}, namely:
\begin{equation}
    I_i= \int_0^1 \frac{dx}{x}\,x^{\frac{\alpha'}{4} (q+k)^2}(1-x)^{-\alpha' k_{v_1}^2}  \bigg( 1, \frac{x}{1-x}, \frac{x^2}{(1-x)^2}, \frac{x}{(1-x)^2}\bigg).\label{TGH8}
\end{equation}
The first integral yields
\begin{equation}
    I_1= \frac{\Gamma(\frac{\alpha'}{4}(q+k)^2)\Gamma(1-\alpha' k_{v_1}^2)}{\Gamma( \frac{\alpha'}{4}(q+k)^2+1-\alpha' k^2_{v_1})} \sim \frac{4}{\alpha'  (k+q)^2}.\label{I11}
\end{equation}
The second integral gives
\begin{equation}
    I_2=\frac{\Gamma(\frac{\alpha'}{4}(q+k)^2 +1)\Gamma(-\alpha' k_{v_1}^2)}{\Gamma( \frac{\alpha'}{4}(q+k)^2+1-\alpha' k^2_{v_1})} \sim- \frac{1}{\alpha'  k^2_{v_1}} \, .
    \label{I22}
\end{equation} 
The third integral gives
\begin{equation}
    I_3= \frac{\Gamma(\frac{\alpha'}{4}(q+k)^2+2)\Gamma(-1-\alpha' k_{v_1}^2)}{\Gamma( \frac{\alpha'}{4}(q+k)^2+1-\alpha' k^2_{v_1})}\sim +\frac{1}{\alpha'   k^2_{v_1}}. \label{I33} 
\end{equation}
Finally, the fourth integral gives
\begin{equation}
    I_4 = \frac{\Gamma (\frac{\alpha'}{4}(k+q)^2+1) \Gamma(-\alpha'k^2_{v_1}-1)}{\Gamma(\frac{\alpha'}{4}(k+q)^2 -\alpha'k^2_{v_1})}
\sim -1+\frac{(q+k)^2}{4k_{v_1}^2}.\label{I44}
\end{equation}

\bibliographystyle{utphys2}
{\small \bibliography{references}{}}

\end{document}